\documentclass[
    aps,
    prl,
    reprint,
    amsmath,amssymb,
    groupedaddress,
    twocolumn,showpacs,
    ]{revtex4-2}
\usepackage{graphicx}
\usepackage{dcolumn}
\usepackage{bm}
\usepackage[mathlines]{lineno}
\usepackage{multirow}
\usepackage{comment}
\usepackage{enumerate}
\usepackage{color}
\usepackage{amsmath}

\newcommand{\mA}{{\mathcal{A}}}

\newcommand{\hatC}{\hat{C}}

\newcommand{\mD}{{\mathcal{D}}}

\newcommand{\bfk}{{\bf k}}
\newcommand{\hatn}{{ \hat{n} }}
\newcommand{\bfq}{{\bf q}}

\newcommand{\mT}{{\mathcal{T}}}
\newcommand{\bfx}{{\bf x}}

\newcommand{\la}{\langle}
\newcommand{\ra}{\rangle}

\newcommand{\mHI}{ {\rm HI} }
\newcommand{\mHII}{ {\rm HII} }
\newcommand{\mHeIII}{ {\rm HeIII} }
\newcommand{\kunit}{{\, h/{\rm Mpc} }}

\begin{document}

\title{Proposal to construct the dark-matter-only counterpart of the observed Universe combining weak lensing and baryon censuses}


\author{Shuren Zhou $^{1,2,3,4}$}
 \email{zhoushuren@sjtu.edu.cn}
\author{Pengjie Zhang $^{2,1,3,4}$}
 \email{zhangpj@sjtu.edu.cn}
\affiliation{
$^{1}$ Tsung-Dao Lee Institute, Shanghai Jiao Tong University, Shanghai 201210, China  \\
$^{2}$ School of Physics and Astronomy, Shanghai Jiao Tong University, Shanghai 200240, China  \\
$^{3}$ State Key Laboratory of Dark Matter Physics, Shanghai 200240, China  \\
$^{4}$ Key Laboratory for Particle Astrophysics and Cosmology (MOE)/Shanghai Key Laboratory for Particle Physics and Cosmology, Shanghai 200240, China  }

\date{\today}


\begin{abstract}
Baryonic effects such as AGN feedback can significantly impact the matter clustering, are harder to model from first principles, and emerge as a severe limiting factor in  weak lensing cosmology. To tackle this issue, we propose a generic relation of mapping the observed matter clustering to its counterpart in a dark-matter-only universe. We verify this relation to be accurate at better than $1\%$ level at $k<1h/$Mpc and $z\in [0,3]$ in both TNG and Illustris simulations, demonstrating its model-independence to the underlying baryonic physics. Implementing this relation in observations will be made possible by the specifically designed cross-correlation statistics and baryon census (ionized diffuse gas through localized fast radio bursts, stellar mass through galaxy surveys, and neutral hydrogen through 21cm mapping). It is capable of correcting the baryonic effect not only in the matter power spectrum, but also at the field level, as demonstrated by tests on the scattering transform statistics. This approach paves the way for constructing the dark-matter-only counterpart of the observed Universe, establishing an ideal cosmological laboratory for probing the dark universe. 

\end{abstract}


\maketitle


{\bf Introduction.--}
Forthcoming surveys will measure weak gravitational lensing and therefore the matter clustering with unprecedented precision, shedding light on fundamental cosmological physics such as dark energy and gravity. However, the scientific returns are limited by baryonic effects  such as gas cooling, AGN and SN feedback. These processes alter the  matter clustering in complicated ways (cold gas, \cite{white2004baryons}; hot gas, \cite{zhan2004effect}; hydrodynamical simulations \cite{jing2006influence, rudd2008effects, semboloni2011quantifying, schaye2023flamingo}) and can reach $\sim 10\%$ at $k=1 \kunit$. Given the expected $S/N\gtrsim 400$ of the weak lensing  measurements by LSST \cite{ivezic2019lsst} and CSST \cite{yao2024csst}, this baryonic effect will become a leading systematic error. It also significantly impacts other large-scale structure probes.

One potential solution for baryonic effects is cosmological hydrodynamical simulations, which incorporate baryon physics in subgrid \cite{le2014towards, nelson2015illustris, schaye2015eagle, mccarthy2016bahamas, chisari2018impact, nelson2019illustristng, schaye2023flamingo, pakmor2023millenniumtng, salcido2023sp, schaye2025colibre}. But due to the lack of a first-principle treatment for baryon processes, different simulations yield diverse results \cite{sharma2025field, bigwood2025kinetic}, and exhibit varying degrees of tension with observations \cite{hadzhiyska2024evidence, popesso2024hot, siegel2025joint}. 
Alternative strategies attempt to post-process dark-matter-only (DMO) simulations to mimic baryonic effects phenomenologically \cite{schneider2015new, zennaro20241, wayland2025calibrating, dai2018gradient, arico2020modelling, arico2021bacco, schneider2022constraining, schneider2025baryonification, bigwood2024weak, pandey2025constraints, chen2023constraining, carlos2024cosmic, giri2021emulation, arico2023y3, bigwood2025confronting}, which require observational calibrations and induce model dependence. 
Observations of the Sunyaev-Zel'dovich effects \cite{chen2023thermal, hadzhiyska2024evidence, guachalla2025backlighting}, X-ray emission \cite{ferreira2024x, bulbul2024srg, popesso2024hot, siegel2025joint}, and various baryon censuses \cite{shull2012baryon, mcquinn2013locating, nicastro2018observations, macquart2020census, driver2021challenge, connor2025gas} have significantly advanced our understanding of baryonic processes. However, each of these probes is sensitive to baryons in specific forms and on limited scales. Either inferences using a single probe or joint fits of multiple probes inevitably rely on modeling baryons, whose uncertainties would eventually propagate to the cosmological inferences. 
 
Given the stringent requirements of probing fundamental physics with weak lensing, it is imperative to develop alternative methods to correct baryonic effects. In this paper, we propose an observation-based method to accurately correct baryonic effects, independent of modeling baryonic physics. As this approach is carried out at the field level, it essentially constructs a DMO universe, the counterpart of the observed Universe.

{\bf A generic relation to correct baryonic effects.--}
The complete treatment of baryonic effects requires a direct relation between the matter clustering of $\delta_m$ and its counterpart of $\delta_{\rm DMO}$ in a hypothetical DMO universe, where the former is in principle an observable and the latter is fully predicted by gravity theory. $\delta_m = \Omega_m^{-1}\left( \Omega_c\delta_c+\Omega_b\delta_b \right)$, where $\delta_{c}$ and $\delta_b$ are the cold dark matter and baryon overdensities in the real Universe, respectively. Although $\delta_b$ can deviate significantly from $\delta_c$ and $\delta_{\rm DMO}$, the difference between $\delta_c$ and $\delta_{\rm DMO}$ should be much smaller because of $\Omega_b\ll \Omega_c$. 

To establish the $\delta_m$-$\delta_{\rm DMO}$ relation, we assume that ({\bf I}) \textit{baryonic effects do not alter the phase of the matter overdensity relative to its DMO counterpart}. If so, the mapping between $\delta_{\rm DMO}$ and $\delta_m$ is fully determined by a transfer function $\mT(k)$ in Fourier space, with $k=|\bfk|$.
\begin{equation}\label{equ:def-T}
    \delta_{\rm DMO}(\bfk) = \mT(k)\, \delta_m(\bfk)  \;.
\end{equation}
Under this assumption, $\mT$ fully encapsulates the baryonic effects on the matter clustering. 

\begin{figure}
\includegraphics[width=0.92\columnwidth]{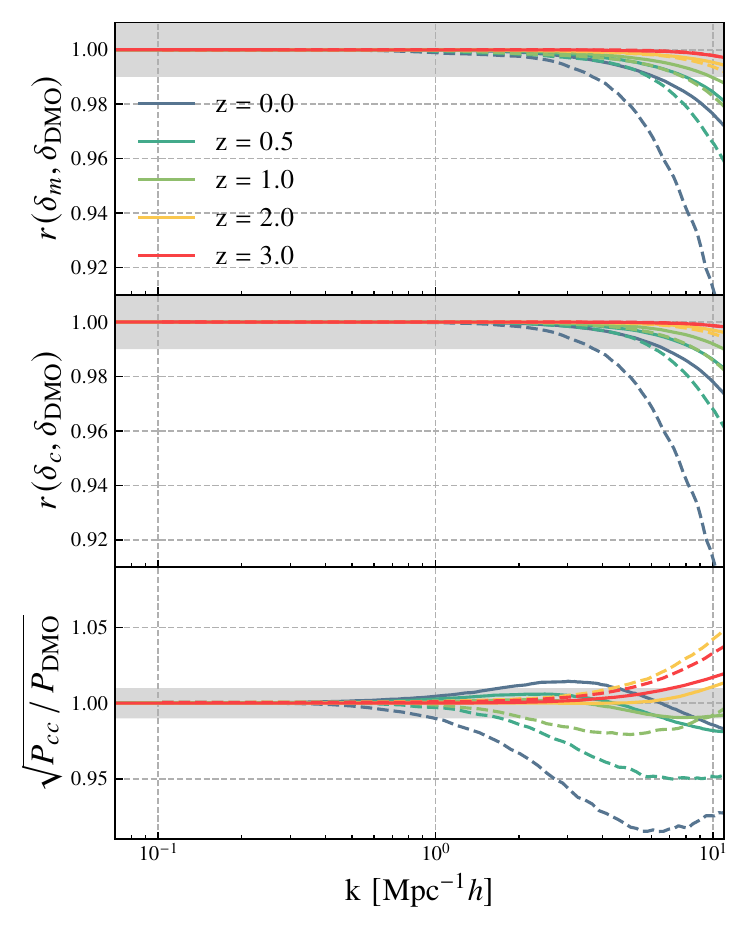}
\caption{ \label{fig:T-r2} 
Validation of the assumptions ({\bf I}) and ({\bf II}) using TNG300-1 (\textit{solid} lines) and Illustris-1 (\textit{dashed} lines) simulations. The shaded regions mark the $\pm 1\%$ deviation. 
\textit{Top}: Cross-correlation coefficients between total matter overdensity $\delta_m$ in hydrodynamical simulation and its DMO counterpart $\delta_{\rm DMO}$. 
\textit{Middle}: Cross-correlation coefficients between cold dark matter overdensity $\delta_c$ and $\delta_{\rm DMO}$. 
\textit{Bottom}: Ratio of the power spectrum $P_{cc}$ of $\delta_c$ to $P_{\rm DMO}$ of $\delta_{\rm DMO}$. 
Despite the different subgrid physics implemented in two simulations, both confirm assumption ({\bf I}) to be $<0.1\%$ accuracy and assumption ({\bf II}) to be $<1\%$ accuracy at scales $k<1\kunit$. 
}
\end{figure}
To solve for $\mT$, we further assume that ({\bf II}) \textit{the $\delta_{\rm DMO}$ field can be approximated by the cold dark matter overdensity field $\delta_c$}. It leads to 
\begin{equation}\label{equ:DMO-c}
\delta_{\rm DMO} \simeq \delta_c = (\Omega_m\delta_m - \Omega_b\delta_b)/\Omega_c  \;.
\end{equation} 
Baryons can be categorized into three major components: ionized diffuse gas, neutral gas, and stellar content. All these components are direct observables in principle, and the total baryon overdensity is thereby \cite{zhou2026cavendish}
\begin{equation}\label{equ:delta_b-xxx}
\delta_b = { 1+\eta \over 1+ {1\over 2}\eta } f_e\delta_e + f_*\delta_* + (1+\eta) f_\mHI\delta_\mHI   \;.
\end{equation}
Here, $f_{*} \equiv \Omega_*/\Omega_b$ and $f_\mHI \equiv \Omega_\mHI/\Omega_b$ are the baryon mass fraction of the species. {\color{black}$\eta=X_{\rm He}/X_{\rm H}$,} where $X_{\rm H}\simeq 0.76$ and $X_{\rm He}\simeq 0.24$ are mass abundances of hydrogen and helium elements.  The ionized electrons overdensity $\delta_e$ is probed by the dispersion measure of fast radio bursts (FRBs), which is the major contribution in the baryon budget of fraction $f_e \equiv f_\mHII + {1\over 2}f_\mHeIII \sim 0.8$. The subdominant components are stars and stellar remnants overdensity $\delta_*$ of fraction $f_*\sim 5\%$, and neutral hydrogen overdensity $\delta_\mHI$ of fraction $f_\mHI\sim 5\%$ traced by 21cm lines \cite{fukugita2004cosmic}.
Substituting Eq.~(\ref{equ:DMO-c})\&(\ref{equ:delta_b-xxx}) into Eq.~(\ref{equ:def-T}), we obtain 
\begin{equation} \label{equ:T}
\hat\mT(k) =  {\Omega_m\over\Omega_c} - {\Omega_b\over\Omega_c}
\left[  { 1+\eta \over 1+ {1\over 2}\eta } f_e b_e + f_*b_* + (1+\eta) f_\mHI b_\mHI  \right]    \,.
\end{equation}
Here $b_i$ denotes the scale-dependent baryon bias of the $i$ species. 

The relations of Eq.~(\ref{equ:T}) together with Eq.~(\ref{equ:def-T}) are the major results of this paper. It maps $\delta_m$ in the real universe to its counterpart $\delta_{\rm DMO}$ in a hypothetical DMO universe, using $\hat{\mT}$ obtained from observational baryon censuses. It not only corrects baryonic effects in the matter power spectrum ($P_{mm}\rightarrow \hat\mT^2P_{mm}$), but also at the field level ($\delta_m({\bf k})\rightarrow \hat\mT(k)\delta_m({\bf k})$).  Next, we validate these relations in hydrodynamical simulations before we discuss their implementation in observations.

\begin{figure}
\includegraphics[width=\columnwidth]{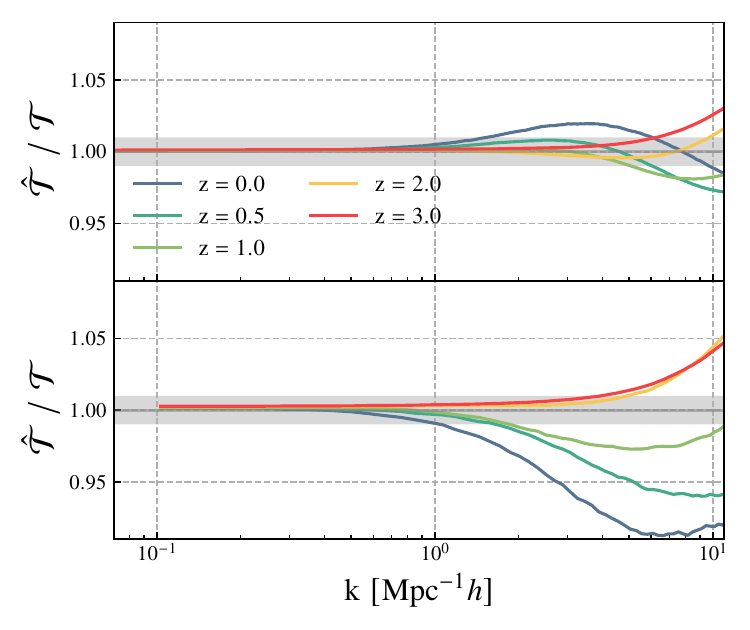}
\caption{ \label{fig:T-ratio} 
Validation of the key relation Eq.~(\ref{equ:T}). We show the ratio of the estimated transfer function $\hat\mT$ to the ground truth $\mT = \sqrt{P_{\rm DMO} /P_{mm}}$, measured in TNG300-1 (\textit{top}) and Illustris-1 (\textit{bottom}) simulations. 
Luminous subhalos are identified as galaxies and cross-correlated with the ionized electrons ($\delta_e$), stars $+$ black holes ($\delta_*$), and neutral hydrogen ($\delta_\mHI$) to measure $\{ f_e b_e, f_* b_*, f_\mHI b_\mHI \}$, and hence $\hat\mT$. 
These results indicate that $\hat\mT$ agrees with the true $\mT$ within $< 1\%$ accuracy on scales $k< 1\kunit$ across redshift range $0<z<3$, regardless of the strength of baryonic effects. 
}
\end{figure}

\begin{figure*}
\includegraphics[width=0.85\textwidth]{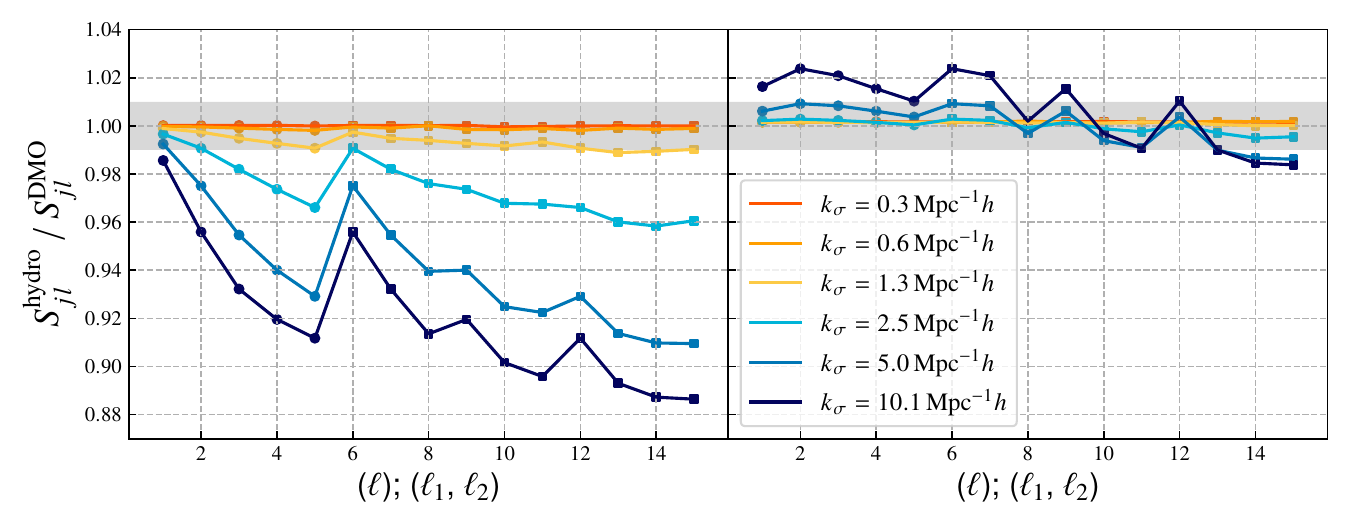}
\caption{ \label{fig:st_coeff}
Validation of the field-level correction capability. We take the scattering transform as an example, which contains higher-order correlations and phase information. 
\textit{Left}: Ratio of coefficients from $\delta_m$ in Illustris-1 simulation at $z=0.5$ to its DMO counterpart $\delta_{\rm DMO}$. 
\textit{Right}: Ratio of coefficients from those with the $\mT$ correction, $\hat\delta_{\rm DMO}(\bfx) = \hat\mT* \delta_m(\bfx)$, to $\delta_{\rm DMO}$. 
We measure the first two order coefficients, $S_{jl}$ and $S_{j_1l_1;j_2l_2}$, and set $j_1=j_2 \;\&\; l_1> l_2$ in $S_{j_1l_1;j_2l_2}$ to reduce data size \cite{cheng2020new}. 
In each panel, different lines represent spatial index $j \in\{ 0, 1, 2, 3, 4, 5\}$, corresponding to the filter size $k_\sigma = 2\pi/\sigma_j$ in Fourier space. 
The horizontal coordinates represent angular indexes $\ell\in\{ 0, 1, 2, 3, 4 \}$, sorted as $\{ {\ell=0},\; {\ell=1},\; \cdots,\; (\ell_1,\ell_2)=(1, 0),\; (\ell_1,\ell_2)=(2, 0),\; \cdots \}$. 
The accurate removal of baryonic effects down to resolution $k_\sigma \lesssim 2\kunit$ demonstrates the effectiveness of the $\hat\mT$ correction at the field level.  
}
\end{figure*}

{\bf Simulation  validations.--}
We validate the assumption ({\bf I}), ({\bf II}) and  Eq.~(\ref{equ:T}) using hydrodynamical simulations TNG300-1 \cite{Springel_2017, Nelson_2017, Pillepich_2017, Naiman_2018, Marinacci_2018}, Illustris-1 \cite{vogelsberger2014properties, vogelsberger2014introducing, genel2014introducing, sijacki2015illustris}, and their DMO counterparts. To accommodate the observational implementation discussed later, the baryon bias $b_i$ in Eq.~(\ref{equ:T}) is measured by the cross-correlations with galaxy overdensity $\delta_g$, i.e., $b_i(k) \equiv P_{\rm gi}(k)/P_{\rm gm}(k)$, where the galaxy sample is chosen as all luminous subhalos in simulations. 
The validity of the assumption ({\bf I}) is quantified by the cross-correlation coefficient $r(\delta_m, \delta_{\rm DMO})$, shown in Fig.~\ref{fig:T-r2}. At the benchmark scale of $k=1\kunit$, the TNG300-1 results show no visible deviation, and the largest deviation in Illustris-1 occurs at $z=0$ where $1-r \simeq 0.1\%$.
In Ref.~\cite{sharma2025field}, the measurements from thousands of hydrodynamical simulations present a more comprehensive justification of the assumption ({\bf I}). In the appendix, we provide {\color{black}an illustrative example demonstrating two contrasting behaviors of matter clustering}: the insensitivity of the phase and the sensitivity of the amplitude to baryonic effects. 

To validate the assumption ({\bf II}), we measure $r(\delta_c, \delta_{\rm DMO})$ and $\sqrt{P_{cc}/P_{\rm DMO}}$ to quantify the phase and amplitude differences between $\delta_c(\bfk)$ and $\delta_{\rm DMO}(\bfk)$ respectively (Fig.~\ref{fig:T-r2}). The phases are nearly identical, with $1 - r < 0.1\%$ at $k<1\kunit$ for both simulations. The amplitudes deviate more significantly, particularly in Illustris-1. This larger discrepancy arises from the stronger AGN feedback implemented in Illustris-1, which substantially modifies the matter distribution, especially baryons residing in halos \cite{genel2014introducing, popesso2024hot}. Nevertheless, even under such aggressive feedback, the maximum deviation remains below $\sim 1\%$ at $z=0$ around $k\sim 1\kunit$, confirming the robustness of the assumption ({\bf II}) on scales $k\lesssim 1\kunit$. 
It is a consequence of $\Omega_c\gg \Omega_b$ and the delayed response of cold dark matter clustering to baryon processes, as their interaction occurs solely through gravitational potential changes, rather than through direct feedback mechanisms. 

The validations of these assumptions already ensure the robustness of the $\hat\mT$ derivation. Furthermore, direct measurements (Fig.~\ref{fig:T-ratio}) in simulations confirm $\hat\mT=\mT$ within $1\%$ accuracy on scales $k< 1\kunit$ and $z\in[0,3]$, despite significantly different feedback strengths adopted in TNG300-1 and Illustris-1. Here  $\mT=\sqrt{P_{\rm DMO}/P_{mm}}$ is the ground truth. It explicitly indicates that the baryonic effect on the matter power spectrum can be corrected accurately at $k<1 \kunit$. On smaller scales $k=1\sim 10\kunit$, deviations increase to $\sim 2\%$ in TNG300-1 and $\sim 8\%$ in Illustris-1 at $z=0$. The primary source of these deviations is the assumption ({\bf II}), since {\color{black} the back-reaction of} baryons indeed modifies the clustering amplitude of cold dark matter on Mpc and smaller scales (Fig.~\ref{fig:T-r2}). 
{\color{black}In the appendix, we further show that massive-neutrino clustering can be incorporated by simply replacing $\Omega_c$ with $\Omega_{c\nu}\equiv \Omega_c+\Omega_\nu$ in Eq.~(\ref{equ:T}), while preserving the validated accuracy in Fig.~\ref{fig:T-ratio}. }

We emphasize that by design, our method removes baryonic effects at the field level. Therefore, baryonic effects in other summarized statistics are removed as well. To illustrate, we take the scattering transform \cite{mallat2012group, bruna2013invariant, eickenberg2018solid, cheng2020new} as an example of beyond 2-point statistics. We adopt the Gaussian-Legendre wavelet filter to extract features of the 3D isotropic field, $\psi_{jl}(\bfx) = \exp[-r^2 /(2\sigma_j^2) ]\, r^l {\rm P}_l (\cos\theta)$. The Gaussian function determines the spatial correlation by scaling $\sigma_j \equiv \sigma/2^j$, while the Legendre polynomials ${\rm P}_{l}$ determine the angular correlation. The scattering transform coefficients of the stochastic field, e.g., $I_0 = \delta_m(\bfx)$, are obtained following, 
\begin{eqnarray}
S_{jl} &=& \la I_{jl} \qquad\;\ra_{\bfx} = \la |I_0 *\psi_{jl}| \ra_{\bfx}     \;, \nonumber\\
S_{j_1l_1;j_2l_2} &=& \la I_{j_1l_1;j_2l_2} \ra_{\bfx} = \la |I_0 *\psi_{j_1l_1}| *\psi_{j_2l_2} | \ra_{\bfx}   \;.
\end{eqnarray}
Here, $*$ denotes the convolution operator, and $\la\cdots\ra_\bfx$ indicates averaging over all real-space pixels. The non-linear operations in the scattering transform enable the extraction of higher-order correlations and phase information. 

In Fig.~\ref{fig:st_coeff}, we present the Illustris-1 results at $z=0.5$, in which baryonic effects suppress the scattering coefficients up to $\sim 10\%$. After applying the transfer function to remap the matter field, i.e., $\hat\delta_{\rm DMO}(\bfx) = \hat\mT* \delta_m(\bfx)$ by Eq.~(\ref{equ:def-T})\&(\ref{equ:T}), the scattering coefficients of the reconstructed field $\hat\delta_{\rm DMO}$ agree with those measured in $\delta_{\rm DMO}$, down to resolution of $k_\sigma\equiv 2\pi/\sigma_j \lesssim 2 \kunit$. For $k_\sigma \gtrsim 5\kunit$, the scattering coefficients incorporate the small-scale modes that are not fully recovered by $\hat\mT$, resulting in increasing deviations with larger $k_\sigma$. It is consistent with the accuracy of $\hat\mT$ shown in Fig.~\ref{fig:T-ratio}. Nevertheless, even for $k_\sigma\simeq 10 \kunit$, the deviation is smaller than $2\%$. This example demonstrates that $\hat\mT$  corrects baryonic impacts on non-Gaussian statistics/phase,  validating its capability to remove baryonic effects at the field level.

{\bf Observational implementation.--} 
In observations, the clustering of matter and baryons is often measured from 2D projected fields on the sky. 
{\color{black}
Within a sufficiently narrow redshift slice, the relations in Eqs.~(\ref{equ:def-T})-(\ref{equ:T}) in Fourier space of $\bfk$ correspond directly to those in harmonic space of $\ell$ by $k = (\ell +{1\over 2})/\chi$, where $\chi$ is the comoving distance to the redshift slice \cite{limber1953analysis, loverde2008extended}.
A robust approach for cosmological inference involves applying the measured $\hat\mT$ as a baryonic correction to the DMO model predictions, where the DMO results can be accurately derived from gravity theory. In this forward-modeling framework, uncertainties in both weak lensing statistics and $\hat\mT$ function propagate directly to the likelihood, while residual contamination at high-$k$ is treated as controlled modeling uncertainty, as in standard analysis pipelines \cite{abbott2026dark, stolzner2025kids, li2023hyper}. 
An alternative application is to map the observed matter distribution to its DMO counterpart. One can first conduct 3D weak lensing to reconstruct the 3D $\delta_m$ fields \cite{taylor2001imaging, hu2002three, bacon2003mapping}, and then the transfer function $\mT$ maps the reconstructed field to its DMO counterpart. After averaging over a chosen redshift range, we obtain the projected matter field in the DMO universe. 
}

To measure $\hat\mT$ in observations, we combine 5 probes $X\in\{\Delta_g, \kappa, \mD, \Sigma_*, T_b\}$: the galaxy overdensity $\Delta_g$, weak lensing convergence $\kappa$, dispersion measure $\mD$ of FRBs, stellar surface density $\Sigma_*$, and surface brightness temperature $T_b$ of 21cm lines. 
They are projected fields and related to the underlying 3D overdensity $Y\in\{\delta_g,\, \delta_m,\, \delta_e, \delta_*, \delta_\mHI\}$ by $X(\hat{n})=\int W_X(\chi)\,Y(\hat{n},\chi)\, d\chi$, where $W_X$ is the kernel function. 
The auxiliary galaxy data $\Delta_g$ is crucial in extracting the redshift information of weak lensing and dispersion measure arising from the intergalactic medium, so its redshift distribution should be sufficiently narrow. The combination $f_ib_i$ in Eq.~(\ref{equ:T}) is therefore measured by 
\begin{equation} \label{equ:fb_esti}
\widehat{f_ib_i}(\ell) = \hat\mA_i\, { \hatC^{gX_i}_\ell \over \hatC^{g\kappa}_\ell }   \;.
\end{equation}
Here the angular power spectra $C_\ell^{AB}$ are defined through  $\la A_\ell B_{\ell'}^*\ra = \delta^D_{\ell\ell'} C^{AB}_\ell$. The normalization factor $\mA_i$ of the $i$ species only relies on the cosmological parameters, and their exact expressions of $\mA_e, \mA_*, \mA_\mHI$ are provided in the appendix. 
Utilizing the tight constraint of cosmological parameters from CMB observations \cite{aghanim2020planck, louis2025atacama} and BAO surveys \cite{alam2021completed, karim2025desi}, we can accurately determine the amplitude $\mA_i$ in Eq.~(\ref{equ:fb_esti}) and the energy fractions $\Omega_i$ in Eq.~(\ref{equ:T}), thereby fully measuring the transfer function $\mT$ in observations, independent of model assumptions.

{\bf Observation prospects.--}
Two auxiliary probes in $\mT$ measurement, galaxy clustering $\Delta_g$ and weak lensing $\kappa$, are being precisely measured in current galaxy surveys such as DESI \cite{adame2024validation} and Rubin \cite{ivezic2019lsst}, and further improved by surveys like CSST \cite{yao2024csst}  and MUST \cite{zhao2024multiplexed}. The limiting factor is baryon census using $\mD$, $\Sigma_*$, and $T_b$.

The dispersion measure of FRBs \cite{zhang2023physics, hussaini2025correlation, wang2025measurement} directly probes free electrons in the ionized diffuse gas, which represents the majority of cosmic baryons \cite{macquart2020census,connor2025gas, leung2025nulling}. Thus $f_e b_e$ is key in determining $\hat\mT$. Its measurement utilizes cross-correlation statistics between $\mD$ and galaxies at lower redshifts. This naturally separates the intergalactic medium contribution from host galaxy contamination. To identify the host galaxy redshifts, FRB localization is required. 
Given the significant event rate \cite{fialkov2017fast} and advances of high angular resolution and wide field-of-view FRB surveys (e.g., DSA-2000 \cite{hallinan2019astro2020} and BURSTT \cite{lin2022burstt}), a sample of $\sim 10^5$ localized FRBs is achievable. Adopting the same redshift distribution in Ref.~\cite{zhou2026cavendish} and bin size $\Delta\log\ell = 0.3$, this moderate estimate of FRB sample size in combination with a DESI-like survey already achieves  cross-correlation signal-to-noise ratio $C^{g\mD}_\ell / \sigma_{C_\ell} \sim 10$ over modes of $0.1\sim 1\kunit$. 
Since ionized diffuse electrons are the dominant contribution in $\hat\mT$, the limited FRB sample size primarily determines the measurement accuracy of $\mT$. $\delta\mT /\mT  \sim \Omega_b/\Omega_c\, \delta(f_e b_e)  \sim \Omega_b/\Omega_c\, \sigma_{C_\ell} / C^{g\mD}_\ell  \sim 0.01$, satisfying the stringent requirement of baryonic effect correction.

The contributions from stars and stellar remnants $f_*b_*$ and neutral hydrogen $f_\mHI b_\mHI$ are subdominant. Given that $f_* +f_\mHI \sim 0.1$ and $b_*, b_\mHI \gtrsim 1$, neglecting these components would introduce a systematic bias at the $\gtrsim 1\%$ level. This shift is negligible in the previous surveys, but exceeds the statistical limit in LSST-like surveys. As argued in Ref.~\cite{zhou2026cavendish} in the context of a cosmological Cavendish experiment, both $f_*b_*$ and $f_\mHI b_\mHI$ would be measured precisely with galaxy surveys and 21cm surveys. We refer the readers to Ref.~\cite{zhou2026cavendish} for a detailed discussion. Together with Ref.~\cite{zhou2026cavendish}, the combination of weak lensing, galaxy surveys, and localized FRBs sheds light on both gravitational and baryonic physics.

{\bf Discussions and conclusions.--}
In summary, we derive a generic relation that maps the observed matter density field to its DMO counterpart. Its implementation in observations will be realized by multiple probes, including weak lensing, galaxy surveys, and localized FRBs. The relation is validated to be accurate at better than $1\%$ level for scales $k<1\kunit$ across the redshift range $0<z<3$, and enables the field-level corrections of baryonic effects independent of modeling baryon physics. This will recover information in the non-linear regime obscured by baryonic physics, and significantly enhance the constraining power of weak lensing cosmology.
{\color{black}
Nevertheless, our methodology implicitly supposes that only the matter component is clustered, so it is not directly applicable to scenarios involving clustered dark energy. Its observational implementation of Eq.~(\ref{equ:fb_esti}) also assumes that the effective gravitational constant of light propagation is equivalent to that governing matter dynamics, which may not hold for specific classes of modified gravity models. 
Extensions to accommodate these scenarios require additional ingredients and are reserved for future work.
}


During the preparation of this work, a proposal employing FRBs to null baryonic effects was published \cite{leung2025nulling}. Both works use FRBs as a primary source of information on baryonic effects and demonstrate the great potential of baryon census enabled by emerging probes. Nonetheless, the two works differ in methodology, observational implementation, and performance, and are therefore complementary. 
{\color{black}The distinguishing merit of the proposed $\hat\mT$ is that it removes baryonic effects at the field level in addition to the power spectrum level, and the cross-correlation technique naturally enables redshift tomography. Moreover, stellar content and neutral hydrogen are included to provide a more complete picture of the baryon census, further reinforcing the observational solution of baryonic systematics in weak lensing cosmology. }

{\bf Acknowledgments.}---
This work is supported by the National Key R\&D Program of China (2023YFA1607800, 2023YFA1607801), 
the National Natural Science Foundation of China (NSFC grant No.12595310), 
and the Fundamental Research Funds for the Central Universities. 
Shuren Zhou is also supported by the T.D. Lee Scholarship. 
This work made use of the Gravity Supercomputer at the Department of Astronomy, Shanghai Jiao Tong University. 
We acknowledge the Virgo Consortium for making their simulation data available. The FLAMINGO simulations were performed using the Durham Memory Intensive system managed by the Institute for Computational Cosmology on behalf of the STFC DiRAC facility (www.dirac.ac.uk).


\bibliographystyle{apsrev4-2}
\bibliography{citations}

\begin{thebibliography}{112}%
\makeatletter
\providecommand \@ifxundefined [1]{%
 \@ifx{#1\undefined}
}%
\providecommand \@ifnum [1]{%
 \ifnum #1\expandafter \@firstoftwo
 \else \expandafter \@secondoftwo
 \fi
}%
\providecommand \@ifx [1]{%
 \ifx #1\expandafter \@firstoftwo
 \else \expandafter \@secondoftwo
 \fi
}%
\providecommand \natexlab [1]{#1}%
\providecommand \enquote  [1]{``#1''}%
\providecommand \bibnamefont  [1]{#1}%
\providecommand \bibfnamefont [1]{#1}%
\providecommand \citenamefont [1]{#1}%
\providecommand \href@noop [0]{\@secondoftwo}%
\providecommand \href [0]{\begingroup \@sanitize@url \@href}%
\providecommand \@href[1]{\@@startlink{#1}\@@href}%
\providecommand \@@href[1]{\endgroup#1\@@endlink}%
\providecommand \@sanitize@url [0]{\catcode `\\12\catcode `\$12\catcode `\&12\catcode `\#12\catcode `\^12\catcode `\_12\catcode `\%12\relax}%
\providecommand \@@startlink[1]{}%
\providecommand \@@endlink[0]{}%
\providecommand \url  [0]{\begingroup\@sanitize@url \@url }%
\providecommand \@url [1]{\endgroup\@href {#1}{\urlprefix }}%
\providecommand \urlprefix  [0]{URL }%
\providecommand \Eprint [0]{\href }%
\providecommand \doibase [0]{https://doi.org/}%
\providecommand \selectlanguage [0]{\@gobble}%
\providecommand \bibinfo  [0]{\@secondoftwo}%
\providecommand \bibfield  [0]{\@secondoftwo}%
\providecommand \translation [1]{[#1]}%
\providecommand \BibitemOpen [0]{}%
\providecommand \bibitemStop [0]{}%
\providecommand \bibitemNoStop [0]{.\EOS\space}%
\providecommand \EOS [0]{\spacefactor3000\relax}%
\providecommand \BibitemShut  [1]{\csname bibitem#1\endcsname}%
\let\auto@bib@innerbib\@empty
\bibitem [{\citenamefont {White}(2004)}]{white2004baryons}%
  \BibitemOpen
  \bibfield  {author} {\bibinfo {author} {\bibfnamefont {M.}~\bibnamefont {White}},\ }\href@noop {} {\bibfield  {journal} {\bibinfo  {journal} {Astroparticle Physics}\ }\textbf {\bibinfo {volume} {22}},\ \bibinfo {pages} {211} (\bibinfo {year} {2004})}\BibitemShut {NoStop}%
\bibitem [{\citenamefont {Zhan}\ and\ \citenamefont {Knox}(2004)}]{zhan2004effect}%
  \BibitemOpen
  \bibfield  {author} {\bibinfo {author} {\bibfnamefont {H.}~\bibnamefont {Zhan}}\ and\ \bibinfo {author} {\bibfnamefont {L.}~\bibnamefont {Knox}},\ }\href@noop {} {\bibfield  {journal} {\bibinfo  {journal} {The Astrophysical Journal}\ }\textbf {\bibinfo {volume} {616}},\ \bibinfo {pages} {L75} (\bibinfo {year} {2004})}\BibitemShut {NoStop}%
\bibitem [{\citenamefont {Jing}\ \emph {et~al.}(2006)\citenamefont {Jing}, \citenamefont {Zhang}, \citenamefont {Lin}, \citenamefont {Gao},\ and\ \citenamefont {Springel}}]{jing2006influence}%
  \BibitemOpen
  \bibfield  {author} {\bibinfo {author} {\bibfnamefont {Y.}~\bibnamefont {Jing}}, \bibinfo {author} {\bibfnamefont {P.}~\bibnamefont {Zhang}}, \bibinfo {author} {\bibfnamefont {W.}~\bibnamefont {Lin}}, \bibinfo {author} {\bibfnamefont {L.}~\bibnamefont {Gao}},\ and\ \bibinfo {author} {\bibfnamefont {V.}~\bibnamefont {Springel}},\ }\href@noop {} {\bibfield  {journal} {\bibinfo  {journal} {The Astrophysical Journal}\ }\textbf {\bibinfo {volume} {640}},\ \bibinfo {pages} {L119} (\bibinfo {year} {2006})}\BibitemShut {NoStop}%
\bibitem [{\citenamefont {Rudd}\ \emph {et~al.}(2008)\citenamefont {Rudd}, \citenamefont {Zentner},\ and\ \citenamefont {Kravtsov}}]{rudd2008effects}%
  \BibitemOpen
  \bibfield  {author} {\bibinfo {author} {\bibfnamefont {D.~H.}\ \bibnamefont {Rudd}}, \bibinfo {author} {\bibfnamefont {A.~R.}\ \bibnamefont {Zentner}},\ and\ \bibinfo {author} {\bibfnamefont {A.~V.}\ \bibnamefont {Kravtsov}},\ }\href@noop {} {\bibfield  {journal} {\bibinfo  {journal} {The Astrophysical Journal}\ }\textbf {\bibinfo {volume} {672}},\ \bibinfo {pages} {19} (\bibinfo {year} {2008})}\BibitemShut {NoStop}%
\bibitem [{\citenamefont {Semboloni}\ \emph {et~al.}(2011)\citenamefont {Semboloni}, \citenamefont {Hoekstra}, \citenamefont {Schaye}, \citenamefont {van Daalen},\ and\ \citenamefont {McCarthy}}]{semboloni2011quantifying}%
  \BibitemOpen
  \bibfield  {author} {\bibinfo {author} {\bibfnamefont {E.}~\bibnamefont {Semboloni}}, \bibinfo {author} {\bibfnamefont {H.}~\bibnamefont {Hoekstra}}, \bibinfo {author} {\bibfnamefont {J.}~\bibnamefont {Schaye}}, \bibinfo {author} {\bibfnamefont {M.~P.}\ \bibnamefont {van Daalen}},\ and\ \bibinfo {author} {\bibfnamefont {I.~G.}\ \bibnamefont {McCarthy}},\ }\href@noop {} {\bibfield  {journal} {\bibinfo  {journal} {Monthly Notices of the Royal Astronomical Society}\ }\textbf {\bibinfo {volume} {417}},\ \bibinfo {pages} {2020} (\bibinfo {year} {2011})}\BibitemShut {NoStop}%
\bibitem [{\citenamefont {Schaye}\ \emph {et~al.}(2023)\citenamefont {Schaye}, \citenamefont {Kugel}, \citenamefont {Schaller}, \citenamefont {Helly}, \citenamefont {Braspenning}, \citenamefont {Elbers}, \citenamefont {McCarthy}, \citenamefont {Van~Daalen}, \citenamefont {Vandenbroucke}, \citenamefont {Frenk} \emph {et~al.}}]{schaye2023flamingo}%
  \BibitemOpen
  \bibfield  {author} {\bibinfo {author} {\bibfnamefont {J.}~\bibnamefont {Schaye}}, \bibinfo {author} {\bibfnamefont {R.}~\bibnamefont {Kugel}}, \bibinfo {author} {\bibfnamefont {M.}~\bibnamefont {Schaller}}, \bibinfo {author} {\bibfnamefont {J.~C.}\ \bibnamefont {Helly}}, \bibinfo {author} {\bibfnamefont {J.}~\bibnamefont {Braspenning}}, \bibinfo {author} {\bibfnamefont {W.}~\bibnamefont {Elbers}}, \bibinfo {author} {\bibfnamefont {I.~G.}\ \bibnamefont {McCarthy}}, \bibinfo {author} {\bibfnamefont {M.~P.}\ \bibnamefont {Van~Daalen}}, \bibinfo {author} {\bibfnamefont {B.}~\bibnamefont {Vandenbroucke}}, \bibinfo {author} {\bibfnamefont {C.~S.}\ \bibnamefont {Frenk}}, \emph {et~al.},\ }\href@noop {} {\bibfield  {journal} {\bibinfo  {journal} {Monthly Notices of the Royal Astronomical Society}\ }\textbf {\bibinfo {volume} {526}},\ \bibinfo {pages} {4978} (\bibinfo {year} {2023})}\BibitemShut {NoStop}%
\bibitem [{\citenamefont {Ivezi{\'c}}\ \emph {et~al.}(2019)\citenamefont {Ivezi{\'c}}, \citenamefont {Kahn}, \citenamefont {Tyson}, \citenamefont {Abel}, \citenamefont {Acosta}, \citenamefont {Allsman}, \citenamefont {Alonso}, \citenamefont {AlSayyad}, \citenamefont {Anderson}, \citenamefont {Andrew} \emph {et~al.}}]{ivezic2019lsst}%
  \BibitemOpen
  \bibfield  {author} {\bibinfo {author} {\bibfnamefont {{\v{Z}}.}~\bibnamefont {Ivezi{\'c}}}, \bibinfo {author} {\bibfnamefont {S.~M.}\ \bibnamefont {Kahn}}, \bibinfo {author} {\bibfnamefont {J.~A.}\ \bibnamefont {Tyson}}, \bibinfo {author} {\bibfnamefont {B.}~\bibnamefont {Abel}}, \bibinfo {author} {\bibfnamefont {E.}~\bibnamefont {Acosta}}, \bibinfo {author} {\bibfnamefont {R.}~\bibnamefont {Allsman}}, \bibinfo {author} {\bibfnamefont {D.}~\bibnamefont {Alonso}}, \bibinfo {author} {\bibfnamefont {Y.}~\bibnamefont {AlSayyad}}, \bibinfo {author} {\bibfnamefont {S.~F.}\ \bibnamefont {Anderson}}, \bibinfo {author} {\bibfnamefont {J.}~\bibnamefont {Andrew}}, \emph {et~al.},\ }\href@noop {} {\bibfield  {journal} {\bibinfo  {journal} {The Astrophysical Journal}\ }\textbf {\bibinfo {volume} {873}},\ \bibinfo {pages} {111} (\bibinfo {year} {2019})}\BibitemShut {NoStop}%
\bibitem [{\citenamefont {Yao}\ \emph {et~al.}(2024)\citenamefont {Yao}, \citenamefont {Shan}, \citenamefont {Li}, \citenamefont {Xu}, \citenamefont {Fan}, \citenamefont {Liu}, \citenamefont {Zhang}, \citenamefont {Yu}, \citenamefont {Wei}, \citenamefont {Hu} \emph {et~al.}}]{yao2024csst}%
  \BibitemOpen
  \bibfield  {author} {\bibinfo {author} {\bibfnamefont {J.}~\bibnamefont {Yao}}, \bibinfo {author} {\bibfnamefont {H.}~\bibnamefont {Shan}}, \bibinfo {author} {\bibfnamefont {R.}~\bibnamefont {Li}}, \bibinfo {author} {\bibfnamefont {Y.}~\bibnamefont {Xu}}, \bibinfo {author} {\bibfnamefont {D.}~\bibnamefont {Fan}}, \bibinfo {author} {\bibfnamefont {D.}~\bibnamefont {Liu}}, \bibinfo {author} {\bibfnamefont {P.}~\bibnamefont {Zhang}}, \bibinfo {author} {\bibfnamefont {Y.}~\bibnamefont {Yu}}, \bibinfo {author} {\bibfnamefont {C.}~\bibnamefont {Wei}}, \bibinfo {author} {\bibfnamefont {B.}~\bibnamefont {Hu}}, \emph {et~al.},\ }\href@noop {} {\bibfield  {journal} {\bibinfo  {journal} {Monthly Notices of the Royal Astronomical Society}\ }\textbf {\bibinfo {volume} {527}},\ \bibinfo {pages} {5206} (\bibinfo {year} {2024})}\BibitemShut {NoStop}%
\bibitem [{\citenamefont {Le~Brun}\ \emph {et~al.}(2014)\citenamefont {Le~Brun}, \citenamefont {McCarthy}, \citenamefont {Schaye},\ and\ \citenamefont {Ponman}}]{le2014towards}%
  \BibitemOpen
  \bibfield  {author} {\bibinfo {author} {\bibfnamefont {A.~M.}\ \bibnamefont {Le~Brun}}, \bibinfo {author} {\bibfnamefont {I.~G.}\ \bibnamefont {McCarthy}}, \bibinfo {author} {\bibfnamefont {J.}~\bibnamefont {Schaye}},\ and\ \bibinfo {author} {\bibfnamefont {T.~J.}\ \bibnamefont {Ponman}},\ }\href@noop {} {\bibfield  {journal} {\bibinfo  {journal} {Monthly Notices of the Royal Astronomical Society}\ }\textbf {\bibinfo {volume} {441}},\ \bibinfo {pages} {1270} (\bibinfo {year} {2014})}\BibitemShut {NoStop}%
\bibitem [{\citenamefont {Nelson}\ \emph {et~al.}(2015)\citenamefont {Nelson}, \citenamefont {Pillepich}, \citenamefont {Genel}, \citenamefont {Vogelsberger}, \citenamefont {Springel}, \citenamefont {Torrey}, \citenamefont {Rodriguez-Gomez}, \citenamefont {Sijacki}, \citenamefont {Snyder}, \citenamefont {Griffen} \emph {et~al.}}]{nelson2015illustris}%
  \BibitemOpen
  \bibfield  {author} {\bibinfo {author} {\bibfnamefont {D.}~\bibnamefont {Nelson}}, \bibinfo {author} {\bibfnamefont {A.}~\bibnamefont {Pillepich}}, \bibinfo {author} {\bibfnamefont {S.}~\bibnamefont {Genel}}, \bibinfo {author} {\bibfnamefont {M.}~\bibnamefont {Vogelsberger}}, \bibinfo {author} {\bibfnamefont {V.}~\bibnamefont {Springel}}, \bibinfo {author} {\bibfnamefont {P.}~\bibnamefont {Torrey}}, \bibinfo {author} {\bibfnamefont {V.}~\bibnamefont {Rodriguez-Gomez}}, \bibinfo {author} {\bibfnamefont {D.}~\bibnamefont {Sijacki}}, \bibinfo {author} {\bibfnamefont {G.~F.}\ \bibnamefont {Snyder}}, \bibinfo {author} {\bibfnamefont {B.}~\bibnamefont {Griffen}}, \emph {et~al.},\ }\href@noop {} {\bibfield  {journal} {\bibinfo  {journal} {Astronomy and Computing}\ }\textbf {\bibinfo {volume} {13}},\ \bibinfo {pages} {12} (\bibinfo {year} {2015})}\BibitemShut {NoStop}%
\bibitem [{\citenamefont {Schaye}\ \emph {et~al.}(2015)\citenamefont {Schaye}, \citenamefont {Crain}, \citenamefont {Bower}, \citenamefont {Furlong}, \citenamefont {Schaller}, \citenamefont {Theuns}, \citenamefont {Dalla~Vecchia}, \citenamefont {Frenk}, \citenamefont {McCarthy}, \citenamefont {Helly} \emph {et~al.}}]{schaye2015eagle}%
  \BibitemOpen
  \bibfield  {author} {\bibinfo {author} {\bibfnamefont {J.}~\bibnamefont {Schaye}}, \bibinfo {author} {\bibfnamefont {R.~A.}\ \bibnamefont {Crain}}, \bibinfo {author} {\bibfnamefont {R.~G.}\ \bibnamefont {Bower}}, \bibinfo {author} {\bibfnamefont {M.}~\bibnamefont {Furlong}}, \bibinfo {author} {\bibfnamefont {M.}~\bibnamefont {Schaller}}, \bibinfo {author} {\bibfnamefont {T.}~\bibnamefont {Theuns}}, \bibinfo {author} {\bibfnamefont {C.}~\bibnamefont {Dalla~Vecchia}}, \bibinfo {author} {\bibfnamefont {C.~S.}\ \bibnamefont {Frenk}}, \bibinfo {author} {\bibfnamefont {I.}~\bibnamefont {McCarthy}}, \bibinfo {author} {\bibfnamefont {J.~C.}\ \bibnamefont {Helly}}, \emph {et~al.},\ }\href@noop {} {\bibfield  {journal} {\bibinfo  {journal} {Monthly Notices of the Royal Astronomical Society}\ }\textbf {\bibinfo {volume} {446}},\ \bibinfo {pages} {521} (\bibinfo {year} {2015})}\BibitemShut {NoStop}%
\bibitem [{\citenamefont {McCarthy}\ \emph {et~al.}(2016)\citenamefont {McCarthy}, \citenamefont {Schaye}, \citenamefont {Bird},\ and\ \citenamefont {Le~Brun}}]{mccarthy2016bahamas}%
  \BibitemOpen
  \bibfield  {author} {\bibinfo {author} {\bibfnamefont {I.~G.}\ \bibnamefont {McCarthy}}, \bibinfo {author} {\bibfnamefont {J.}~\bibnamefont {Schaye}}, \bibinfo {author} {\bibfnamefont {S.}~\bibnamefont {Bird}},\ and\ \bibinfo {author} {\bibfnamefont {A.~M.~C.}\ \bibnamefont {Le~Brun}},\ }\href@noop {} {\bibfield  {journal} {\bibinfo  {journal} {Monthly Notices of the Royal Astronomical Society}\ ,\ \bibinfo {pages} {stw2792}} (\bibinfo {year} {2016})}\BibitemShut {NoStop}%
\bibitem [{\citenamefont {Chisari}\ \emph {et~al.}(2018)\citenamefont {Chisari}, \citenamefont {Richardson}, \citenamefont {Devriendt}, \citenamefont {Dubois}, \citenamefont {Schneider}, \citenamefont {Le~Brun}, \citenamefont {Beckmann}, \citenamefont {Peirani}, \citenamefont {Slyz},\ and\ \citenamefont {Pichon}}]{chisari2018impact}%
  \BibitemOpen
  \bibfield  {author} {\bibinfo {author} {\bibfnamefont {N.~E.}\ \bibnamefont {Chisari}}, \bibinfo {author} {\bibfnamefont {M.~L.}\ \bibnamefont {Richardson}}, \bibinfo {author} {\bibfnamefont {J.}~\bibnamefont {Devriendt}}, \bibinfo {author} {\bibfnamefont {Y.}~\bibnamefont {Dubois}}, \bibinfo {author} {\bibfnamefont {A.}~\bibnamefont {Schneider}}, \bibinfo {author} {\bibfnamefont {A.~M.}\ \bibnamefont {Le~Brun}}, \bibinfo {author} {\bibfnamefont {R.~S.}\ \bibnamefont {Beckmann}}, \bibinfo {author} {\bibfnamefont {S.}~\bibnamefont {Peirani}}, \bibinfo {author} {\bibfnamefont {A.}~\bibnamefont {Slyz}},\ and\ \bibinfo {author} {\bibfnamefont {C.}~\bibnamefont {Pichon}},\ }\href@noop {} {\bibfield  {journal} {\bibinfo  {journal} {Monthly Notices of the Royal Astronomical Society}\ }\textbf {\bibinfo {volume} {480}},\ \bibinfo {pages} {3962} (\bibinfo {year} {2018})}\BibitemShut {NoStop}%
\bibitem [{\citenamefont {Nelson}\ \emph {et~al.}(2019)\citenamefont {Nelson}, \citenamefont {Springel}, \citenamefont {Pillepich}, \citenamefont {Rodriguez-Gomez}, \citenamefont {Torrey}, \citenamefont {Genel}, \citenamefont {Vogelsberger}, \citenamefont {Pakmor}, \citenamefont {Marinacci}, \citenamefont {Weinberger} \emph {et~al.}}]{nelson2019illustristng}%
  \BibitemOpen
  \bibfield  {author} {\bibinfo {author} {\bibfnamefont {D.}~\bibnamefont {Nelson}}, \bibinfo {author} {\bibfnamefont {V.}~\bibnamefont {Springel}}, \bibinfo {author} {\bibfnamefont {A.}~\bibnamefont {Pillepich}}, \bibinfo {author} {\bibfnamefont {V.}~\bibnamefont {Rodriguez-Gomez}}, \bibinfo {author} {\bibfnamefont {P.}~\bibnamefont {Torrey}}, \bibinfo {author} {\bibfnamefont {S.}~\bibnamefont {Genel}}, \bibinfo {author} {\bibfnamefont {M.}~\bibnamefont {Vogelsberger}}, \bibinfo {author} {\bibfnamefont {R.}~\bibnamefont {Pakmor}}, \bibinfo {author} {\bibfnamefont {F.}~\bibnamefont {Marinacci}}, \bibinfo {author} {\bibfnamefont {R.}~\bibnamefont {Weinberger}}, \emph {et~al.},\ }\href@noop {} {\bibfield  {journal} {\bibinfo  {journal} {Computational Astrophysics and Cosmology}\ }\textbf {\bibinfo {volume} {6}},\ \bibinfo {pages} {2} (\bibinfo {year} {2019})}\BibitemShut {NoStop}%
\bibitem [{\citenamefont {Pakmor}\ \emph {et~al.}(2023)\citenamefont {Pakmor}, \citenamefont {Springel}, \citenamefont {Coles}, \citenamefont {Guillet}, \citenamefont {Pfrommer}, \citenamefont {Bose}, \citenamefont {Barrera}, \citenamefont {Delgado}, \citenamefont {Ferlito}, \citenamefont {Frenk} \emph {et~al.}}]{pakmor2023millenniumtng}%
  \BibitemOpen
  \bibfield  {author} {\bibinfo {author} {\bibfnamefont {R.}~\bibnamefont {Pakmor}}, \bibinfo {author} {\bibfnamefont {V.}~\bibnamefont {Springel}}, \bibinfo {author} {\bibfnamefont {J.~P.}\ \bibnamefont {Coles}}, \bibinfo {author} {\bibfnamefont {T.}~\bibnamefont {Guillet}}, \bibinfo {author} {\bibfnamefont {C.}~\bibnamefont {Pfrommer}}, \bibinfo {author} {\bibfnamefont {S.}~\bibnamefont {Bose}}, \bibinfo {author} {\bibfnamefont {M.}~\bibnamefont {Barrera}}, \bibinfo {author} {\bibfnamefont {A.~M.}\ \bibnamefont {Delgado}}, \bibinfo {author} {\bibfnamefont {F.}~\bibnamefont {Ferlito}}, \bibinfo {author} {\bibfnamefont {C.}~\bibnamefont {Frenk}}, \emph {et~al.},\ }\href@noop {} {\bibfield  {journal} {\bibinfo  {journal} {Monthly Notices of the Royal Astronomical Society}\ }\textbf {\bibinfo {volume} {524}},\ \bibinfo {pages} {2539} (\bibinfo {year} {2023})}\BibitemShut {NoStop}%
\bibitem [{\citenamefont {Salcido}\ \emph {et~al.}(2023)\citenamefont {Salcido}, \citenamefont {McCarthy}, \citenamefont {Kwan}, \citenamefont {Upadhye},\ and\ \citenamefont {Font}}]{salcido2023sp}%
  \BibitemOpen
  \bibfield  {author} {\bibinfo {author} {\bibfnamefont {J.}~\bibnamefont {Salcido}}, \bibinfo {author} {\bibfnamefont {I.~G.}\ \bibnamefont {McCarthy}}, \bibinfo {author} {\bibfnamefont {J.}~\bibnamefont {Kwan}}, \bibinfo {author} {\bibfnamefont {A.}~\bibnamefont {Upadhye}},\ and\ \bibinfo {author} {\bibfnamefont {A.~S.}\ \bibnamefont {Font}},\ }\href@noop {} {\bibfield  {journal} {\bibinfo  {journal} {Monthly Notices of the Royal Astronomical Society}\ }\textbf {\bibinfo {volume} {523}},\ \bibinfo {pages} {2247} (\bibinfo {year} {2023})}\BibitemShut {NoStop}%
\bibitem [{\citenamefont {Schaye}\ \emph {et~al.}(2025)\citenamefont {Schaye}, \citenamefont {Chaikin}, \citenamefont {Schaller}, \citenamefont {Ploeckinger}, \citenamefont {Hu{\v{s}}ko}, \citenamefont {McGibbon}, \citenamefont {Trayford}, \citenamefont {Ben{\'\i}tez-Llambay}, \citenamefont {Correa}, \citenamefont {Frenk} \emph {et~al.}}]{schaye2025colibre}%
  \BibitemOpen
  \bibfield  {author} {\bibinfo {author} {\bibfnamefont {J.}~\bibnamefont {Schaye}}, \bibinfo {author} {\bibfnamefont {E.}~\bibnamefont {Chaikin}}, \bibinfo {author} {\bibfnamefont {M.}~\bibnamefont {Schaller}}, \bibinfo {author} {\bibfnamefont {S.}~\bibnamefont {Ploeckinger}}, \bibinfo {author} {\bibfnamefont {F.}~\bibnamefont {Hu{\v{s}}ko}}, \bibinfo {author} {\bibfnamefont {R.}~\bibnamefont {McGibbon}}, \bibinfo {author} {\bibfnamefont {J.~W.}\ \bibnamefont {Trayford}}, \bibinfo {author} {\bibfnamefont {A.}~\bibnamefont {Ben{\'\i}tez-Llambay}}, \bibinfo {author} {\bibfnamefont {C.}~\bibnamefont {Correa}}, \bibinfo {author} {\bibfnamefont {C.~S.}\ \bibnamefont {Frenk}}, \emph {et~al.},\ }\href@noop {} {\bibfield  {journal} {\bibinfo  {journal} {arXiv preprint arXiv:2508.21126}\ } (\bibinfo {year} {2025})}\BibitemShut {NoStop}%
\bibitem [{\citenamefont {Sharma}\ \emph {et~al.}(2025)\citenamefont {Sharma}, \citenamefont {Dai}, \citenamefont {Villaescusa-Navarro},\ and\ \citenamefont {Seljak}}]{sharma2025field}%
  \BibitemOpen
  \bibfield  {author} {\bibinfo {author} {\bibfnamefont {D.}~\bibnamefont {Sharma}}, \bibinfo {author} {\bibfnamefont {B.}~\bibnamefont {Dai}}, \bibinfo {author} {\bibfnamefont {F.}~\bibnamefont {Villaescusa-Navarro}},\ and\ \bibinfo {author} {\bibfnamefont {U.}~\bibnamefont {Seljak}},\ }\href@noop {} {\bibfield  {journal} {\bibinfo  {journal} {Monthly Notices of the Royal Astronomical Society}\ }\textbf {\bibinfo {volume} {538}},\ \bibinfo {pages} {1415} (\bibinfo {year} {2025})}\BibitemShut {NoStop}%
\bibitem [{\citenamefont {Bigwood}\ \emph {et~al.}(2025{\natexlab{a}})\citenamefont {Bigwood}, \citenamefont {Yamamoto}, \citenamefont {Siegel}, \citenamefont {Amon}, \citenamefont {McCarthy}, \citenamefont {Dave}, \citenamefont {Salcido}, \citenamefont {Schaller}, \citenamefont {Schaye},\ and\ \citenamefont {Yang}}]{bigwood2025kinetic}%
  \BibitemOpen
  \bibfield  {author} {\bibinfo {author} {\bibfnamefont {L.}~\bibnamefont {Bigwood}}, \bibinfo {author} {\bibfnamefont {M.}~\bibnamefont {Yamamoto}}, \bibinfo {author} {\bibfnamefont {J.}~\bibnamefont {Siegel}}, \bibinfo {author} {\bibfnamefont {A.}~\bibnamefont {Amon}}, \bibinfo {author} {\bibfnamefont {I.~G.}\ \bibnamefont {McCarthy}}, \bibinfo {author} {\bibfnamefont {R.}~\bibnamefont {Dave}}, \bibinfo {author} {\bibfnamefont {J.}~\bibnamefont {Salcido}}, \bibinfo {author} {\bibfnamefont {M.}~\bibnamefont {Schaller}}, \bibinfo {author} {\bibfnamefont {J.}~\bibnamefont {Schaye}},\ and\ \bibinfo {author} {\bibfnamefont {T.}~\bibnamefont {Yang}},\ }\href@noop {} {\bibfield  {journal} {\bibinfo  {journal} {arXiv preprint arXiv:2510.15822}\ } (\bibinfo {year} {2025}{\natexlab{a}})}\BibitemShut {NoStop}%
\bibitem [{\citenamefont {Hadzhiyska}\ \emph {et~al.}(2024)\citenamefont {Hadzhiyska}, \citenamefont {Ferraro}, \citenamefont {Guachalla}, \citenamefont {Schaan}, \citenamefont {Aguilar}, \citenamefont {Battaglia}, \citenamefont {Bond}, \citenamefont {Brooks}, \citenamefont {Calabrese}, \citenamefont {Choi} \emph {et~al.}}]{hadzhiyska2024evidence}%
  \BibitemOpen
  \bibfield  {author} {\bibinfo {author} {\bibfnamefont {B.}~\bibnamefont {Hadzhiyska}}, \bibinfo {author} {\bibfnamefont {S.}~\bibnamefont {Ferraro}}, \bibinfo {author} {\bibfnamefont {B.~R.}\ \bibnamefont {Guachalla}}, \bibinfo {author} {\bibfnamefont {E.}~\bibnamefont {Schaan}}, \bibinfo {author} {\bibfnamefont {J.}~\bibnamefont {Aguilar}}, \bibinfo {author} {\bibfnamefont {N.}~\bibnamefont {Battaglia}}, \bibinfo {author} {\bibfnamefont {J.}~\bibnamefont {Bond}}, \bibinfo {author} {\bibfnamefont {D.}~\bibnamefont {Brooks}}, \bibinfo {author} {\bibfnamefont {E.}~\bibnamefont {Calabrese}}, \bibinfo {author} {\bibfnamefont {S.}~\bibnamefont {Choi}}, \emph {et~al.},\ }\href@noop {} {\bibfield  {journal} {\bibinfo  {journal} {arXiv preprint arXiv:2407.07152}\ } (\bibinfo {year} {2024})}\BibitemShut {NoStop}%
\bibitem [{\citenamefont {Popesso}\ \emph {et~al.}(2024)\citenamefont {Popesso}, \citenamefont {Biviano}, \citenamefont {Marini}, \citenamefont {Dolag}, \citenamefont {Vladutescu-Zopp}, \citenamefont {Csizi}, \citenamefont {Biffi}, \citenamefont {Lamer}, \citenamefont {Robothan}, \citenamefont {Bravo} \emph {et~al.}}]{popesso2024hot}%
  \BibitemOpen
  \bibfield  {author} {\bibinfo {author} {\bibfnamefont {P.}~\bibnamefont {Popesso}}, \bibinfo {author} {\bibfnamefont {A.}~\bibnamefont {Biviano}}, \bibinfo {author} {\bibfnamefont {I.}~\bibnamefont {Marini}}, \bibinfo {author} {\bibfnamefont {K.}~\bibnamefont {Dolag}}, \bibinfo {author} {\bibfnamefont {S.}~\bibnamefont {Vladutescu-Zopp}}, \bibinfo {author} {\bibfnamefont {B.}~\bibnamefont {Csizi}}, \bibinfo {author} {\bibfnamefont {V.}~\bibnamefont {Biffi}}, \bibinfo {author} {\bibfnamefont {G.}~\bibnamefont {Lamer}}, \bibinfo {author} {\bibfnamefont {A.}~\bibnamefont {Robothan}}, \bibinfo {author} {\bibfnamefont {M.}~\bibnamefont {Bravo}}, \emph {et~al.},\ }\href@noop {} {\bibfield  {journal} {\bibinfo  {journal} {arXiv preprint arXiv:2411.16555}\ } (\bibinfo {year} {2024})}\BibitemShut {NoStop}%
\bibitem [{\citenamefont {Siegel}\ \emph {et~al.}(2025{\natexlab{a}})\citenamefont {Siegel}, \citenamefont {Amon}, \citenamefont {McCarthy}, \citenamefont {Bigwood}, \citenamefont {Yamamoto}, \citenamefont {Bulbul}, \citenamefont {Greene}, \citenamefont {McCullough}, \citenamefont {Schaller},\ and\ \citenamefont {Schaye}}]{siegel2025joint}%
  \BibitemOpen
  \bibfield  {author} {\bibinfo {author} {\bibfnamefont {J.}~\bibnamefont {Siegel}}, \bibinfo {author} {\bibfnamefont {A.}~\bibnamefont {Amon}}, \bibinfo {author} {\bibfnamefont {I.~G.}\ \bibnamefont {McCarthy}}, \bibinfo {author} {\bibfnamefont {L.}~\bibnamefont {Bigwood}}, \bibinfo {author} {\bibfnamefont {M.}~\bibnamefont {Yamamoto}}, \bibinfo {author} {\bibfnamefont {E.}~\bibnamefont {Bulbul}}, \bibinfo {author} {\bibfnamefont {J.~E.}\ \bibnamefont {Greene}}, \bibinfo {author} {\bibfnamefont {J.}~\bibnamefont {McCullough}}, \bibinfo {author} {\bibfnamefont {M.}~\bibnamefont {Schaller}},\ and\ \bibinfo {author} {\bibfnamefont {J.}~\bibnamefont {Schaye}},\ }\href@noop {} {\bibfield  {journal} {\bibinfo  {journal} {arXiv preprint arXiv:2509.10455}\ } (\bibinfo {year} {2025}{\natexlab{a}})}\BibitemShut {NoStop}%
\bibitem [{\citenamefont {Schneider}\ and\ \citenamefont {Teyssier}(2015)}]{schneider2015new}%
  \BibitemOpen
  \bibfield  {author} {\bibinfo {author} {\bibfnamefont {A.}~\bibnamefont {Schneider}}\ and\ \bibinfo {author} {\bibfnamefont {R.}~\bibnamefont {Teyssier}},\ }\href@noop {} {\bibfield  {journal} {\bibinfo  {journal} {Journal of Cosmology and Astroparticle Physics}\ }\textbf {\bibinfo {volume} {2015}}\bibinfo  {number} { (12)},\ \bibinfo {pages} {049}}\BibitemShut {NoStop}%
\bibitem [{\citenamefont {Zennaro}\ \emph {et~al.}(2024)\citenamefont {Zennaro}, \citenamefont {Aric{\`o}}, \citenamefont {Garc{\'\i}a-Garc{\'\i}a}, \citenamefont {Angulo}, \citenamefont {Ondaro-Mallea}, \citenamefont {Contreras}, \citenamefont {Nicola}, \citenamefont {Schaller},\ and\ \citenamefont {Schaye}}]{zennaro20241}%
  \BibitemOpen
\bibfield  {number} {  }\bibfield  {author} {\bibinfo {author} {\bibfnamefont {M.}~\bibnamefont {Zennaro}}, \bibinfo {author} {\bibfnamefont {G.}~\bibnamefont {Aric{\`o}}}, \bibinfo {author} {\bibfnamefont {C.}~\bibnamefont {Garc{\'\i}a-Garc{\'\i}a}}, \bibinfo {author} {\bibfnamefont {R.~E.}\ \bibnamefont {Angulo}}, \bibinfo {author} {\bibfnamefont {L.}~\bibnamefont {Ondaro-Mallea}}, \bibinfo {author} {\bibfnamefont {S.}~\bibnamefont {Contreras}}, \bibinfo {author} {\bibfnamefont {A.}~\bibnamefont {Nicola}}, \bibinfo {author} {\bibfnamefont {M.}~\bibnamefont {Schaller}},\ and\ \bibinfo {author} {\bibfnamefont {J.}~\bibnamefont {Schaye}},\ }\href@noop {} {\bibfield  {journal} {\bibinfo  {journal} {arXiv preprint arXiv:2412.08623}\ } (\bibinfo {year} {2024})}\BibitemShut {NoStop}%
\bibitem [{\citenamefont {Wayland}\ \emph {et~al.}(2025)\citenamefont {Wayland}, \citenamefont {Alonso},\ and\ \citenamefont {Zennaro}}]{wayland2025calibrating}%
  \BibitemOpen
  \bibfield  {author} {\bibinfo {author} {\bibfnamefont {A.}~\bibnamefont {Wayland}}, \bibinfo {author} {\bibfnamefont {D.}~\bibnamefont {Alonso}},\ and\ \bibinfo {author} {\bibfnamefont {M.}~\bibnamefont {Zennaro}},\ }\href@noop {} {\bibfield  {journal} {\bibinfo  {journal} {Monthly Notices of the Royal Astronomical Society}\ }\textbf {\bibinfo {volume} {543}},\ \bibinfo {pages} {1518} (\bibinfo {year} {2025})}\BibitemShut {NoStop}%
\bibitem [{\citenamefont {Dai}\ \emph {et~al.}(2018)\citenamefont {Dai}, \citenamefont {Feng},\ and\ \citenamefont {Seljak}}]{dai2018gradient}%
  \BibitemOpen
  \bibfield  {author} {\bibinfo {author} {\bibfnamefont {B.}~\bibnamefont {Dai}}, \bibinfo {author} {\bibfnamefont {Y.}~\bibnamefont {Feng}},\ and\ \bibinfo {author} {\bibfnamefont {U.}~\bibnamefont {Seljak}},\ }\href@noop {} {\bibfield  {journal} {\bibinfo  {journal} {Journal of Cosmology and Astroparticle Physics}\ }\textbf {\bibinfo {volume} {2018}}\bibinfo  {number} { (11)},\ \bibinfo {pages} {009}}\BibitemShut {NoStop}%
\bibitem [{\citenamefont {Aric{\`o}}\ \emph {et~al.}(2020)\citenamefont {Aric{\`o}}, \citenamefont {Angulo}, \citenamefont {Hern{\'a}ndez-Monteagudo}, \citenamefont {Contreras}, \citenamefont {Zennaro}, \citenamefont {Pellejero-Iba{\~n}ez},\ and\ \citenamefont {Rosas-Guevara}}]{arico2020modelling}%
  \BibitemOpen
\bibfield  {number} {  }\bibfield  {author} {\bibinfo {author} {\bibfnamefont {G.}~\bibnamefont {Aric{\`o}}}, \bibinfo {author} {\bibfnamefont {R.~E.}\ \bibnamefont {Angulo}}, \bibinfo {author} {\bibfnamefont {C.}~\bibnamefont {Hern{\'a}ndez-Monteagudo}}, \bibinfo {author} {\bibfnamefont {S.}~\bibnamefont {Contreras}}, \bibinfo {author} {\bibfnamefont {M.}~\bibnamefont {Zennaro}}, \bibinfo {author} {\bibfnamefont {M.}~\bibnamefont {Pellejero-Iba{\~n}ez}},\ and\ \bibinfo {author} {\bibfnamefont {Y.}~\bibnamefont {Rosas-Guevara}},\ }\href@noop {} {\bibfield  {journal} {\bibinfo  {journal} {Monthly Notices of the Royal Astronomical Society}\ }\textbf {\bibinfo {volume} {495}},\ \bibinfo {pages} {4800} (\bibinfo {year} {2020})}\BibitemShut {NoStop}%
\bibitem [{\citenamefont {Aric{\`o}}\ \emph {et~al.}(2021)\citenamefont {Aric{\`o}}, \citenamefont {Angulo}, \citenamefont {Contreras}, \citenamefont {Ondaro-Mallea}, \citenamefont {Pellejero-Ibanez},\ and\ \citenamefont {Zennaro}}]{arico2021bacco}%
  \BibitemOpen
  \bibfield  {author} {\bibinfo {author} {\bibfnamefont {G.}~\bibnamefont {Aric{\`o}}}, \bibinfo {author} {\bibfnamefont {R.~E.}\ \bibnamefont {Angulo}}, \bibinfo {author} {\bibfnamefont {S.}~\bibnamefont {Contreras}}, \bibinfo {author} {\bibfnamefont {L.}~\bibnamefont {Ondaro-Mallea}}, \bibinfo {author} {\bibfnamefont {M.}~\bibnamefont {Pellejero-Ibanez}},\ and\ \bibinfo {author} {\bibfnamefont {M.}~\bibnamefont {Zennaro}},\ }\href@noop {} {\bibfield  {journal} {\bibinfo  {journal} {Monthly Notices of the Royal Astronomical Society}\ }\textbf {\bibinfo {volume} {506}},\ \bibinfo {pages} {4070} (\bibinfo {year} {2021})}\BibitemShut {NoStop}%
\bibitem [{\citenamefont {Schneider}\ \emph {et~al.}(2022)\citenamefont {Schneider}, \citenamefont {Giri}, \citenamefont {Amodeo},\ and\ \citenamefont {Refregier}}]{schneider2022constraining}%
  \BibitemOpen
  \bibfield  {author} {\bibinfo {author} {\bibfnamefont {A.}~\bibnamefont {Schneider}}, \bibinfo {author} {\bibfnamefont {S.~K.}\ \bibnamefont {Giri}}, \bibinfo {author} {\bibfnamefont {S.}~\bibnamefont {Amodeo}},\ and\ \bibinfo {author} {\bibfnamefont {A.}~\bibnamefont {Refregier}},\ }\href@noop {} {\bibfield  {journal} {\bibinfo  {journal} {Monthly Notices of the Royal Astronomical Society}\ }\textbf {\bibinfo {volume} {514}},\ \bibinfo {pages} {3802} (\bibinfo {year} {2022})}\BibitemShut {NoStop}%
\bibitem [{\citenamefont {Schneider}\ \emph {et~al.}(2025)\citenamefont {Schneider}, \citenamefont {Kova{\v{c}}}, \citenamefont {Bucko}, \citenamefont {Nicola}, \citenamefont {Reischke}, \citenamefont {Giri}, \citenamefont {Teyssier}, \citenamefont {Tr{\"o}ster}, \citenamefont {Refregier}, \citenamefont {Schaller} \emph {et~al.}}]{schneider2025baryonification}%
  \BibitemOpen
  \bibfield  {author} {\bibinfo {author} {\bibfnamefont {A.}~\bibnamefont {Schneider}}, \bibinfo {author} {\bibfnamefont {M.}~\bibnamefont {Kova{\v{c}}}}, \bibinfo {author} {\bibfnamefont {J.}~\bibnamefont {Bucko}}, \bibinfo {author} {\bibfnamefont {A.}~\bibnamefont {Nicola}}, \bibinfo {author} {\bibfnamefont {R.}~\bibnamefont {Reischke}}, \bibinfo {author} {\bibfnamefont {S.~K.}\ \bibnamefont {Giri}}, \bibinfo {author} {\bibfnamefont {R.}~\bibnamefont {Teyssier}}, \bibinfo {author} {\bibfnamefont {T.}~\bibnamefont {Tr{\"o}ster}}, \bibinfo {author} {\bibfnamefont {A.}~\bibnamefont {Refregier}}, \bibinfo {author} {\bibfnamefont {M.}~\bibnamefont {Schaller}}, \emph {et~al.},\ }\href@noop {} {\bibfield  {journal} {\bibinfo  {journal} {Journal of Cosmology and Astroparticle Physics}\ }\textbf {\bibinfo {volume} {2025}}\bibinfo  {number} { (12)},\ \bibinfo {pages} {043}}\BibitemShut {NoStop}%
\bibitem [{\citenamefont {Bigwood}\ \emph {et~al.}(2024)\citenamefont {Bigwood}, \citenamefont {Amon}, \citenamefont {Schneider}, \citenamefont {Salcido}, \citenamefont {McCarthy}, \citenamefont {Preston}, \citenamefont {Sanchez}, \citenamefont {Sijacki}, \citenamefont {Schaan}, \citenamefont {Ferraro} \emph {et~al.}}]{bigwood2024weak}%
  \BibitemOpen
\bibfield  {number} {  }\bibfield  {author} {\bibinfo {author} {\bibfnamefont {L.}~\bibnamefont {Bigwood}}, \bibinfo {author} {\bibfnamefont {A.}~\bibnamefont {Amon}}, \bibinfo {author} {\bibfnamefont {A.}~\bibnamefont {Schneider}}, \bibinfo {author} {\bibfnamefont {J.}~\bibnamefont {Salcido}}, \bibinfo {author} {\bibfnamefont {I.~G.}\ \bibnamefont {McCarthy}}, \bibinfo {author} {\bibfnamefont {C.}~\bibnamefont {Preston}}, \bibinfo {author} {\bibfnamefont {D.}~\bibnamefont {Sanchez}}, \bibinfo {author} {\bibfnamefont {D.}~\bibnamefont {Sijacki}}, \bibinfo {author} {\bibfnamefont {E.}~\bibnamefont {Schaan}}, \bibinfo {author} {\bibfnamefont {S.}~\bibnamefont {Ferraro}}, \emph {et~al.},\ }\href@noop {} {\bibfield  {journal} {\bibinfo  {journal} {Monthly Notices of the Royal Astronomical Society}\ }\textbf {\bibinfo {volume} {534}},\ \bibinfo {pages} {655} (\bibinfo {year} {2024})}\BibitemShut {NoStop}%
\bibitem [{\citenamefont {Pandey}\ \emph {et~al.}(2025)\citenamefont {Pandey}, \citenamefont {Hill}, \citenamefont {Alarcon}, \citenamefont {Alves}, \citenamefont {Amon}, \citenamefont {Anbajagane}, \citenamefont {Andrade-Oliveira}, \citenamefont {Battaglia}, \citenamefont {Baxter}, \citenamefont {Bechtol} \emph {et~al.}}]{pandey2025constraints}%
  \BibitemOpen
  \bibfield  {author} {\bibinfo {author} {\bibfnamefont {S.}~\bibnamefont {Pandey}}, \bibinfo {author} {\bibfnamefont {J.}~\bibnamefont {Hill}}, \bibinfo {author} {\bibfnamefont {A.}~\bibnamefont {Alarcon}}, \bibinfo {author} {\bibfnamefont {O.}~\bibnamefont {Alves}}, \bibinfo {author} {\bibfnamefont {A.}~\bibnamefont {Amon}}, \bibinfo {author} {\bibfnamefont {D.}~\bibnamefont {Anbajagane}}, \bibinfo {author} {\bibfnamefont {F.}~\bibnamefont {Andrade-Oliveira}}, \bibinfo {author} {\bibfnamefont {N.}~\bibnamefont {Battaglia}}, \bibinfo {author} {\bibfnamefont {E.}~\bibnamefont {Baxter}}, \bibinfo {author} {\bibfnamefont {K.}~\bibnamefont {Bechtol}}, \emph {et~al.},\ }\href@noop {} {\bibfield  {journal} {\bibinfo  {journal} {arXiv preprint arXiv:2506.07432}\ } (\bibinfo {year} {2025})}\BibitemShut {NoStop}%
\bibitem [{\citenamefont {Chen}\ \emph {et~al.}(2023{\natexlab{a}})\citenamefont {Chen}, \citenamefont {Aric{\`o}}, \citenamefont {Huterer}, \citenamefont {Angulo}, \citenamefont {Weaverdyck}, \citenamefont {Friedrich}, \citenamefont {Secco}, \citenamefont {Hern{\'a}ndez-Monteagudo}, \citenamefont {Alarcon}, \citenamefont {Alves} \emph {et~al.}}]{chen2023constraining}%
  \BibitemOpen
  \bibfield  {author} {\bibinfo {author} {\bibfnamefont {A.}~\bibnamefont {Chen}}, \bibinfo {author} {\bibfnamefont {G.}~\bibnamefont {Aric{\`o}}}, \bibinfo {author} {\bibfnamefont {D.}~\bibnamefont {Huterer}}, \bibinfo {author} {\bibfnamefont {R.~E.}\ \bibnamefont {Angulo}}, \bibinfo {author} {\bibfnamefont {N.}~\bibnamefont {Weaverdyck}}, \bibinfo {author} {\bibfnamefont {O.}~\bibnamefont {Friedrich}}, \bibinfo {author} {\bibfnamefont {L.~F.}\ \bibnamefont {Secco}}, \bibinfo {author} {\bibfnamefont {C.}~\bibnamefont {Hern{\'a}ndez-Monteagudo}}, \bibinfo {author} {\bibfnamefont {A.}~\bibnamefont {Alarcon}}, \bibinfo {author} {\bibfnamefont {O.}~\bibnamefont {Alves}}, \emph {et~al.},\ }\href@noop {} {\bibfield  {journal} {\bibinfo  {journal} {Monthly Notices of the Royal Astronomical Society}\ }\textbf {\bibinfo {volume} {518}},\ \bibinfo {pages} {5340} (\bibinfo {year} {2023}{\natexlab{a}})}\BibitemShut {NoStop}%
\bibitem [{\citenamefont {Carlos}\ \emph {et~al.}(2024)\citenamefont {Carlos}, \citenamefont {Matteo}, \citenamefont {Giovanni}, \citenamefont {David} \emph {et~al.}}]{carlos2024cosmic}%
  \BibitemOpen
  \bibfield  {author} {\bibinfo {author} {\bibfnamefont {G.-G.}\ \bibnamefont {Carlos}}, \bibinfo {author} {\bibfnamefont {Z.}~\bibnamefont {Matteo}}, \bibinfo {author} {\bibfnamefont {A.}~\bibnamefont {Giovanni}}, \bibinfo {author} {\bibfnamefont {A.}~\bibnamefont {David}}, \emph {et~al.},\ }\href@noop {} {\bibfield  {journal} {\bibinfo  {journal} {JCAP}\ }\textbf {\bibinfo {volume} {8}}}\BibitemShut {NoStop}%
\bibitem [{\citenamefont {Giri}\ and\ \citenamefont {Schneider}(2021)}]{giri2021emulation}%
  \BibitemOpen
  \bibfield  {author} {\bibinfo {author} {\bibfnamefont {S.~K.}\ \bibnamefont {Giri}}\ and\ \bibinfo {author} {\bibfnamefont {A.}~\bibnamefont {Schneider}},\ }\href@noop {} {\bibfield  {journal} {\bibinfo  {journal} {Journal of Cosmology and Astroparticle Physics}\ }\textbf {\bibinfo {volume} {2021}}\bibinfo  {number} { (12)},\ \bibinfo {pages} {046}}\BibitemShut {NoStop}%
\bibitem [{\citenamefont {Aric{\`o}}\ \emph {et~al.}(2023)\citenamefont {Aric{\`o}}, \citenamefont {Angulo}, \citenamefont {Zennaro}, \citenamefont {Contreras}, \citenamefont {Chen},\ and\ \citenamefont {Hern{\'a}ndez-Monteagudo}}]{arico2023y3}%
  \BibitemOpen
\bibfield  {number} {  }\bibfield  {author} {\bibinfo {author} {\bibfnamefont {G.}~\bibnamefont {Aric{\`o}}}, \bibinfo {author} {\bibfnamefont {R.~E.}\ \bibnamefont {Angulo}}, \bibinfo {author} {\bibfnamefont {M.}~\bibnamefont {Zennaro}}, \bibinfo {author} {\bibfnamefont {S.}~\bibnamefont {Contreras}}, \bibinfo {author} {\bibfnamefont {A.}~\bibnamefont {Chen}},\ and\ \bibinfo {author} {\bibfnamefont {C.}~\bibnamefont {Hern{\'a}ndez-Monteagudo}},\ }\href@noop {} {\bibfield  {journal} {\bibinfo  {journal} {Astronomy \& Astrophysics}\ }\textbf {\bibinfo {volume} {678}},\ \bibinfo {pages} {A109} (\bibinfo {year} {2023})}\BibitemShut {NoStop}%
\bibitem [{\citenamefont {Bigwood}\ \emph {et~al.}(2025{\natexlab{b}})\citenamefont {Bigwood}, \citenamefont {McCullough}, \citenamefont {Siegel}, \citenamefont {Amon}, \citenamefont {Efstathiou}, \citenamefont {Sanchez-Cid}, \citenamefont {Legnani}, \citenamefont {Gruen}, \citenamefont {Blazek}, \citenamefont {Doux} \emph {et~al.}}]{bigwood2025confronting}%
  \BibitemOpen
  \bibfield  {author} {\bibinfo {author} {\bibfnamefont {L.}~\bibnamefont {Bigwood}}, \bibinfo {author} {\bibfnamefont {J.}~\bibnamefont {McCullough}}, \bibinfo {author} {\bibfnamefont {J.}~\bibnamefont {Siegel}}, \bibinfo {author} {\bibfnamefont {A.}~\bibnamefont {Amon}}, \bibinfo {author} {\bibfnamefont {G.}~\bibnamefont {Efstathiou}}, \bibinfo {author} {\bibfnamefont {D.}~\bibnamefont {Sanchez-Cid}}, \bibinfo {author} {\bibfnamefont {E.}~\bibnamefont {Legnani}}, \bibinfo {author} {\bibfnamefont {D.}~\bibnamefont {Gruen}}, \bibinfo {author} {\bibfnamefont {J.}~\bibnamefont {Blazek}}, \bibinfo {author} {\bibfnamefont {C.}~\bibnamefont {Doux}}, \emph {et~al.},\ }\href@noop {} {\bibfield  {journal} {\bibinfo  {journal} {arXiv preprint arXiv:2512.04209}\ } (\bibinfo {year} {2025}{\natexlab{b}})}\BibitemShut {NoStop}%
\bibitem [{\citenamefont {Chen}\ \emph {et~al.}(2023{\natexlab{b}})\citenamefont {Chen}, \citenamefont {Zhang},\ and\ \citenamefont {Yang}}]{chen2023thermal}%
  \BibitemOpen
  \bibfield  {author} {\bibinfo {author} {\bibfnamefont {Z.}~\bibnamefont {Chen}}, \bibinfo {author} {\bibfnamefont {P.}~\bibnamefont {Zhang}},\ and\ \bibinfo {author} {\bibfnamefont {X.}~\bibnamefont {Yang}},\ }\href@noop {} {\bibfield  {journal} {\bibinfo  {journal} {The Astrophysical Journal}\ }\textbf {\bibinfo {volume} {953}},\ \bibinfo {pages} {188} (\bibinfo {year} {2023}{\natexlab{b}})}\BibitemShut {NoStop}%
\bibitem [{\citenamefont {Guachalla}\ \emph {et~al.}(2025)\citenamefont {Guachalla}, \citenamefont {Schaan}, \citenamefont {Hadzhiyska}, \citenamefont {Ferraro}, \citenamefont {Aguilar}, \citenamefont {Ahlen}, \citenamefont {Battaglia}, \citenamefont {Bianchi}, \citenamefont {Bond}, \citenamefont {Brooks} \emph {et~al.}}]{guachalla2025backlighting}%
  \BibitemOpen
  \bibfield  {author} {\bibinfo {author} {\bibfnamefont {B.~R.}\ \bibnamefont {Guachalla}}, \bibinfo {author} {\bibfnamefont {E.}~\bibnamefont {Schaan}}, \bibinfo {author} {\bibfnamefont {B.}~\bibnamefont {Hadzhiyska}}, \bibinfo {author} {\bibfnamefont {S.}~\bibnamefont {Ferraro}}, \bibinfo {author} {\bibfnamefont {J.~N.}\ \bibnamefont {Aguilar}}, \bibinfo {author} {\bibfnamefont {S.}~\bibnamefont {Ahlen}}, \bibinfo {author} {\bibfnamefont {N.}~\bibnamefont {Battaglia}}, \bibinfo {author} {\bibfnamefont {D.}~\bibnamefont {Bianchi}}, \bibinfo {author} {\bibfnamefont {R.}~\bibnamefont {Bond}}, \bibinfo {author} {\bibfnamefont {D.}~\bibnamefont {Brooks}}, \emph {et~al.},\ }\href@noop {} {\bibfield  {journal} {\bibinfo  {journal} {arXiv preprint arXiv:2503.19870}\ } (\bibinfo {year} {2025})}\BibitemShut {NoStop}%
\bibitem [{\citenamefont {Ferreira}\ \emph {et~al.}(2024)\citenamefont {Ferreira}, \citenamefont {Alonso}, \citenamefont {Garcia-Garcia},\ and\ \citenamefont {Chisari}}]{ferreira2024x}%
  \BibitemOpen
  \bibfield  {author} {\bibinfo {author} {\bibfnamefont {T.}~\bibnamefont {Ferreira}}, \bibinfo {author} {\bibfnamefont {D.}~\bibnamefont {Alonso}}, \bibinfo {author} {\bibfnamefont {C.}~\bibnamefont {Garcia-Garcia}},\ and\ \bibinfo {author} {\bibfnamefont {N.~E.}\ \bibnamefont {Chisari}},\ }\href@noop {} {\bibfield  {journal} {\bibinfo  {journal} {Physical Review Letters}\ }\textbf {\bibinfo {volume} {133}},\ \bibinfo {pages} {051001} (\bibinfo {year} {2024})}\BibitemShut {NoStop}%
\bibitem [{\citenamefont {Bulbul}\ \emph {et~al.}(2024)\citenamefont {Bulbul}, \citenamefont {Liu}, \citenamefont {Kluge}, \citenamefont {Zhang}, \citenamefont {Sanders}, \citenamefont {Bahar}, \citenamefont {Ghirardini}, \citenamefont {Artis}, \citenamefont {Seppi}, \citenamefont {Garrel} \emph {et~al.}}]{bulbul2024srg}%
  \BibitemOpen
  \bibfield  {author} {\bibinfo {author} {\bibfnamefont {E.}~\bibnamefont {Bulbul}}, \bibinfo {author} {\bibfnamefont {A.}~\bibnamefont {Liu}}, \bibinfo {author} {\bibfnamefont {M.}~\bibnamefont {Kluge}}, \bibinfo {author} {\bibfnamefont {X.}~\bibnamefont {Zhang}}, \bibinfo {author} {\bibfnamefont {J.}~\bibnamefont {Sanders}}, \bibinfo {author} {\bibfnamefont {Y.}~\bibnamefont {Bahar}}, \bibinfo {author} {\bibfnamefont {V.}~\bibnamefont {Ghirardini}}, \bibinfo {author} {\bibfnamefont {E.}~\bibnamefont {Artis}}, \bibinfo {author} {\bibfnamefont {R.}~\bibnamefont {Seppi}}, \bibinfo {author} {\bibfnamefont {C.}~\bibnamefont {Garrel}}, \emph {et~al.},\ }\href@noop {} {\bibfield  {journal} {\bibinfo  {journal} {Astronomy \& Astrophysics}\ }\textbf {\bibinfo {volume} {685}},\ \bibinfo {pages} {A106} (\bibinfo {year} {2024})}\BibitemShut {NoStop}%
\bibitem [{\citenamefont {Shull}\ \emph {et~al.}(2012)\citenamefont {Shull}, \citenamefont {Smith},\ and\ \citenamefont {Danforth}}]{shull2012baryon}%
  \BibitemOpen
  \bibfield  {author} {\bibinfo {author} {\bibfnamefont {J.~M.}\ \bibnamefont {Shull}}, \bibinfo {author} {\bibfnamefont {B.~D.}\ \bibnamefont {Smith}},\ and\ \bibinfo {author} {\bibfnamefont {C.~W.}\ \bibnamefont {Danforth}},\ }\href@noop {} {\bibfield  {journal} {\bibinfo  {journal} {The Astrophysical Journal}\ }\textbf {\bibinfo {volume} {759}},\ \bibinfo {pages} {23} (\bibinfo {year} {2012})}\BibitemShut {NoStop}%
\bibitem [{\citenamefont {McQuinn}(2013)}]{mcquinn2013locating}%
  \BibitemOpen
  \bibfield  {author} {\bibinfo {author} {\bibfnamefont {M.}~\bibnamefont {McQuinn}},\ }\href@noop {} {\bibfield  {journal} {\bibinfo  {journal} {The Astrophysical Journal Letters}\ }\textbf {\bibinfo {volume} {780}},\ \bibinfo {pages} {L33} (\bibinfo {year} {2013})}\BibitemShut {NoStop}%
\bibitem [{\citenamefont {Nicastro}\ \emph {et~al.}(2018)\citenamefont {Nicastro}, \citenamefont {Kaastra}, \citenamefont {Krongold}, \citenamefont {Borgani}, \citenamefont {Branchini}, \citenamefont {Cen}, \citenamefont {Dadina}, \citenamefont {Danforth}, \citenamefont {Elvis}, \citenamefont {Fiore} \emph {et~al.}}]{nicastro2018observations}%
  \BibitemOpen
  \bibfield  {author} {\bibinfo {author} {\bibfnamefont {F.}~\bibnamefont {Nicastro}}, \bibinfo {author} {\bibfnamefont {J.}~\bibnamefont {Kaastra}}, \bibinfo {author} {\bibfnamefont {Y.}~\bibnamefont {Krongold}}, \bibinfo {author} {\bibfnamefont {S.}~\bibnamefont {Borgani}}, \bibinfo {author} {\bibfnamefont {E.}~\bibnamefont {Branchini}}, \bibinfo {author} {\bibfnamefont {R.}~\bibnamefont {Cen}}, \bibinfo {author} {\bibfnamefont {M.}~\bibnamefont {Dadina}}, \bibinfo {author} {\bibfnamefont {C.}~\bibnamefont {Danforth}}, \bibinfo {author} {\bibfnamefont {M.}~\bibnamefont {Elvis}}, \bibinfo {author} {\bibfnamefont {F.}~\bibnamefont {Fiore}}, \emph {et~al.},\ }\href@noop {} {\bibfield  {journal} {\bibinfo  {journal} {Nature}\ }\textbf {\bibinfo {volume} {558}},\ \bibinfo {pages} {406} (\bibinfo {year} {2018})}\BibitemShut {NoStop}%
\bibitem [{\citenamefont {Macquart}\ \emph {et~al.}(2020)\citenamefont {Macquart}, \citenamefont {Prochaska}, \citenamefont {McQuinn}, \citenamefont {Bannister}, \citenamefont {Bhandari}, \citenamefont {Day}, \citenamefont {Deller}, \citenamefont {Ekers}, \citenamefont {James}, \citenamefont {Marnoch} \emph {et~al.}}]{macquart2020census}%
  \BibitemOpen
  \bibfield  {author} {\bibinfo {author} {\bibfnamefont {J.-P.}\ \bibnamefont {Macquart}}, \bibinfo {author} {\bibfnamefont {J.}~\bibnamefont {Prochaska}}, \bibinfo {author} {\bibfnamefont {M.}~\bibnamefont {McQuinn}}, \bibinfo {author} {\bibfnamefont {K.}~\bibnamefont {Bannister}}, \bibinfo {author} {\bibfnamefont {S.}~\bibnamefont {Bhandari}}, \bibinfo {author} {\bibfnamefont {C.}~\bibnamefont {Day}}, \bibinfo {author} {\bibfnamefont {A.}~\bibnamefont {Deller}}, \bibinfo {author} {\bibfnamefont {R.}~\bibnamefont {Ekers}}, \bibinfo {author} {\bibfnamefont {C.}~\bibnamefont {James}}, \bibinfo {author} {\bibfnamefont {L.}~\bibnamefont {Marnoch}}, \emph {et~al.},\ }\href@noop {} {\bibfield  {journal} {\bibinfo  {journal} {Nature}\ }\textbf {\bibinfo {volume} {581}},\ \bibinfo {pages} {391} (\bibinfo {year} {2020})}\BibitemShut {NoStop}%
\bibitem [{\citenamefont {Driver}(2021)}]{driver2021challenge}%
  \BibitemOpen
  \bibfield  {author} {\bibinfo {author} {\bibfnamefont {S.}~\bibnamefont {Driver}},\ }\href@noop {} {\bibfield  {journal} {\bibinfo  {journal} {Nature Astronomy}\ }\textbf {\bibinfo {volume} {5}},\ \bibinfo {pages} {852} (\bibinfo {year} {2021})}\BibitemShut {NoStop}%
\bibitem [{\citenamefont {Connor}\ \emph {et~al.}(2025)\citenamefont {Connor}, \citenamefont {Ravi}, \citenamefont {Sharma}, \citenamefont {Ocker}, \citenamefont {Faber}, \citenamefont {Hallinan}, \citenamefont {Harnach}, \citenamefont {Hellbourg}, \citenamefont {Hobbs}, \citenamefont {Hodge} \emph {et~al.}}]{connor2025gas}%
  \BibitemOpen
  \bibfield  {author} {\bibinfo {author} {\bibfnamefont {L.}~\bibnamefont {Connor}}, \bibinfo {author} {\bibfnamefont {V.}~\bibnamefont {Ravi}}, \bibinfo {author} {\bibfnamefont {K.}~\bibnamefont {Sharma}}, \bibinfo {author} {\bibfnamefont {S.~K.}\ \bibnamefont {Ocker}}, \bibinfo {author} {\bibfnamefont {J.}~\bibnamefont {Faber}}, \bibinfo {author} {\bibfnamefont {G.}~\bibnamefont {Hallinan}}, \bibinfo {author} {\bibfnamefont {C.}~\bibnamefont {Harnach}}, \bibinfo {author} {\bibfnamefont {G.}~\bibnamefont {Hellbourg}}, \bibinfo {author} {\bibfnamefont {R.}~\bibnamefont {Hobbs}}, \bibinfo {author} {\bibfnamefont {D.}~\bibnamefont {Hodge}}, \emph {et~al.},\ }\href@noop {} {\bibfield  {journal} {\bibinfo  {journal} {Nature Astronomy}\ ,\ \bibinfo {pages} {1}} (\bibinfo {year} {2025})}\BibitemShut {NoStop}%
\bibitem [{\citenamefont {Zhou}\ and\ \citenamefont {Zhang}(2026)}]{zhou2026cavendish}%
  \BibitemOpen
  \bibfield  {author} {\bibinfo {author} {\bibfnamefont {S.}~\bibnamefont {Zhou}}\ and\ \bibinfo {author} {\bibfnamefont {P.}~\bibnamefont {Zhang}},\ }\href@noop {} {\bibfield  {journal} {\bibinfo  {journal} {Physical Review D}\ }\textbf {\bibinfo {volume} {113}},\ \bibinfo {pages} {103516} (\bibinfo {year} {2026})}\BibitemShut {NoStop}%
\bibitem [{\citenamefont {Fukugita}\ and\ \citenamefont {Peebles}(2004)}]{fukugita2004cosmic}%
  \BibitemOpen
  \bibfield  {author} {\bibinfo {author} {\bibfnamefont {M.}~\bibnamefont {Fukugita}}\ and\ \bibinfo {author} {\bibfnamefont {P.~J.~E.}\ \bibnamefont {Peebles}},\ }\href@noop {} {\bibfield  {journal} {\bibinfo  {journal} {The Astrophysical Journal}\ }\textbf {\bibinfo {volume} {616}},\ \bibinfo {pages} {643} (\bibinfo {year} {2004})}\BibitemShut {NoStop}%
\bibitem [{\citenamefont {Cheng}\ \emph {et~al.}(2020)\citenamefont {Cheng}, \citenamefont {Ting}, \citenamefont {M{\'e}nard},\ and\ \citenamefont {Bruna}}]{cheng2020new}%
  \BibitemOpen
  \bibfield  {author} {\bibinfo {author} {\bibfnamefont {S.}~\bibnamefont {Cheng}}, \bibinfo {author} {\bibfnamefont {Y.-S.}\ \bibnamefont {Ting}}, \bibinfo {author} {\bibfnamefont {B.}~\bibnamefont {M{\'e}nard}},\ and\ \bibinfo {author} {\bibfnamefont {J.}~\bibnamefont {Bruna}},\ }\href@noop {} {\bibfield  {journal} {\bibinfo  {journal} {Monthly Notices of the Royal Astronomical Society}\ }\textbf {\bibinfo {volume} {499}},\ \bibinfo {pages} {5902} (\bibinfo {year} {2020})}\BibitemShut {NoStop}%
\bibitem [{\citenamefont {Springel}\ \emph {et~al.}(2017)\citenamefont {Springel}, \citenamefont {Pakmor}, \citenamefont {Pillepich}, \citenamefont {Weinberger}, \citenamefont {Nelson}, \citenamefont {Hernquist}, \citenamefont {Vogelsberger}, \citenamefont {Genel}, \citenamefont {Torrey}, \citenamefont {Marinacci},\ and\ \citenamefont {Naiman}}]{Springel_2017}%
  \BibitemOpen
  \bibfield  {author} {\bibinfo {author} {\bibfnamefont {V.}~\bibnamefont {Springel}}, \bibinfo {author} {\bibfnamefont {R.}~\bibnamefont {Pakmor}}, \bibinfo {author} {\bibfnamefont {A.}~\bibnamefont {Pillepich}}, \bibinfo {author} {\bibfnamefont {R.}~\bibnamefont {Weinberger}}, \bibinfo {author} {\bibfnamefont {D.}~\bibnamefont {Nelson}}, \bibinfo {author} {\bibfnamefont {L.}~\bibnamefont {Hernquist}}, \bibinfo {author} {\bibfnamefont {M.}~\bibnamefont {Vogelsberger}}, \bibinfo {author} {\bibfnamefont {S.}~\bibnamefont {Genel}}, \bibinfo {author} {\bibfnamefont {P.}~\bibnamefont {Torrey}}, \bibinfo {author} {\bibfnamefont {F.}~\bibnamefont {Marinacci}},\ and\ \bibinfo {author} {\bibfnamefont {J.}~\bibnamefont {Naiman}},\ }\href {https://doi.org/10.1093/mnras/stx3304} {\bibfield  {journal} {\bibinfo  {journal} {Monthly Notices of the Royal Astronomical Society}\ }\textbf {\bibinfo {volume} {475}},\ \bibinfo {pages} {676} (\bibinfo {year} {2017})}\BibitemShut {NoStop}%
\bibitem [{\citenamefont {Nelson}\ \emph {et~al.}(2017)\citenamefont {Nelson}, \citenamefont {Pillepich}, \citenamefont {Springel}, \citenamefont {Weinberger}, \citenamefont {Hernquist}, \citenamefont {Pakmor}, \citenamefont {Genel}, \citenamefont {Torrey}, \citenamefont {Vogelsberger}, \citenamefont {Kauffmann}, \citenamefont {Marinacci},\ and\ \citenamefont {Naiman}}]{Nelson_2017}%
  \BibitemOpen
  \bibfield  {author} {\bibinfo {author} {\bibfnamefont {D.}~\bibnamefont {Nelson}}, \bibinfo {author} {\bibfnamefont {A.}~\bibnamefont {Pillepich}}, \bibinfo {author} {\bibfnamefont {V.}~\bibnamefont {Springel}}, \bibinfo {author} {\bibfnamefont {R.}~\bibnamefont {Weinberger}}, \bibinfo {author} {\bibfnamefont {L.}~\bibnamefont {Hernquist}}, \bibinfo {author} {\bibfnamefont {R.}~\bibnamefont {Pakmor}}, \bibinfo {author} {\bibfnamefont {S.}~\bibnamefont {Genel}}, \bibinfo {author} {\bibfnamefont {P.}~\bibnamefont {Torrey}}, \bibinfo {author} {\bibfnamefont {M.}~\bibnamefont {Vogelsberger}}, \bibinfo {author} {\bibfnamefont {G.}~\bibnamefont {Kauffmann}}, \bibinfo {author} {\bibfnamefont {F.}~\bibnamefont {Marinacci}},\ and\ \bibinfo {author} {\bibfnamefont {J.}~\bibnamefont {Naiman}},\ }\href {https://doi.org/10.1093/mnras/stx3040} {\bibfield  {journal} {\bibinfo  {journal} {Monthly Notices of the Royal Astronomical Society}\ }\textbf {\bibinfo {volume} {475}},\ \bibinfo {pages} {624} (\bibinfo {year} {2017})}\BibitemShut {NoStop}%
\bibitem [{\citenamefont {Pillepich}\ \emph {et~al.}(2017)\citenamefont {Pillepich}, \citenamefont {Nelson}, \citenamefont {Hernquist}, \citenamefont {Springel}, \citenamefont {Pakmor}, \citenamefont {Torrey}, \citenamefont {Weinberger}, \citenamefont {Genel}, \citenamefont {Naiman}, \citenamefont {Marinacci},\ and\ \citenamefont {Vogelsberger}}]{Pillepich_2017}%
  \BibitemOpen
  \bibfield  {author} {\bibinfo {author} {\bibfnamefont {A.}~\bibnamefont {Pillepich}}, \bibinfo {author} {\bibfnamefont {D.}~\bibnamefont {Nelson}}, \bibinfo {author} {\bibfnamefont {L.}~\bibnamefont {Hernquist}}, \bibinfo {author} {\bibfnamefont {V.}~\bibnamefont {Springel}}, \bibinfo {author} {\bibfnamefont {R.}~\bibnamefont {Pakmor}}, \bibinfo {author} {\bibfnamefont {P.}~\bibnamefont {Torrey}}, \bibinfo {author} {\bibfnamefont {R.}~\bibnamefont {Weinberger}}, \bibinfo {author} {\bibfnamefont {S.}~\bibnamefont {Genel}}, \bibinfo {author} {\bibfnamefont {J.~P.}\ \bibnamefont {Naiman}}, \bibinfo {author} {\bibfnamefont {F.}~\bibnamefont {Marinacci}},\ and\ \bibinfo {author} {\bibfnamefont {M.}~\bibnamefont {Vogelsberger}},\ }\href {https://doi.org/10.1093/mnras/stx3112} {\bibfield  {journal} {\bibinfo  {journal} {Monthly Notices of the Royal Astronomical Society}\ }\textbf {\bibinfo {volume} {475}},\ \bibinfo {pages} {648} (\bibinfo {year} {2017})}\BibitemShut {NoStop}%
\bibitem [{\citenamefont {Naiman}\ \emph {et~al.}(2018)\citenamefont {Naiman}, \citenamefont {Pillepich}, \citenamefont {Springel}, \citenamefont {Ramirez-Ruiz}, \citenamefont {Torrey}, \citenamefont {Vogelsberger}, \citenamefont {Pakmor}, \citenamefont {Nelson}, \citenamefont {Marinacci}, \citenamefont {Hernquist}, \citenamefont {Weinberger},\ and\ \citenamefont {Genel}}]{Naiman_2018}%
  \BibitemOpen
  \bibfield  {author} {\bibinfo {author} {\bibfnamefont {J.~P.}\ \bibnamefont {Naiman}}, \bibinfo {author} {\bibfnamefont {A.}~\bibnamefont {Pillepich}}, \bibinfo {author} {\bibfnamefont {V.}~\bibnamefont {Springel}}, \bibinfo {author} {\bibfnamefont {E.}~\bibnamefont {Ramirez-Ruiz}}, \bibinfo {author} {\bibfnamefont {P.}~\bibnamefont {Torrey}}, \bibinfo {author} {\bibfnamefont {M.}~\bibnamefont {Vogelsberger}}, \bibinfo {author} {\bibfnamefont {R.}~\bibnamefont {Pakmor}}, \bibinfo {author} {\bibfnamefont {D.}~\bibnamefont {Nelson}}, \bibinfo {author} {\bibfnamefont {F.}~\bibnamefont {Marinacci}}, \bibinfo {author} {\bibfnamefont {L.}~\bibnamefont {Hernquist}}, \bibinfo {author} {\bibfnamefont {R.}~\bibnamefont {Weinberger}},\ and\ \bibinfo {author} {\bibfnamefont {S.}~\bibnamefont {Genel}},\ }\href {https://doi.org/10.1093/mnras/sty618} {\bibfield  {journal} {\bibinfo  {journal} {Monthly Notices of the Royal Astronomical Society}\ }\textbf {\bibinfo {volume} {477}},\ \bibinfo {pages} {1206} (\bibinfo {year} {2018})}\BibitemShut {NoStop}%
\bibitem [{\citenamefont {Marinacci}\ \emph {et~al.}(2018)\citenamefont {Marinacci}, \citenamefont {Vogelsberger}, \citenamefont {Pakmor}, \citenamefont {Torrey}, \citenamefont {Springel}, \citenamefont {Hernquist}, \citenamefont {Nelson}, \citenamefont {Weinberger}, \citenamefont {Pillepich}, \citenamefont {Naiman},\ and\ \citenamefont {Genel}}]{Marinacci_2018}%
  \BibitemOpen
  \bibfield  {author} {\bibinfo {author} {\bibfnamefont {F.}~\bibnamefont {Marinacci}}, \bibinfo {author} {\bibfnamefont {M.}~\bibnamefont {Vogelsberger}}, \bibinfo {author} {\bibfnamefont {R.}~\bibnamefont {Pakmor}}, \bibinfo {author} {\bibfnamefont {P.}~\bibnamefont {Torrey}}, \bibinfo {author} {\bibfnamefont {V.}~\bibnamefont {Springel}}, \bibinfo {author} {\bibfnamefont {L.}~\bibnamefont {Hernquist}}, \bibinfo {author} {\bibfnamefont {D.}~\bibnamefont {Nelson}}, \bibinfo {author} {\bibfnamefont {R.}~\bibnamefont {Weinberger}}, \bibinfo {author} {\bibfnamefont {A.}~\bibnamefont {Pillepich}}, \bibinfo {author} {\bibfnamefont {J.}~\bibnamefont {Naiman}},\ and\ \bibinfo {author} {\bibfnamefont {S.}~\bibnamefont {Genel}},\ }\bibfield  {journal} {\bibinfo  {journal} {Monthly Notices of the Royal Astronomical Society}\ }\href {https://doi.org/10.1093/mnras/sty2206} {10.1093/mnras/sty2206} (\bibinfo {year} {2018})\BibitemShut {NoStop}%
\bibitem [{\citenamefont {Vogelsberger}\ \emph {et~al.}(2014{\natexlab{a}})\citenamefont {Vogelsberger}, \citenamefont {Genel}, \citenamefont {Springel}, \citenamefont {Torrey}, \citenamefont {Sijacki}, \citenamefont {Xu}, \citenamefont {Snyder}, \citenamefont {Bird}, \citenamefont {Nelson},\ and\ \citenamefont {Hernquist}}]{vogelsberger2014properties}%
  \BibitemOpen
  \bibfield  {author} {\bibinfo {author} {\bibfnamefont {M.}~\bibnamefont {Vogelsberger}}, \bibinfo {author} {\bibfnamefont {S.}~\bibnamefont {Genel}}, \bibinfo {author} {\bibfnamefont {V.}~\bibnamefont {Springel}}, \bibinfo {author} {\bibfnamefont {P.}~\bibnamefont {Torrey}}, \bibinfo {author} {\bibfnamefont {D.}~\bibnamefont {Sijacki}}, \bibinfo {author} {\bibfnamefont {D.}~\bibnamefont {Xu}}, \bibinfo {author} {\bibfnamefont {G.}~\bibnamefont {Snyder}}, \bibinfo {author} {\bibfnamefont {S.}~\bibnamefont {Bird}}, \bibinfo {author} {\bibfnamefont {D.}~\bibnamefont {Nelson}},\ and\ \bibinfo {author} {\bibfnamefont {L.}~\bibnamefont {Hernquist}},\ }\href@noop {} {\bibfield  {journal} {\bibinfo  {journal} {Nature}\ }\textbf {\bibinfo {volume} {509}},\ \bibinfo {pages} {177} (\bibinfo {year} {2014}{\natexlab{a}})}\BibitemShut {NoStop}%
\bibitem [{\citenamefont {Vogelsberger}\ \emph {et~al.}(2014{\natexlab{b}})\citenamefont {Vogelsberger}, \citenamefont {Genel}, \citenamefont {Springel}, \citenamefont {Torrey}, \citenamefont {Sijacki}, \citenamefont {Xu}, \citenamefont {Snyder}, \citenamefont {Nelson},\ and\ \citenamefont {Hernquist}}]{vogelsberger2014introducing}%
  \BibitemOpen
  \bibfield  {author} {\bibinfo {author} {\bibfnamefont {M.}~\bibnamefont {Vogelsberger}}, \bibinfo {author} {\bibfnamefont {S.}~\bibnamefont {Genel}}, \bibinfo {author} {\bibfnamefont {V.}~\bibnamefont {Springel}}, \bibinfo {author} {\bibfnamefont {P.}~\bibnamefont {Torrey}}, \bibinfo {author} {\bibfnamefont {D.}~\bibnamefont {Sijacki}}, \bibinfo {author} {\bibfnamefont {D.}~\bibnamefont {Xu}}, \bibinfo {author} {\bibfnamefont {G.}~\bibnamefont {Snyder}}, \bibinfo {author} {\bibfnamefont {D.}~\bibnamefont {Nelson}},\ and\ \bibinfo {author} {\bibfnamefont {L.}~\bibnamefont {Hernquist}},\ }\href@noop {} {\bibfield  {journal} {\bibinfo  {journal} {Monthly Notices of the Royal Astronomical Society}\ }\textbf {\bibinfo {volume} {444}},\ \bibinfo {pages} {1518} (\bibinfo {year} {2014}{\natexlab{b}})}\BibitemShut {NoStop}%
\bibitem [{\citenamefont {Genel}\ \emph {et~al.}(2014)\citenamefont {Genel}, \citenamefont {Vogelsberger}, \citenamefont {Springel}, \citenamefont {Sijacki}, \citenamefont {Nelson}, \citenamefont {Snyder}, \citenamefont {Rodriguez-Gomez}, \citenamefont {Torrey},\ and\ \citenamefont {Hernquist}}]{genel2014introducing}%
  \BibitemOpen
  \bibfield  {author} {\bibinfo {author} {\bibfnamefont {S.}~\bibnamefont {Genel}}, \bibinfo {author} {\bibfnamefont {M.}~\bibnamefont {Vogelsberger}}, \bibinfo {author} {\bibfnamefont {V.}~\bibnamefont {Springel}}, \bibinfo {author} {\bibfnamefont {D.}~\bibnamefont {Sijacki}}, \bibinfo {author} {\bibfnamefont {D.}~\bibnamefont {Nelson}}, \bibinfo {author} {\bibfnamefont {G.}~\bibnamefont {Snyder}}, \bibinfo {author} {\bibfnamefont {V.}~\bibnamefont {Rodriguez-Gomez}}, \bibinfo {author} {\bibfnamefont {P.}~\bibnamefont {Torrey}},\ and\ \bibinfo {author} {\bibfnamefont {L.}~\bibnamefont {Hernquist}},\ }\href@noop {} {\bibfield  {journal} {\bibinfo  {journal} {Monthly Notices of the Royal Astronomical Society}\ }\textbf {\bibinfo {volume} {445}},\ \bibinfo {pages} {175} (\bibinfo {year} {2014})}\BibitemShut {NoStop}%
\bibitem [{\citenamefont {Sijacki}\ \emph {et~al.}(2015)\citenamefont {Sijacki}, \citenamefont {Vogelsberger}, \citenamefont {Genel}, \citenamefont {Springel}, \citenamefont {Torrey}, \citenamefont {Snyder}, \citenamefont {Nelson},\ and\ \citenamefont {Hernquist}}]{sijacki2015illustris}%
  \BibitemOpen
  \bibfield  {author} {\bibinfo {author} {\bibfnamefont {D.}~\bibnamefont {Sijacki}}, \bibinfo {author} {\bibfnamefont {M.}~\bibnamefont {Vogelsberger}}, \bibinfo {author} {\bibfnamefont {S.}~\bibnamefont {Genel}}, \bibinfo {author} {\bibfnamefont {V.}~\bibnamefont {Springel}}, \bibinfo {author} {\bibfnamefont {P.}~\bibnamefont {Torrey}}, \bibinfo {author} {\bibfnamefont {G.~F.}\ \bibnamefont {Snyder}}, \bibinfo {author} {\bibfnamefont {D.}~\bibnamefont {Nelson}},\ and\ \bibinfo {author} {\bibfnamefont {L.}~\bibnamefont {Hernquist}},\ }\href@noop {} {\bibfield  {journal} {\bibinfo  {journal} {Monthly Notices of the Royal Astronomical Society}\ }\textbf {\bibinfo {volume} {452}},\ \bibinfo {pages} {575} (\bibinfo {year} {2015})}\BibitemShut {NoStop}%
\bibitem [{\citenamefont {Mallat}(2012)}]{mallat2012group}%
  \BibitemOpen
  \bibfield  {author} {\bibinfo {author} {\bibfnamefont {S.}~\bibnamefont {Mallat}},\ }\href@noop {} {\bibfield  {journal} {\bibinfo  {journal} {Communications on Pure and Applied Mathematics}\ }\textbf {\bibinfo {volume} {65}},\ \bibinfo {pages} {1331} (\bibinfo {year} {2012})}\BibitemShut {NoStop}%
\bibitem [{\citenamefont {Bruna}\ and\ \citenamefont {Mallat}(2013)}]{bruna2013invariant}%
  \BibitemOpen
  \bibfield  {author} {\bibinfo {author} {\bibfnamefont {J.}~\bibnamefont {Bruna}}\ and\ \bibinfo {author} {\bibfnamefont {S.}~\bibnamefont {Mallat}},\ }\href@noop {} {\bibfield  {journal} {\bibinfo  {journal} {IEEE transactions on pattern analysis and machine intelligence}\ }\textbf {\bibinfo {volume} {35}},\ \bibinfo {pages} {1872} (\bibinfo {year} {2013})}\BibitemShut {NoStop}%
\bibitem [{\citenamefont {Eickenberg}\ \emph {et~al.}(2018)\citenamefont {Eickenberg}, \citenamefont {Exarchakis}, \citenamefont {Hirn}, \citenamefont {Mallat},\ and\ \citenamefont {Thiry}}]{eickenberg2018solid}%
  \BibitemOpen
  \bibfield  {author} {\bibinfo {author} {\bibfnamefont {M.}~\bibnamefont {Eickenberg}}, \bibinfo {author} {\bibfnamefont {G.}~\bibnamefont {Exarchakis}}, \bibinfo {author} {\bibfnamefont {M.}~\bibnamefont {Hirn}}, \bibinfo {author} {\bibfnamefont {S.}~\bibnamefont {Mallat}},\ and\ \bibinfo {author} {\bibfnamefont {L.}~\bibnamefont {Thiry}},\ }\href@noop {} {\bibfield  {journal} {\bibinfo  {journal} {The Journal of chemical physics}\ }\textbf {\bibinfo {volume} {148}} (\bibinfo {year} {2018})}\BibitemShut {NoStop}%
\bibitem [{\citenamefont {Limber}(1953)}]{limber1953analysis}%
  \BibitemOpen
  \bibfield  {author} {\bibinfo {author} {\bibfnamefont {D.~N.}\ \bibnamefont {Limber}},\ }\href@noop {} {\bibfield  {journal} {\bibinfo  {journal} {Astrophysical Journal, vol. 117, p. 134}\ }\textbf {\bibinfo {volume} {117}},\ \bibinfo {pages} {134} (\bibinfo {year} {1953})}\BibitemShut {NoStop}%
\bibitem [{\citenamefont {LoVerde}\ and\ \citenamefont {Afshordi}(2008)}]{loverde2008extended}%
  \BibitemOpen
  \bibfield  {author} {\bibinfo {author} {\bibfnamefont {M.}~\bibnamefont {LoVerde}}\ and\ \bibinfo {author} {\bibfnamefont {N.}~\bibnamefont {Afshordi}},\ }\href@noop {} {\bibfield  {journal} {\bibinfo  {journal} {Physical Review D—Particles, Fields, Gravitation, and Cosmology}\ }\textbf {\bibinfo {volume} {78}},\ \bibinfo {pages} {123506} (\bibinfo {year} {2008})}\BibitemShut {NoStop}%
\bibitem [{\citenamefont {Abbott}\ \emph {et~al.}(2026)\citenamefont {Abbott}, \citenamefont {Aguena}, \citenamefont {Alarcon}, \citenamefont {Alves}, \citenamefont {Amon}, \citenamefont {Anbajagane}, \citenamefont {Andrade-Oliveira}, \citenamefont {d'Assignies}, \citenamefont {Avila}, \citenamefont {Bacon} \emph {et~al.}}]{abbott2026dark}%
  \BibitemOpen
  \bibfield  {author} {\bibinfo {author} {\bibfnamefont {T.}~\bibnamefont {Abbott}}, \bibinfo {author} {\bibfnamefont {M.}~\bibnamefont {Aguena}}, \bibinfo {author} {\bibfnamefont {A.}~\bibnamefont {Alarcon}}, \bibinfo {author} {\bibfnamefont {O.}~\bibnamefont {Alves}}, \bibinfo {author} {\bibfnamefont {A.}~\bibnamefont {Amon}}, \bibinfo {author} {\bibfnamefont {D.}~\bibnamefont {Anbajagane}}, \bibinfo {author} {\bibfnamefont {F.}~\bibnamefont {Andrade-Oliveira}}, \bibinfo {author} {\bibfnamefont {W.}~\bibnamefont {d'Assignies}}, \bibinfo {author} {\bibfnamefont {S.}~\bibnamefont {Avila}}, \bibinfo {author} {\bibfnamefont {D.}~\bibnamefont {Bacon}}, \emph {et~al.},\ }\href@noop {} {\bibfield  {journal} {\bibinfo  {journal} {arXiv preprint arXiv:2602.10065}\ } (\bibinfo {year} {2026})}\BibitemShut {NoStop}%
\bibitem [{\citenamefont {St{\"o}lzner}\ \emph {et~al.}(2025)\citenamefont {St{\"o}lzner}, \citenamefont {Wright}, \citenamefont {Asgari}, \citenamefont {Heymans}, \citenamefont {Hildebrandt}, \citenamefont {Hoekstra}, \citenamefont {Joachimi}, \citenamefont {Kuijken}, \citenamefont {Li}, \citenamefont {Mahony} \emph {et~al.}}]{stolzner2025kids}%
  \BibitemOpen
  \bibfield  {author} {\bibinfo {author} {\bibfnamefont {B.}~\bibnamefont {St{\"o}lzner}}, \bibinfo {author} {\bibfnamefont {A.~H.}\ \bibnamefont {Wright}}, \bibinfo {author} {\bibfnamefont {M.}~\bibnamefont {Asgari}}, \bibinfo {author} {\bibfnamefont {C.}~\bibnamefont {Heymans}}, \bibinfo {author} {\bibfnamefont {H.}~\bibnamefont {Hildebrandt}}, \bibinfo {author} {\bibfnamefont {H.}~\bibnamefont {Hoekstra}}, \bibinfo {author} {\bibfnamefont {B.}~\bibnamefont {Joachimi}}, \bibinfo {author} {\bibfnamefont {K.}~\bibnamefont {Kuijken}}, \bibinfo {author} {\bibfnamefont {S.-S.}\ \bibnamefont {Li}}, \bibinfo {author} {\bibfnamefont {C.}~\bibnamefont {Mahony}}, \emph {et~al.},\ }\href@noop {} {\bibfield  {journal} {\bibinfo  {journal} {Astronomy \& Astrophysics}\ }\textbf {\bibinfo {volume} {702}},\ \bibinfo {pages} {A169} (\bibinfo {year} {2025})}\BibitemShut {NoStop}%
\bibitem [{\citenamefont {Li}\ \emph {et~al.}(2023)\citenamefont {Li}, \citenamefont {Zhang}, \citenamefont {Sugiyama}, \citenamefont {Dalal}, \citenamefont {Terasawa}, \citenamefont {Rau}, \citenamefont {Mandelbaum}, \citenamefont {Takada}, \citenamefont {More}, \citenamefont {Strauss} \emph {et~al.}}]{li2023hyper}%
  \BibitemOpen
  \bibfield  {author} {\bibinfo {author} {\bibfnamefont {X.}~\bibnamefont {Li}}, \bibinfo {author} {\bibfnamefont {T.}~\bibnamefont {Zhang}}, \bibinfo {author} {\bibfnamefont {S.}~\bibnamefont {Sugiyama}}, \bibinfo {author} {\bibfnamefont {R.}~\bibnamefont {Dalal}}, \bibinfo {author} {\bibfnamefont {R.}~\bibnamefont {Terasawa}}, \bibinfo {author} {\bibfnamefont {M.~M.}\ \bibnamefont {Rau}}, \bibinfo {author} {\bibfnamefont {R.}~\bibnamefont {Mandelbaum}}, \bibinfo {author} {\bibfnamefont {M.}~\bibnamefont {Takada}}, \bibinfo {author} {\bibfnamefont {S.}~\bibnamefont {More}}, \bibinfo {author} {\bibfnamefont {M.~A.}\ \bibnamefont {Strauss}}, \emph {et~al.},\ }\href@noop {} {\bibfield  {journal} {\bibinfo  {journal} {Physical Review D}\ }\textbf {\bibinfo {volume} {108}},\ \bibinfo {pages} {123518} (\bibinfo {year} {2023})}\BibitemShut {NoStop}%
\bibitem [{\citenamefont {Taylor}(2001)}]{taylor2001imaging}%
  \BibitemOpen
  \bibfield  {author} {\bibinfo {author} {\bibfnamefont {A.}~\bibnamefont {Taylor}},\ }\href@noop {} {\bibfield  {journal} {\bibinfo  {journal} {Arxiv preprint astro-ph/0111605}\ } (\bibinfo {year} {2001})}\BibitemShut {NoStop}%
\bibitem [{\citenamefont {Hu}\ and\ \citenamefont {Keeton}(2002)}]{hu2002three}%
  \BibitemOpen
  \bibfield  {author} {\bibinfo {author} {\bibfnamefont {W.}~\bibnamefont {Hu}}\ and\ \bibinfo {author} {\bibfnamefont {C.~R.}\ \bibnamefont {Keeton}},\ }\href@noop {} {\bibfield  {journal} {\bibinfo  {journal} {Physical Review D}\ }\textbf {\bibinfo {volume} {66}},\ \bibinfo {pages} {063506} (\bibinfo {year} {2002})}\BibitemShut {NoStop}%
\bibitem [{\citenamefont {Bacon}\ and\ \citenamefont {Taylor}(2003)}]{bacon2003mapping}%
  \BibitemOpen
  \bibfield  {author} {\bibinfo {author} {\bibfnamefont {D.}~\bibnamefont {Bacon}}\ and\ \bibinfo {author} {\bibfnamefont {A.}~\bibnamefont {Taylor}},\ }\href@noop {} {\bibfield  {journal} {\bibinfo  {journal} {Monthly Notices of the Royal Astronomical Society}\ }\textbf {\bibinfo {volume} {344}},\ \bibinfo {pages} {1307} (\bibinfo {year} {2003})}\BibitemShut {NoStop}%
\bibitem [{\citenamefont {Aghanim}\ \emph {et~al.}(2020)\citenamefont {Aghanim} \emph {et~al.}}]{aghanim2020planck}%
  \BibitemOpen
  \bibfield  {author} {\bibinfo {author} {\bibfnamefont {N.}~\bibnamefont {Aghanim}} \emph {et~al.},\ }\href@noop {} {\bibfield  {journal} {\bibinfo  {journal} {Astron. Astrophys}\ }\textbf {\bibinfo {volume} {641}},\ \bibinfo {pages} {A6} (\bibinfo {year} {2020})}\BibitemShut {NoStop}%
\bibitem [{\citenamefont {Louis}\ \emph {et~al.}(2025)\citenamefont {Louis}, \citenamefont {La~Posta}, \citenamefont {Atkins}, \citenamefont {Jense}, \citenamefont {Abril-Cabezas}, \citenamefont {Addison}, \citenamefont {Ade}, \citenamefont {Aiola}, \citenamefont {Alford}, \citenamefont {Alonso} \emph {et~al.}}]{louis2025atacama}%
  \BibitemOpen
  \bibfield  {author} {\bibinfo {author} {\bibfnamefont {T.}~\bibnamefont {Louis}}, \bibinfo {author} {\bibfnamefont {A.}~\bibnamefont {La~Posta}}, \bibinfo {author} {\bibfnamefont {Z.}~\bibnamefont {Atkins}}, \bibinfo {author} {\bibfnamefont {H.~T.}\ \bibnamefont {Jense}}, \bibinfo {author} {\bibfnamefont {I.}~\bibnamefont {Abril-Cabezas}}, \bibinfo {author} {\bibfnamefont {G.~E.}\ \bibnamefont {Addison}}, \bibinfo {author} {\bibfnamefont {P.~A.}\ \bibnamefont {Ade}}, \bibinfo {author} {\bibfnamefont {S.}~\bibnamefont {Aiola}}, \bibinfo {author} {\bibfnamefont {T.}~\bibnamefont {Alford}}, \bibinfo {author} {\bibfnamefont {D.}~\bibnamefont {Alonso}}, \emph {et~al.},\ }\href@noop {} {\bibfield  {journal} {\bibinfo  {journal} {arXiv preprint arXiv:2503.14452}\ } (\bibinfo {year} {2025})}\BibitemShut {NoStop}%
\bibitem [{\citenamefont {Alam}\ \emph {et~al.}(2021)\citenamefont {Alam}, \citenamefont {Aubert}, \citenamefont {Avila}, \citenamefont {Balland}, \citenamefont {Bautista}, \citenamefont {Bershady}, \citenamefont {Bizyaev}, \citenamefont {Blanton}, \citenamefont {Bolton}, \citenamefont {Bovy} \emph {et~al.}}]{alam2021completed}%
  \BibitemOpen
  \bibfield  {author} {\bibinfo {author} {\bibfnamefont {S.}~\bibnamefont {Alam}}, \bibinfo {author} {\bibfnamefont {M.}~\bibnamefont {Aubert}}, \bibinfo {author} {\bibfnamefont {S.}~\bibnamefont {Avila}}, \bibinfo {author} {\bibfnamefont {C.}~\bibnamefont {Balland}}, \bibinfo {author} {\bibfnamefont {J.~E.}\ \bibnamefont {Bautista}}, \bibinfo {author} {\bibfnamefont {M.~A.}\ \bibnamefont {Bershady}}, \bibinfo {author} {\bibfnamefont {D.}~\bibnamefont {Bizyaev}}, \bibinfo {author} {\bibfnamefont {M.~R.}\ \bibnamefont {Blanton}}, \bibinfo {author} {\bibfnamefont {A.~S.}\ \bibnamefont {Bolton}}, \bibinfo {author} {\bibfnamefont {J.}~\bibnamefont {Bovy}}, \emph {et~al.},\ }\href@noop {} {\bibfield  {journal} {\bibinfo  {journal} {Physical Review D}\ }\textbf {\bibinfo {volume} {103}},\ \bibinfo {pages} {083533} (\bibinfo {year} {2021})}\BibitemShut {NoStop}%
\bibitem [{\citenamefont {Karim}\ \emph {et~al.}(2025)\citenamefont {Karim}, \citenamefont {Aguilar}, \citenamefont {Ahlen}, \citenamefont {Alam}, \citenamefont {Allen}, \citenamefont {Allende~Prieto}, \citenamefont {Alves}, \citenamefont {Anand}, \citenamefont {Andrade}, \citenamefont {Armengaud} \emph {et~al.}}]{karim2025desi}%
  \BibitemOpen
  \bibfield  {author} {\bibinfo {author} {\bibfnamefont {M.~A.}\ \bibnamefont {Karim}}, \bibinfo {author} {\bibfnamefont {J.}~\bibnamefont {Aguilar}}, \bibinfo {author} {\bibfnamefont {S.}~\bibnamefont {Ahlen}}, \bibinfo {author} {\bibfnamefont {S.}~\bibnamefont {Alam}}, \bibinfo {author} {\bibfnamefont {L.}~\bibnamefont {Allen}}, \bibinfo {author} {\bibfnamefont {C.}~\bibnamefont {Allende~Prieto}}, \bibinfo {author} {\bibfnamefont {O.}~\bibnamefont {Alves}}, \bibinfo {author} {\bibfnamefont {A.}~\bibnamefont {Anand}}, \bibinfo {author} {\bibfnamefont {U.}~\bibnamefont {Andrade}}, \bibinfo {author} {\bibfnamefont {E.}~\bibnamefont {Armengaud}}, \emph {et~al.},\ }\href@noop {} {\bibfield  {journal} {\bibinfo  {journal} {arXiv e-prints}\ ,\ \bibinfo {pages} {arXiv}} (\bibinfo {year} {2025})}\BibitemShut {NoStop}%
\bibitem [{\citenamefont {Adame}\ \emph {et~al.}(2024)\citenamefont {Adame}, \citenamefont {Aguilar}, \citenamefont {Ahlen}, \citenamefont {Alam}, \citenamefont {Aldering}, \citenamefont {Alexander}, \citenamefont {Alfarsy}, \citenamefont {Prieto}, \citenamefont {Alvarez}, \citenamefont {Alves} \emph {et~al.}}]{adame2024validation}%
  \BibitemOpen
  \bibfield  {author} {\bibinfo {author} {\bibfnamefont {A.}~\bibnamefont {Adame}}, \bibinfo {author} {\bibfnamefont {J.}~\bibnamefont {Aguilar}}, \bibinfo {author} {\bibfnamefont {S.}~\bibnamefont {Ahlen}}, \bibinfo {author} {\bibfnamefont {S.}~\bibnamefont {Alam}}, \bibinfo {author} {\bibfnamefont {G.}~\bibnamefont {Aldering}}, \bibinfo {author} {\bibfnamefont {D.}~\bibnamefont {Alexander}}, \bibinfo {author} {\bibfnamefont {R.}~\bibnamefont {Alfarsy}}, \bibinfo {author} {\bibfnamefont {C.~A.}\ \bibnamefont {Prieto}}, \bibinfo {author} {\bibfnamefont {M.}~\bibnamefont {Alvarez}}, \bibinfo {author} {\bibfnamefont {O.}~\bibnamefont {Alves}}, \emph {et~al.},\ }\href@noop {} {\bibfield  {journal} {\bibinfo  {journal} {The Astronomical Journal}\ }\textbf {\bibinfo {volume} {167}},\ \bibinfo {pages} {62} (\bibinfo {year} {2024})}\BibitemShut {NoStop}%
\bibitem [{\citenamefont {Zhao}\ \emph {et~al.}(2024)\citenamefont {Zhao}, \citenamefont {Huang}, \citenamefont {He}, \citenamefont {Montero-Camacho}, \citenamefont {Liu}, \citenamefont {Renard}, \citenamefont {Tang}, \citenamefont {Verdier}, \citenamefont {Xu}, \citenamefont {Yang} \emph {et~al.}}]{zhao2024multiplexed}%
  \BibitemOpen
  \bibfield  {author} {\bibinfo {author} {\bibfnamefont {C.}~\bibnamefont {Zhao}}, \bibinfo {author} {\bibfnamefont {S.}~\bibnamefont {Huang}}, \bibinfo {author} {\bibfnamefont {M.}~\bibnamefont {He}}, \bibinfo {author} {\bibfnamefont {P.}~\bibnamefont {Montero-Camacho}}, \bibinfo {author} {\bibfnamefont {Y.}~\bibnamefont {Liu}}, \bibinfo {author} {\bibfnamefont {P.}~\bibnamefont {Renard}}, \bibinfo {author} {\bibfnamefont {Y.}~\bibnamefont {Tang}}, \bibinfo {author} {\bibfnamefont {A.}~\bibnamefont {Verdier}}, \bibinfo {author} {\bibfnamefont {W.}~\bibnamefont {Xu}}, \bibinfo {author} {\bibfnamefont {X.}~\bibnamefont {Yang}}, \emph {et~al.},\ }\href@noop {} {\bibfield  {journal} {\bibinfo  {journal} {arXiv preprint arXiv:2411.07970}\ } (\bibinfo {year} {2024})}\BibitemShut {NoStop}%
\bibitem [{\citenamefont {Zhang}(2023)}]{zhang2023physics}%
  \BibitemOpen
  \bibfield  {author} {\bibinfo {author} {\bibfnamefont {B.}~\bibnamefont {Zhang}},\ }\href@noop {} {\bibfield  {journal} {\bibinfo  {journal} {Reviews of Modern Physics}\ }\textbf {\bibinfo {volume} {95}},\ \bibinfo {pages} {035005} (\bibinfo {year} {2023})}\BibitemShut {NoStop}%
\bibitem [{\citenamefont {Hussaini}\ \emph {et~al.}(2025)\citenamefont {Hussaini}, \citenamefont {Connor}, \citenamefont {Konietzka}, \citenamefont {Ravi}, \citenamefont {Faber}, \citenamefont {Sharma},\ and\ \citenamefont {Sherman}}]{hussaini2025correlation}%
  \BibitemOpen
  \bibfield  {author} {\bibinfo {author} {\bibfnamefont {M.}~\bibnamefont {Hussaini}}, \bibinfo {author} {\bibfnamefont {L.}~\bibnamefont {Connor}}, \bibinfo {author} {\bibfnamefont {R.~M.}\ \bibnamefont {Konietzka}}, \bibinfo {author} {\bibfnamefont {V.}~\bibnamefont {Ravi}}, \bibinfo {author} {\bibfnamefont {J.}~\bibnamefont {Faber}}, \bibinfo {author} {\bibfnamefont {K.}~\bibnamefont {Sharma}},\ and\ \bibinfo {author} {\bibfnamefont {M.}~\bibnamefont {Sherman}},\ }\href@noop {} {\bibfield  {journal} {\bibinfo  {journal} {arXiv preprint arXiv:2506.04186}\ } (\bibinfo {year} {2025})}\BibitemShut {NoStop}%
\bibitem [{\citenamefont {Wang}\ \emph {et~al.}(2025)\citenamefont {Wang}, \citenamefont {Masui}, \citenamefont {Andrew}, \citenamefont {Fonseca}, \citenamefont {Gaensler}, \citenamefont {Joseph}, \citenamefont {Kaspi}, \citenamefont {Kharel}, \citenamefont {Lanman}, \citenamefont {Leung} \emph {et~al.}}]{wang2025measurement}%
  \BibitemOpen
  \bibfield  {author} {\bibinfo {author} {\bibfnamefont {H.}~\bibnamefont {Wang}}, \bibinfo {author} {\bibfnamefont {K.}~\bibnamefont {Masui}}, \bibinfo {author} {\bibfnamefont {S.}~\bibnamefont {Andrew}}, \bibinfo {author} {\bibfnamefont {E.}~\bibnamefont {Fonseca}}, \bibinfo {author} {\bibfnamefont {B.}~\bibnamefont {Gaensler}}, \bibinfo {author} {\bibfnamefont {R.}~\bibnamefont {Joseph}}, \bibinfo {author} {\bibfnamefont {V.~M.}\ \bibnamefont {Kaspi}}, \bibinfo {author} {\bibfnamefont {B.}~\bibnamefont {Kharel}}, \bibinfo {author} {\bibfnamefont {A.~E.}\ \bibnamefont {Lanman}}, \bibinfo {author} {\bibfnamefont {C.}~\bibnamefont {Leung}}, \emph {et~al.},\ }\href@noop {} {\bibfield  {journal} {\bibinfo  {journal} {arXiv preprint arXiv:2506.08932}\ } (\bibinfo {year} {2025})}\BibitemShut {NoStop}%
\bibitem [{\citenamefont {Leung}\ \emph {et~al.}(2025)\citenamefont {Leung}, \citenamefont {Borrow}, \citenamefont {Masui}, \citenamefont {Andrew}, \citenamefont {Chen}, \citenamefont {Schaye},\ and\ \citenamefont {Schaller}}]{leung2025nulling}%
  \BibitemOpen
  \bibfield  {author} {\bibinfo {author} {\bibfnamefont {C.}~\bibnamefont {Leung}}, \bibinfo {author} {\bibfnamefont {J.}~\bibnamefont {Borrow}}, \bibinfo {author} {\bibfnamefont {K.~W.}\ \bibnamefont {Masui}}, \bibinfo {author} {\bibfnamefont {S.}~\bibnamefont {Andrew}}, \bibinfo {author} {\bibfnamefont {K.-F.}\ \bibnamefont {Chen}}, \bibinfo {author} {\bibfnamefont {J.}~\bibnamefont {Schaye}},\ and\ \bibinfo {author} {\bibfnamefont {M.}~\bibnamefont {Schaller}},\ }\href@noop {} {\bibfield  {journal} {\bibinfo  {journal} {arXiv preprint arXiv:2509.19514}\ } (\bibinfo {year} {2025})}\BibitemShut {NoStop}%
\bibitem [{\citenamefont {Fialkov}\ and\ \citenamefont {Loeb}(2017)}]{fialkov2017fast}%
  \BibitemOpen
  \bibfield  {author} {\bibinfo {author} {\bibfnamefont {A.}~\bibnamefont {Fialkov}}\ and\ \bibinfo {author} {\bibfnamefont {A.}~\bibnamefont {Loeb}},\ }\href@noop {} {\bibfield  {journal} {\bibinfo  {journal} {The Astrophysical Journal Letters}\ }\textbf {\bibinfo {volume} {846}},\ \bibinfo {pages} {L27} (\bibinfo {year} {2017})}\BibitemShut {NoStop}%
\bibitem [{\citenamefont {Hallinan}\ \emph {et~al.}(2019)\citenamefont {Hallinan}, \citenamefont {Ravi}, \citenamefont {Weinreb}, \citenamefont {Kocz}, \citenamefont {Huang}, \citenamefont {Woody}, \citenamefont {Lamb}, \citenamefont {D’Addario}, \citenamefont {Catha}, \citenamefont {Shi} \emph {et~al.}}]{hallinan2019astro2020}%
  \BibitemOpen
  \bibfield  {author} {\bibinfo {author} {\bibfnamefont {G.}~\bibnamefont {Hallinan}}, \bibinfo {author} {\bibfnamefont {V.}~\bibnamefont {Ravi}}, \bibinfo {author} {\bibfnamefont {S.}~\bibnamefont {Weinreb}}, \bibinfo {author} {\bibfnamefont {J.}~\bibnamefont {Kocz}}, \bibinfo {author} {\bibfnamefont {Y.}~\bibnamefont {Huang}}, \bibinfo {author} {\bibfnamefont {D.}~\bibnamefont {Woody}}, \bibinfo {author} {\bibfnamefont {J.}~\bibnamefont {Lamb}}, \bibinfo {author} {\bibfnamefont {L.}~\bibnamefont {D’Addario}}, \bibinfo {author} {\bibfnamefont {M.}~\bibnamefont {Catha}}, \bibinfo {author} {\bibfnamefont {J.}~\bibnamefont {Shi}}, \emph {et~al.},\ }\href@noop {} {\bibfield  {journal} {\bibinfo  {journal} {arXiv preprint arXiv:1907.07648}\ } (\bibinfo {year} {2019})}\BibitemShut {NoStop}%
\bibitem [{\citenamefont {Lin}\ \emph {et~al.}(2022)\citenamefont {Lin}, \citenamefont {Lin}, \citenamefont {Li}, \citenamefont {Tseng}, \citenamefont {Jiang}, \citenamefont {Wang}, \citenamefont {Cheng}, \citenamefont {Pen}, \citenamefont {Chen}, \citenamefont {Chen} \emph {et~al.}}]{lin2022burstt}%
  \BibitemOpen
  \bibfield  {author} {\bibinfo {author} {\bibfnamefont {H.-H.}\ \bibnamefont {Lin}}, \bibinfo {author} {\bibfnamefont {K.-y.}\ \bibnamefont {Lin}}, \bibinfo {author} {\bibfnamefont {C.-T.}\ \bibnamefont {Li}}, \bibinfo {author} {\bibfnamefont {Y.-H.}\ \bibnamefont {Tseng}}, \bibinfo {author} {\bibfnamefont {H.}~\bibnamefont {Jiang}}, \bibinfo {author} {\bibfnamefont {J.-H.}\ \bibnamefont {Wang}}, \bibinfo {author} {\bibfnamefont {J.-C.}\ \bibnamefont {Cheng}}, \bibinfo {author} {\bibfnamefont {U.-L.}\ \bibnamefont {Pen}}, \bibinfo {author} {\bibfnamefont {M.-T.}\ \bibnamefont {Chen}}, \bibinfo {author} {\bibfnamefont {P.}~\bibnamefont {Chen}}, \emph {et~al.},\ }\href@noop {} {\bibfield  {journal} {\bibinfo  {journal} {Publications of the Astronomical Society of the Pacific}\ }\textbf {\bibinfo {volume} {134}},\ \bibinfo {pages} {094106} (\bibinfo {year} {2022})}\BibitemShut {NoStop}%
\bibitem [{\citenamefont {Vlah}\ \emph {et~al.}(2016)\citenamefont {Vlah}, \citenamefont {Castorina},\ and\ \citenamefont {White}}]{vlah2016gaussian}%
  \BibitemOpen
  \bibfield  {author} {\bibinfo {author} {\bibfnamefont {Z.}~\bibnamefont {Vlah}}, \bibinfo {author} {\bibfnamefont {E.}~\bibnamefont {Castorina}},\ and\ \bibinfo {author} {\bibfnamefont {M.}~\bibnamefont {White}},\ }\href@noop {} {\bibfield  {journal} {\bibinfo  {journal} {Journal of Cosmology and Astroparticle Physics}\ }\textbf {\bibinfo {volume} {2016}}\bibinfo  {number} { (12)},\ \bibinfo {pages} {007}}\BibitemShut {NoStop}%
\bibitem [{\citenamefont {Vlah}\ \emph {et~al.}(2015)\citenamefont {Vlah}, \citenamefont {Seljak},\ and\ \citenamefont {Baldauf}}]{vlah2015lagrangian}%
  \BibitemOpen
\bibfield  {number} {  }\bibfield  {author} {\bibinfo {author} {\bibfnamefont {Z.}~\bibnamefont {Vlah}}, \bibinfo {author} {\bibfnamefont {U.}~\bibnamefont {Seljak}},\ and\ \bibinfo {author} {\bibfnamefont {T.}~\bibnamefont {Baldauf}},\ }\href@noop {} {\bibfield  {journal} {\bibinfo  {journal} {Physical Review D}\ }\textbf {\bibinfo {volume} {91}},\ \bibinfo {pages} {023508} (\bibinfo {year} {2015})}\BibitemShut {NoStop}%
\bibitem [{\citenamefont {White}(2014)}]{white2014zel}%
  \BibitemOpen
  \bibfield  {author} {\bibinfo {author} {\bibfnamefont {M.}~\bibnamefont {White}},\ }\href@noop {} {\bibfield  {journal} {\bibinfo  {journal} {Monthly Notices of the Royal Astronomical Society}\ }\textbf {\bibinfo {volume} {439}},\ \bibinfo {pages} {3630} (\bibinfo {year} {2014})}\BibitemShut {NoStop}%
\bibitem [{\citenamefont {Kokron}\ \emph {et~al.}(2022)\citenamefont {Kokron}, \citenamefont {Chen}, \citenamefont {White}, \citenamefont {DeRose},\ and\ \citenamefont {Maus}}]{kokron2022accurate}%
  \BibitemOpen
  \bibfield  {author} {\bibinfo {author} {\bibfnamefont {N.}~\bibnamefont {Kokron}}, \bibinfo {author} {\bibfnamefont {S.-F.}\ \bibnamefont {Chen}}, \bibinfo {author} {\bibfnamefont {M.}~\bibnamefont {White}}, \bibinfo {author} {\bibfnamefont {J.}~\bibnamefont {DeRose}},\ and\ \bibinfo {author} {\bibfnamefont {M.}~\bibnamefont {Maus}},\ }\href@noop {} {\bibfield  {journal} {\bibinfo  {journal} {Journal of Cosmology and Astroparticle Physics}\ }\textbf {\bibinfo {volume} {2022}}\bibinfo  {number} { (09)},\ \bibinfo {pages} {059}}\BibitemShut {NoStop}%
\bibitem [{\citenamefont {Chen}\ \emph {et~al.}(2020)\citenamefont {Chen}, \citenamefont {Vlah},\ and\ \citenamefont {White}}]{chen2020consistent}%
  \BibitemOpen
\bibfield  {number} {  }\bibfield  {author} {\bibinfo {author} {\bibfnamefont {S.-F.}\ \bibnamefont {Chen}}, \bibinfo {author} {\bibfnamefont {Z.}~\bibnamefont {Vlah}},\ and\ \bibinfo {author} {\bibfnamefont {M.}~\bibnamefont {White}},\ }\href@noop {} {\bibfield  {journal} {\bibinfo  {journal} {Journal of Cosmology and Astroparticle Physics}\ }\textbf {\bibinfo {volume} {2020}}\bibinfo  {number} { (07)},\ \bibinfo {pages} {062}}\BibitemShut {NoStop}%
\bibitem [{\citenamefont {Bruzual}\ and\ \citenamefont {Charlot}(2003)}]{bruzual2003stellar}%
  \BibitemOpen
\bibfield  {number} {  }\bibfield  {author} {\bibinfo {author} {\bibfnamefont {G.}~\bibnamefont {Bruzual}}\ and\ \bibinfo {author} {\bibfnamefont {S.}~\bibnamefont {Charlot}},\ }\href@noop {} {\bibfield  {journal} {\bibinfo  {journal} {Monthly Notices of the Royal Astronomical Society}\ }\textbf {\bibinfo {volume} {344}},\ \bibinfo {pages} {1000} (\bibinfo {year} {2003})}\BibitemShut {NoStop}%
\bibitem [{\citenamefont {Kauffmann}\ \emph {et~al.}(2003)\citenamefont {Kauffmann}, \citenamefont {Heckman}, \citenamefont {White}, \citenamefont {Charlot}, \citenamefont {Tremonti}, \citenamefont {Brinchmann}, \citenamefont {Bruzual}, \citenamefont {Peng}, \citenamefont {Seibert}, \citenamefont {Bernardi} \emph {et~al.}}]{kauffmann2003stellar}%
  \BibitemOpen
  \bibfield  {author} {\bibinfo {author} {\bibfnamefont {G.}~\bibnamefont {Kauffmann}}, \bibinfo {author} {\bibfnamefont {T.~M.}\ \bibnamefont {Heckman}}, \bibinfo {author} {\bibfnamefont {S.~D.}\ \bibnamefont {White}}, \bibinfo {author} {\bibfnamefont {S.}~\bibnamefont {Charlot}}, \bibinfo {author} {\bibfnamefont {C.}~\bibnamefont {Tremonti}}, \bibinfo {author} {\bibfnamefont {J.}~\bibnamefont {Brinchmann}}, \bibinfo {author} {\bibfnamefont {G.}~\bibnamefont {Bruzual}}, \bibinfo {author} {\bibfnamefont {E.~W.}\ \bibnamefont {Peng}}, \bibinfo {author} {\bibfnamefont {M.}~\bibnamefont {Seibert}}, \bibinfo {author} {\bibfnamefont {M.}~\bibnamefont {Bernardi}}, \emph {et~al.},\ }\href@noop {} {\bibfield  {journal} {\bibinfo  {journal} {Monthly Notices of the Royal Astronomical Society}\ }\textbf {\bibinfo {volume} {341}},\ \bibinfo {pages} {33} (\bibinfo {year} {2003})}\BibitemShut {NoStop}%
\bibitem [{\citenamefont {Zu}\ and\ \citenamefont {Mandelbaum}(2015)}]{zu2015mapping}%
  \BibitemOpen
  \bibfield  {author} {\bibinfo {author} {\bibfnamefont {Y.}~\bibnamefont {Zu}}\ and\ \bibinfo {author} {\bibfnamefont {R.}~\bibnamefont {Mandelbaum}},\ }\href@noop {} {\bibfield  {journal} {\bibinfo  {journal} {Monthly Notices of the Royal Astronomical Society}\ }\textbf {\bibinfo {volume} {454}},\ \bibinfo {pages} {1161} (\bibinfo {year} {2015})}\BibitemShut {NoStop}%
\bibitem [{\citenamefont {Battye}\ \emph {et~al.}(2013)\citenamefont {Battye}, \citenamefont {Browne}, \citenamefont {Dickinson}, \citenamefont {Heron}, \citenamefont {Maffei},\ and\ \citenamefont {Pourtsidou}}]{battye2013h}%
  \BibitemOpen
  \bibfield  {author} {\bibinfo {author} {\bibfnamefont {R.}~\bibnamefont {Battye}}, \bibinfo {author} {\bibfnamefont {I.}~\bibnamefont {Browne}}, \bibinfo {author} {\bibfnamefont {C.}~\bibnamefont {Dickinson}}, \bibinfo {author} {\bibfnamefont {G.}~\bibnamefont {Heron}}, \bibinfo {author} {\bibfnamefont {B.}~\bibnamefont {Maffei}},\ and\ \bibinfo {author} {\bibfnamefont {A.}~\bibnamefont {Pourtsidou}},\ }\href@noop {} {\bibfield  {journal} {\bibinfo  {journal} {Monthly Notices of the Royal Astronomical Society}\ }\textbf {\bibinfo {volume} {434}},\ \bibinfo {pages} {1239} (\bibinfo {year} {2013})}\BibitemShut {NoStop}%
\bibitem [{\citenamefont {Bull}\ \emph {et~al.}(2015)\citenamefont {Bull}, \citenamefont {Ferreira}, \citenamefont {Patel},\ and\ \citenamefont {Santos}}]{bull2015late}%
  \BibitemOpen
  \bibfield  {author} {\bibinfo {author} {\bibfnamefont {P.}~\bibnamefont {Bull}}, \bibinfo {author} {\bibfnamefont {P.~G.}\ \bibnamefont {Ferreira}}, \bibinfo {author} {\bibfnamefont {P.}~\bibnamefont {Patel}},\ and\ \bibinfo {author} {\bibfnamefont {M.~G.}\ \bibnamefont {Santos}},\ }\href@noop {} {\bibfield  {journal} {\bibinfo  {journal} {The Astrophysical Journal}\ }\textbf {\bibinfo {volume} {803}},\ \bibinfo {pages} {21} (\bibinfo {year} {2015})}\BibitemShut {NoStop}%
\bibitem [{\citenamefont {Osato}\ \emph {et~al.}(2021)\citenamefont {Osato}, \citenamefont {Liu},\ and\ \citenamefont {Haiman}}]{osato2021kappatng}%
  \BibitemOpen
  \bibfield  {author} {\bibinfo {author} {\bibfnamefont {K.}~\bibnamefont {Osato}}, \bibinfo {author} {\bibfnamefont {J.}~\bibnamefont {Liu}},\ and\ \bibinfo {author} {\bibfnamefont {Z.}~\bibnamefont {Haiman}},\ }\href@noop {} {\bibfield  {journal} {\bibinfo  {journal} {Monthly Notices of the Royal Astronomical Society}\ }\textbf {\bibinfo {volume} {502}},\ \bibinfo {pages} {5593} (\bibinfo {year} {2021})}\BibitemShut {NoStop}%
\bibitem [{\citenamefont {DeRose}\ and\ \citenamefont {Chen}(2025)}]{derose2025lensing}%
  \BibitemOpen
  \bibfield  {author} {\bibinfo {author} {\bibfnamefont {J.}~\bibnamefont {DeRose}}\ and\ \bibinfo {author} {\bibfnamefont {S.-F.}\ \bibnamefont {Chen}},\ }\href@noop {} {\bibfield  {journal} {\bibinfo  {journal} {arXiv preprint arXiv:2510.18981}\ } (\bibinfo {year} {2025})}\BibitemShut {NoStop}%
\bibitem [{\citenamefont {Sanchez-Cid}\ \emph {et~al.}(2026)\citenamefont {Sanchez-Cid}, \citenamefont {Fert{\'e}}, \citenamefont {Blazek}, \citenamefont {Samuroff}, \citenamefont {Amon}, \citenamefont {Andrade-Oliveira}, \citenamefont {Coloma-Nadal}, \citenamefont {Muir}, \citenamefont {Porredon}, \citenamefont {Prat} \emph {et~al.}}]{sanchez2026dark}%
  \BibitemOpen
  \bibfield  {author} {\bibinfo {author} {\bibfnamefont {D.}~\bibnamefont {Sanchez-Cid}}, \bibinfo {author} {\bibfnamefont {A.}~\bibnamefont {Fert{\'e}}}, \bibinfo {author} {\bibfnamefont {J.}~\bibnamefont {Blazek}}, \bibinfo {author} {\bibfnamefont {S.}~\bibnamefont {Samuroff}}, \bibinfo {author} {\bibfnamefont {A.}~\bibnamefont {Amon}}, \bibinfo {author} {\bibfnamefont {F.}~\bibnamefont {Andrade-Oliveira}}, \bibinfo {author} {\bibfnamefont {J.}~\bibnamefont {Coloma-Nadal}}, \bibinfo {author} {\bibfnamefont {J.}~\bibnamefont {Muir}}, \bibinfo {author} {\bibfnamefont {A.}~\bibnamefont {Porredon}}, \bibinfo {author} {\bibfnamefont {J.}~\bibnamefont {Prat}}, \emph {et~al.},\ }\href@noop {} {\bibfield  {journal} {\bibinfo  {journal} {arXiv preprint arXiv:2601.14859}\ } (\bibinfo {year} {2026})}\BibitemShut {NoStop}%
\bibitem [{\citenamefont {Kugel}\ \emph {et~al.}(2023)\citenamefont {Kugel}, \citenamefont {Schaye}, \citenamefont {Schaller}, \citenamefont {Helly}, \citenamefont {Braspenning}, \citenamefont {Elbers}, \citenamefont {Frenk}, \citenamefont {McCarthy}, \citenamefont {Kwan}, \citenamefont {Salcido} \emph {et~al.}}]{kugel2023flamingo}%
  \BibitemOpen
  \bibfield  {author} {\bibinfo {author} {\bibfnamefont {R.}~\bibnamefont {Kugel}}, \bibinfo {author} {\bibfnamefont {J.}~\bibnamefont {Schaye}}, \bibinfo {author} {\bibfnamefont {M.}~\bibnamefont {Schaller}}, \bibinfo {author} {\bibfnamefont {J.~C.}\ \bibnamefont {Helly}}, \bibinfo {author} {\bibfnamefont {J.}~\bibnamefont {Braspenning}}, \bibinfo {author} {\bibfnamefont {W.}~\bibnamefont {Elbers}}, \bibinfo {author} {\bibfnamefont {C.~S.}\ \bibnamefont {Frenk}}, \bibinfo {author} {\bibfnamefont {I.~G.}\ \bibnamefont {McCarthy}}, \bibinfo {author} {\bibfnamefont {J.}~\bibnamefont {Kwan}}, \bibinfo {author} {\bibfnamefont {J.}~\bibnamefont {Salcido}}, \emph {et~al.},\ }\href@noop {} {\bibfield  {journal} {\bibinfo  {journal} {Monthly Notices of the Royal Astronomical Society}\ }\textbf {\bibinfo {volume} {526}},\ \bibinfo {pages} {6103} (\bibinfo {year} {2023})}\BibitemShut {NoStop}%
\bibitem [{\citenamefont {Helly}\ \emph {et~al.}(2026)\citenamefont {Helly}, \citenamefont {McGibbon}, \citenamefont {Schaye}, \citenamefont {Schaller}, \citenamefont {McDonald}, \citenamefont {Braspenning}, \citenamefont {Broxterman}, \citenamefont {Costello}, \citenamefont {Elbers}, \citenamefont {Frenk} \emph {et~al.}}]{helly2026flamingo}%
  \BibitemOpen
  \bibfield  {author} {\bibinfo {author} {\bibfnamefont {J.~C.}\ \bibnamefont {Helly}}, \bibinfo {author} {\bibfnamefont {R.~J.}\ \bibnamefont {McGibbon}}, \bibinfo {author} {\bibfnamefont {J.}~\bibnamefont {Schaye}}, \bibinfo {author} {\bibfnamefont {M.}~\bibnamefont {Schaller}}, \bibinfo {author} {\bibfnamefont {W.}~\bibnamefont {McDonald}}, \bibinfo {author} {\bibfnamefont {J.}~\bibnamefont {Braspenning}}, \bibinfo {author} {\bibfnamefont {J.~C.}\ \bibnamefont {Broxterman}}, \bibinfo {author} {\bibfnamefont {E.~E.}\ \bibnamefont {Costello}}, \bibinfo {author} {\bibfnamefont {W.}~\bibnamefont {Elbers}}, \bibinfo {author} {\bibfnamefont {C.~S.}\ \bibnamefont {Frenk}}, \emph {et~al.},\ }\href@noop {} {\bibfield  {journal} {\bibinfo  {journal} {arXiv preprint arXiv:2604.24324}\ } (\bibinfo {year} {2026})}\BibitemShut {NoStop}%
\bibitem [{\citenamefont {Elbers}\ \emph {et~al.}(2021)\citenamefont {Elbers}, \citenamefont {Frenk}, \citenamefont {Jenkins}, \citenamefont {Li},\ and\ \citenamefont {Pascoli}}]{elbers2021optimal}%
  \BibitemOpen
  \bibfield  {author} {\bibinfo {author} {\bibfnamefont {W.}~\bibnamefont {Elbers}}, \bibinfo {author} {\bibfnamefont {C.~S.}\ \bibnamefont {Frenk}}, \bibinfo {author} {\bibfnamefont {A.}~\bibnamefont {Jenkins}}, \bibinfo {author} {\bibfnamefont {B.}~\bibnamefont {Li}},\ and\ \bibinfo {author} {\bibfnamefont {S.}~\bibnamefont {Pascoli}},\ }\href@noop {} {\bibfield  {journal} {\bibinfo  {journal} {Monthly Notices of the Royal Astronomical Society}\ }\textbf {\bibinfo {volume} {507}},\ \bibinfo {pages} {2614} (\bibinfo {year} {2021})}\BibitemShut {NoStop}%
\bibitem [{\citenamefont {van Loon}\ and\ \citenamefont {van Daalen}(2024)}]{van2024contribution}%
  \BibitemOpen
  \bibfield  {author} {\bibinfo {author} {\bibfnamefont {M.}~\bibnamefont {van Loon}}\ and\ \bibinfo {author} {\bibfnamefont {M.~P.}\ \bibnamefont {van Daalen}},\ }\href@noop {} {\bibfield  {journal} {\bibinfo  {journal} {Monthly Notices of the Royal Astronomical Society}\ }\textbf {\bibinfo {volume} {528}},\ \bibinfo {pages} {4623} (\bibinfo {year} {2024})}\BibitemShut {NoStop}%
\bibitem [{\citenamefont {van Daalen}\ \emph {et~al.}(2026)\citenamefont {van Daalen}, \citenamefont {Koutalios}, \citenamefont {Broxterman}, \citenamefont {Wolfs}, \citenamefont {Helly}, \citenamefont {Schaller},\ and\ \citenamefont {Schaye}}]{van2026resummation}%
  \BibitemOpen
  \bibfield  {author} {\bibinfo {author} {\bibfnamefont {M.~P.}\ \bibnamefont {van Daalen}}, \bibinfo {author} {\bibfnamefont {I.}~\bibnamefont {Koutalios}}, \bibinfo {author} {\bibfnamefont {J.~C.}\ \bibnamefont {Broxterman}}, \bibinfo {author} {\bibfnamefont {B.~J.}\ \bibnamefont {Wolfs}}, \bibinfo {author} {\bibfnamefont {J.~C.}\ \bibnamefont {Helly}}, \bibinfo {author} {\bibfnamefont {M.}~\bibnamefont {Schaller}},\ and\ \bibinfo {author} {\bibfnamefont {J.}~\bibnamefont {Schaye}},\ }\href@noop {} {\bibfield  {journal} {\bibinfo  {journal} {Monthly Notices of the Royal Astronomical Society}\ }\textbf {\bibinfo {volume} {545}},\ \bibinfo {pages} {staf2086} (\bibinfo {year} {2026})}\BibitemShut {NoStop}%
\bibitem [{\citenamefont {Abbott}\ \emph {et~al.}(2022)\citenamefont {Abbott}, \citenamefont {Aguena}, \citenamefont {Alarcon}, \citenamefont {Allam}, \citenamefont {Alves}, \citenamefont {Amon}, \citenamefont {Andrade-Oliveira}, \citenamefont {Annis}, \citenamefont {Avila}, \citenamefont {Bacon} \emph {et~al.}}]{abbott2022dark}%
  \BibitemOpen
  \bibfield  {author} {\bibinfo {author} {\bibfnamefont {T.~M.}\ \bibnamefont {Abbott}}, \bibinfo {author} {\bibfnamefont {M.}~\bibnamefont {Aguena}}, \bibinfo {author} {\bibfnamefont {A.}~\bibnamefont {Alarcon}}, \bibinfo {author} {\bibfnamefont {S.}~\bibnamefont {Allam}}, \bibinfo {author} {\bibfnamefont {O.}~\bibnamefont {Alves}}, \bibinfo {author} {\bibfnamefont {A.}~\bibnamefont {Amon}}, \bibinfo {author} {\bibfnamefont {F.}~\bibnamefont {Andrade-Oliveira}}, \bibinfo {author} {\bibfnamefont {J.}~\bibnamefont {Annis}}, \bibinfo {author} {\bibfnamefont {S.}~\bibnamefont {Avila}}, \bibinfo {author} {\bibfnamefont {D.}~\bibnamefont {Bacon}}, \emph {et~al.},\ }\href@noop {} {\bibfield  {journal} {\bibinfo  {journal} {Physical Review D}\ }\textbf {\bibinfo {volume} {105}},\ \bibinfo {pages} {023520} (\bibinfo {year} {2022})}\BibitemShut {NoStop}%
\bibitem [{\citenamefont {Elbers}\ \emph {et~al.}(2025)\citenamefont {Elbers}, \citenamefont {Frenk}, \citenamefont {Jenkins}, \citenamefont {Li}, \citenamefont {Helly}, \citenamefont {Kugel}, \citenamefont {Schaller}, \citenamefont {Schaye}, \citenamefont {Braspenning}, \citenamefont {Kwan} \emph {et~al.}}]{elbers2025flamingo}%
  \BibitemOpen
  \bibfield  {author} {\bibinfo {author} {\bibfnamefont {W.}~\bibnamefont {Elbers}}, \bibinfo {author} {\bibfnamefont {C.~S.}\ \bibnamefont {Frenk}}, \bibinfo {author} {\bibfnamefont {A.}~\bibnamefont {Jenkins}}, \bibinfo {author} {\bibfnamefont {B.}~\bibnamefont {Li}}, \bibinfo {author} {\bibfnamefont {J.~C.}\ \bibnamefont {Helly}}, \bibinfo {author} {\bibfnamefont {R.}~\bibnamefont {Kugel}}, \bibinfo {author} {\bibfnamefont {M.}~\bibnamefont {Schaller}}, \bibinfo {author} {\bibfnamefont {J.}~\bibnamefont {Schaye}}, \bibinfo {author} {\bibfnamefont {J.}~\bibnamefont {Braspenning}}, \bibinfo {author} {\bibfnamefont {J.}~\bibnamefont {Kwan}}, \emph {et~al.},\ }\href@noop {} {\bibfield  {journal} {\bibinfo  {journal} {Monthly Notices of the Royal Astronomical Society}\ }\textbf {\bibinfo {volume} {537}},\ \bibinfo {pages} {2160} (\bibinfo {year} {2025})}\BibitemShut {NoStop}%
\bibitem [{\citenamefont {McCarthy}\ \emph {et~al.}(2025)\citenamefont {McCarthy}, \citenamefont {Amon}, \citenamefont {Schaye}, \citenamefont {Schaan}, \citenamefont {Angulo}, \citenamefont {Salcido}, \citenamefont {Schaller}, \citenamefont {Bigwood}, \citenamefont {Elbers}, \citenamefont {Kugel} \emph {et~al.}}]{mccarthy2025flamingo}%
  \BibitemOpen
  \bibfield  {author} {\bibinfo {author} {\bibfnamefont {I.~G.}\ \bibnamefont {McCarthy}}, \bibinfo {author} {\bibfnamefont {A.}~\bibnamefont {Amon}}, \bibinfo {author} {\bibfnamefont {J.}~\bibnamefont {Schaye}}, \bibinfo {author} {\bibfnamefont {E.}~\bibnamefont {Schaan}}, \bibinfo {author} {\bibfnamefont {R.~E.}\ \bibnamefont {Angulo}}, \bibinfo {author} {\bibfnamefont {J.}~\bibnamefont {Salcido}}, \bibinfo {author} {\bibfnamefont {M.}~\bibnamefont {Schaller}}, \bibinfo {author} {\bibfnamefont {L.}~\bibnamefont {Bigwood}}, \bibinfo {author} {\bibfnamefont {W.}~\bibnamefont {Elbers}}, \bibinfo {author} {\bibfnamefont {R.}~\bibnamefont {Kugel}}, \emph {et~al.},\ }\href@noop {} {\bibfield  {journal} {\bibinfo  {journal} {Monthly Notices of the Royal Astronomical Society}\ }\textbf {\bibinfo {volume} {540}},\ \bibinfo {pages} {143} (\bibinfo {year} {2025})}\BibitemShut {NoStop}%
\bibitem [{\citenamefont {Schaller}\ \emph {et~al.}(2025)\citenamefont {Schaller}, \citenamefont {Schaye}, \citenamefont {Kugel}, \citenamefont {Broxterman},\ and\ \citenamefont {van Daalen}}]{schaller2025flamingo}%
  \BibitemOpen
  \bibfield  {author} {\bibinfo {author} {\bibfnamefont {M.}~\bibnamefont {Schaller}}, \bibinfo {author} {\bibfnamefont {J.}~\bibnamefont {Schaye}}, \bibinfo {author} {\bibfnamefont {R.}~\bibnamefont {Kugel}}, \bibinfo {author} {\bibfnamefont {J.~C.}\ \bibnamefont {Broxterman}},\ and\ \bibinfo {author} {\bibfnamefont {M.~P.}\ \bibnamefont {van Daalen}},\ }\href@noop {} {\bibfield  {journal} {\bibinfo  {journal} {Monthly Notices of the Royal Astronomical Society}\ }\textbf {\bibinfo {volume} {539}},\ \bibinfo {pages} {1337} (\bibinfo {year} {2025})}\BibitemShut {NoStop}%
\bibitem [{\citenamefont {Siegel}\ \emph {et~al.}(2025{\natexlab{b}})\citenamefont {Siegel}, \citenamefont {Bigwood}, \citenamefont {Amon}, \citenamefont {McCullough}, \citenamefont {Yamamoto}, \citenamefont {McCarthy}, \citenamefont {Schaller}, \citenamefont {Schneider},\ and\ \citenamefont {Schaye}}]{siegel2025suppression}%
  \BibitemOpen
  \bibfield  {author} {\bibinfo {author} {\bibfnamefont {J.}~\bibnamefont {Siegel}}, \bibinfo {author} {\bibfnamefont {L.}~\bibnamefont {Bigwood}}, \bibinfo {author} {\bibfnamefont {A.}~\bibnamefont {Amon}}, \bibinfo {author} {\bibfnamefont {J.}~\bibnamefont {McCullough}}, \bibinfo {author} {\bibfnamefont {M.}~\bibnamefont {Yamamoto}}, \bibinfo {author} {\bibfnamefont {I.~G.}\ \bibnamefont {McCarthy}}, \bibinfo {author} {\bibfnamefont {M.}~\bibnamefont {Schaller}}, \bibinfo {author} {\bibfnamefont {A.}~\bibnamefont {Schneider}},\ and\ \bibinfo {author} {\bibfnamefont {J.}~\bibnamefont {Schaye}},\ }\href@noop {} {\bibfield  {journal} {\bibinfo  {journal} {arXiv preprint arXiv:2512.02954}\ } (\bibinfo {year} {2025}{\natexlab{b}})}\BibitemShut {NoStop}%
\bibitem [{\citenamefont {Lin}\ \emph {et~al.}(2006)\citenamefont {Lin}, \citenamefont {Jing}, \citenamefont {Mao}, \citenamefont {Gao},\ and\ \citenamefont {McCarthy}}]{lin2006influence}%
  \BibitemOpen
  \bibfield  {author} {\bibinfo {author} {\bibfnamefont {W.-P.}\ \bibnamefont {Lin}}, \bibinfo {author} {\bibfnamefont {Y.}~\bibnamefont {Jing}}, \bibinfo {author} {\bibfnamefont {S.}~\bibnamefont {Mao}}, \bibinfo {author} {\bibfnamefont {L.}~\bibnamefont {Gao}},\ and\ \bibinfo {author} {\bibfnamefont {I.}~\bibnamefont {McCarthy}},\ }\href@noop {} {\bibfield  {journal} {\bibinfo  {journal} {The Astrophysical Journal}\ }\textbf {\bibinfo {volume} {651}},\ \bibinfo {pages} {636} (\bibinfo {year} {2006})}\BibitemShut {NoStop}%
\bibitem [{\citenamefont {Han}\ \emph {et~al.}(2018)\citenamefont {Han}, \citenamefont {Cole}, \citenamefont {Frenk}, \citenamefont {Benitez-Llambay},\ and\ \citenamefont {Helly}}]{han2018hbtPlus}%
  \BibitemOpen
  \bibfield  {author} {\bibinfo {author} {\bibfnamefont {J.}~\bibnamefont {Han}}, \bibinfo {author} {\bibfnamefont {S.}~\bibnamefont {Cole}}, \bibinfo {author} {\bibfnamefont {C.~S.}\ \bibnamefont {Frenk}}, \bibinfo {author} {\bibfnamefont {A.}~\bibnamefont {Benitez-Llambay}},\ and\ \bibinfo {author} {\bibfnamefont {J.}~\bibnamefont {Helly}},\ }\href@noop {} {\bibfield  {journal} {\bibinfo  {journal} {Monthly Notices of the Royal Astronomical Society}\ }\textbf {\bibinfo {volume} {474}},\ \bibinfo {pages} {604} (\bibinfo {year} {2018})}\BibitemShut {NoStop}%
\bibitem [{\citenamefont {Forouhar~Moreno}\ \emph {et~al.}(2025)\citenamefont {Forouhar~Moreno}, \citenamefont {Helly}, \citenamefont {McGibbon}, \citenamefont {Schaye}, \citenamefont {Schaller}, \citenamefont {Han}, \citenamefont {Kugel},\ and\ \citenamefont {Bah{\'e}}}]{forouhar2025assessing}%
  \BibitemOpen
  \bibfield  {author} {\bibinfo {author} {\bibfnamefont {V.~J.}\ \bibnamefont {Forouhar~Moreno}}, \bibinfo {author} {\bibfnamefont {J.}~\bibnamefont {Helly}}, \bibinfo {author} {\bibfnamefont {R.}~\bibnamefont {McGibbon}}, \bibinfo {author} {\bibfnamefont {J.}~\bibnamefont {Schaye}}, \bibinfo {author} {\bibfnamefont {M.}~\bibnamefont {Schaller}}, \bibinfo {author} {\bibfnamefont {J.}~\bibnamefont {Han}}, \bibinfo {author} {\bibfnamefont {R.}~\bibnamefont {Kugel}},\ and\ \bibinfo {author} {\bibfnamefont {Y.~M.}\ \bibnamefont {Bah{\'e}}},\ }\href@noop {} {\bibfield  {journal} {\bibinfo  {journal} {Monthly Notices of the Royal Astronomical Society}\ }\textbf {\bibinfo {volume} {543}},\ \bibinfo {pages} {1339} (\bibinfo {year} {2025})}\BibitemShut {NoStop}%
\bibitem [{\citenamefont {McGibbon}\ \emph {et~al.}(2025)\citenamefont {McGibbon}, \citenamefont {Helly}, \citenamefont {Schaye}, \citenamefont {Schaller},\ and\ \citenamefont {Vandenbroucke}}]{mcgibbon2025soap}%
  \BibitemOpen
  \bibfield  {author} {\bibinfo {author} {\bibfnamefont {R.}~\bibnamefont {McGibbon}}, \bibinfo {author} {\bibfnamefont {J.~C.}\ \bibnamefont {Helly}}, \bibinfo {author} {\bibfnamefont {J.}~\bibnamefont {Schaye}}, \bibinfo {author} {\bibfnamefont {M.}~\bibnamefont {Schaller}},\ and\ \bibinfo {author} {\bibfnamefont {B.}~\bibnamefont {Vandenbroucke}},\ }\href@noop {} {\bibfield  {journal} {\bibinfo  {journal} {arXiv preprint arXiv:2507.22669}\ } (\bibinfo {year} {2025})}\BibitemShut {NoStop}%
\bibitem [{\citenamefont {Villaescusa-Navarro}\ \emph {et~al.}(2014)\citenamefont {Villaescusa-Navarro}, \citenamefont {Marulli}, \citenamefont {Viel}, \citenamefont {Branchini}, \citenamefont {Castorina}, \citenamefont {Sefusatti},\ and\ \citenamefont {Saito}}]{villaescusa2014cosmology}%
  \BibitemOpen
  \bibfield  {author} {\bibinfo {author} {\bibfnamefont {F.}~\bibnamefont {Villaescusa-Navarro}}, \bibinfo {author} {\bibfnamefont {F.}~\bibnamefont {Marulli}}, \bibinfo {author} {\bibfnamefont {M.}~\bibnamefont {Viel}}, \bibinfo {author} {\bibfnamefont {E.}~\bibnamefont {Branchini}}, \bibinfo {author} {\bibfnamefont {E.}~\bibnamefont {Castorina}}, \bibinfo {author} {\bibfnamefont {E.}~\bibnamefont {Sefusatti}},\ and\ \bibinfo {author} {\bibfnamefont {S.}~\bibnamefont {Saito}},\ }\href@noop {} {\bibfield  {journal} {\bibinfo  {journal} {Journal of Cosmology and Astroparticle Physics}\ }\textbf {\bibinfo {volume} {2014}}\bibinfo  {number} { (03)},\ \bibinfo {pages} {011}}\BibitemShut {NoStop}%
\bibitem [{\citenamefont {Castorina}\ \emph {et~al.}(2014)\citenamefont {Castorina}, \citenamefont {Sefusatti}, \citenamefont {Sheth}, \citenamefont {Villaescusa-Navarro},\ and\ \citenamefont {Viel}}]{castorina2014cosmology}%
  \BibitemOpen
\bibfield  {number} {  }\bibfield  {author} {\bibinfo {author} {\bibfnamefont {E.}~\bibnamefont {Castorina}}, \bibinfo {author} {\bibfnamefont {E.}~\bibnamefont {Sefusatti}}, \bibinfo {author} {\bibfnamefont {R.~K.}\ \bibnamefont {Sheth}}, \bibinfo {author} {\bibfnamefont {F.}~\bibnamefont {Villaescusa-Navarro}},\ and\ \bibinfo {author} {\bibfnamefont {M.}~\bibnamefont {Viel}},\ }\href@noop {} {\bibfield  {journal} {\bibinfo  {journal} {Journal of Cosmology and Astroparticle Physics}\ }\textbf {\bibinfo {volume} {2014}}\bibinfo  {number} { (02)},\ \bibinfo {pages} {049}}\BibitemShut {NoStop}%
\end{thebibliography}%


\onecolumngrid
\clearpage

\appendix
\section*{Appendix}
\addcontentsline{toc}{section}{Supplemental Materials}

\section{\color{black}Baryonic effects in a toy-model analysis}

Assumption ({\bf I}) states that baryonic feedback predominantly affects the amplitude rather than the phase of the matter field. This behavior is verified by measurements in TNG300-1 and Illustris-1 simulations (Fig.~\ref{fig:T-r2}), and further supported by Ref.~\cite{sharma2025field}, which reaches the same conclusion based on thousands of hydrodynamical simulations. 
A physical intuition is embedded in the contrast between the suppressed amplitude and yet tight cross-correlation with the DMO counterpart. At a fixed scale, baryonic effects can suppress the clustering strength, but struggle to disrupt the overall similarity of structures between two universes. Consequently, the sensitivity of matter clustering to baryonic effects is much larger in the amplitude, quantified by $\sqrt{P_{mm}/P_{\rm DMO}}$, than in the phase, the structural similarity quantified by $r(\delta_m, \delta_{\rm DMO})$. 
{\color{black}Beyond numerical validation using hydrodynamical simulations, we can gain further physical intuition through an illustrative example. }

In the Lagrangian picture, a fluid element initially located at position $\bfq$ in Lagrangian space evolves dynamically to the Eulerian position $\bfx(\bfq, z) = \bfq +{\bf\Psi}(\bfq, z)$ at redshift $z$.
{Mass conservation yields the matter overdensity field as
\begin{equation}  \label{equ:LPT-deltam}
1+\delta_m(\bfx, z) = \int_{\bfq}\,\delta^D(\bfx-\bfq-{\bf\Psi}(\bfq, z))
\end{equation}
where $\int_\bfq\, \cdots \equiv \int d^3\bfq\, \cdots$ denotes the integral over spatial coordinates. 
The displacement field ${\bf\Psi}(\bfq, z)$ maps the particle position from Lagrangian position $\bfq$ to Eulerian position $\bfx$ at redshift $z$, which is solely induced by gravity in a DMO universe. 
For simplicity, we adopt the Zel'dovich approximation ${\bf\Psi}(\bfk,z) \simeq {\bf\Psi}^Z(\bfk,z) = {i\bfk}\,k^{-2}\delta(\bfk,z)$ and further assume that $\delta(\bfx,z)$ is a Gaussian field. The matter power spectrum with $k\neq 0$ is therefore simplified to \cite{vlah2016gaussian,vlah2015lagrangian, white2014zel, kokron2022accurate, chen2020consistent}
\begin{equation}  \label{equ:LPT-Pmm-ZA}
P_{\rm zz}(k, z) = \int_\bfq e^{-i\bfk\cdot\bfq}\, e^{-{1\over 2} \la\Delta\Delta\ra_c}
\end{equation}
in the Zel'dovich universe. Here $\bfq=\bfq_1-\bfq_2\,$, $\Delta \equiv \bfk\cdot\left[{{\bf\Psi}(\bfq_1,z) - {\bf\Psi}(\bfq_2,z)}\right]$, and $\la...\ra_c$ is the cumulant of the stochastic field. 
}

{\color{black}We consider another scenario in which the universe evolves {\color{black}under both gravity and baryon physics}.} The primary impacts of baryonic effects on matter clustering, including gas cooling, star formation, and AGN feedback, are complex but highly local. These processes mainly alter the internal structures of halos, for instance, by expanding or redistributing matter within them, rather than directly modifying large-scale perturbations. Consequently, during the coevolution of baryons and cold dark matter, baryonic effects can be regarded as an additional short-range force acting on the matter motion relative to its DMO counterpart. 
In the Lagrangian framework, this short-range force from baryons induces an extra displacement field ${\bf\Psi}_s$, which dominates small-scale structure formation but remains subdominant to gravity on large scales, satisfying $\bf|\Psi_s(k)| \ll |\Psi(k)|$ for $|\bfk|\ll R_{\rm b}^{-1}$, where $R_{\rm b} \sim 1\,{\rm Mpc}$ characterizes the scale below which baryonic physics are significant. 
By substituting ${\bf\Psi} \rightarrow {\bf\Psi} + {\bf\Psi}_s$ into Eq.~(\ref{equ:LPT-deltam}) \& Eq.~(\ref{equ:LPT-Pmm-ZA}), we obtain the modified matter overdensity field and matter power spectrum in the presence of baryons. 

{\color{black}
In this scenario, the matter overdensity field with the additional baryonic shift ${\bf\Psi}_s(\bfq, z)$ is 
\begin{equation}  \label{equ:LPT-deltam-b}
1 + \delta_{\rm s}(\bfx, z) = 
\int_\bfq  \delta^D\left( \bfx-\bfq - {\bf\Psi}(\bfq, z) - {\bf\Psi}_s(\bfq, z) \right)  \;.
\end{equation}
By adopting a similar approximation in which ${\bf\Psi}_s$ is also the gradient of a Gaussian field, the power spectrum of $\delta_{\rm s}$ is obtained as
\begin{align}
\label{equ:LPT-Pmbm}
P_{\rm ss}(k,z) & = \int_\bfq e^{-i\bfk\cdot\bfq}\, e^{-{1\over 2} \la(\Delta+\Delta_s)^2\ra_c}    \\
&= \int_\bfq e^{-i\bfk\cdot\bfq}\, e^{-{1\over 2} \la\Delta\Delta\ra_c} 
\left[  1 - \la\Delta\Delta_s\ra_c + \mathcal{O}\left(k^2\Psi_s^2\right) \; \right]    \\
&= P_{\rm zz}(k,z) + I_s(k,z) + \mathcal{O}\left(k^2\Psi_s^2\right)   \;,     
\end{align}
where $\Delta_s \equiv \bfk\cdot\left[ {\bf\Psi}_s(\bfq_1,z) - {\bf\Psi}_s(\bfq_2,z) \right]$ and the leading-order baryonic contribution is 
\begin{equation}
I_s(k,z) = - \int_\bfq e^{-i\bfk\cdot\bfq}\, e^{-{1\over 2} \la\Delta\Delta\ra_c}\, \la\Delta\Delta_s\ra_c 
\end{equation}
}Similarly, the cross power spectrum between $\delta_{\rm z}$ and $\delta_{\rm s}$ is
\begin{equation}
\label{equ:LPT-Pmbmb}
P_{\rm zs}(k) = P_{\rm zz}(k) + {1\over 2} I_s(k) + \mathcal{O}\left(k^2\Psi_s^2\right)   \;. 
\end{equation}
Here, $P_{\rm ss}$, $P_{\rm zz}$, and $P_{\rm zs}$ denote the matter power spectrum with baryonic effects, its DMO counterpart, and their cross-correlation, respectively. Hence, the cross-correlation coefficient between $\delta_{\rm z}$ and $\delta_{\rm s}$ is
\begin{equation} \label{equ:ZA_r2}
r_{\rm zs}(k) \equiv {  P_{\rm zs}(k) \over \sqrt{ P_{\rm zz}(k) P_{\rm ss}(k) } } 
\simeq 1 + \mathcal{O}\left( \Psi_s^2\over \Psi^2 \right)    \;,
\end{equation}
Here $\Psi = |{\bf\Psi}|$, $\Psi_s = |{\bf\Psi}_s|$ and $\Psi_s/\Psi \ll 1$ on large scales. The suppression of the matter power spectrum resulting from the baryonic shift is 
\begin{equation} \label{equ:ZA_Sk}
S(k) \equiv \sqrt{P_{\rm ss}(k) \over P_{\rm zz}(k) } 
= 1 + {1\over 2} {I_s(k)\over P_{\rm zz}(k) } + \mathcal{O}\left( \Psi_s^2\over \Psi^2 \right) 
\simeq 1 + \mathcal{O}\left( \Psi_s\over \Psi \right)    \;.
\end{equation}
where $I_s(k) \sim \mathcal{O}(k\Psi_s)$. Therefore, the leading-order contribution of baryonic effects to the cross-correlation coefficient $r^2_{\rm zs}$ is canceled, resulting in deviations from unity only at the order $\sim \left(\Psi_s/\Psi\right)^2$. In contrast, the matter power spectrum is suppressed by a proportion $I_s/P_{\rm zz} \sim \Psi_s/\Psi$, which is larger by one order of {\color{black}$\Psi_s/\Psi$}. 
In this case, we demonstrate that the matter clustering is more sensitive to baryonic effects in amplitude than in phase.

\section{Baryon tracers and tomographic measurements}
\label{appendix:tomographic}

We decompose the total baryon overdensity $\delta_b$ into three contributions: the free ionized electrons $\delta_e$ traced by the dispersion measure of FRBs, the clustering of stellar content $\delta_*$ estimated from galaxy surveys, and the neutral hydrogen $\delta_\mHI$ traced by the 21 cm line intensity mapping. The validity of this decomposition is trivial, as it is merely a classification of major baryons, while unaccounted components such as metals contribute negligibly to total baryon content. 
{\color{black}
The dominant component is the ionized electron, of which the FRB dispersion measure provides an unbiased observation. As a comparison, the kinetic Sunyaev-Zel'dovich (kSZ) effect is also sensitive to the total free electron distribution. However, its detection and interpretation are complicated by an intrinsic degeneracy between the electron density and the bulk velocity field, requiring a template of the large-scale velocity field reconstructed through the linear continuity equation. 
}

To deproject the mixed Fourier modes, we utilize galaxy samples with a narrow redshift width $\Delta z_g \ll 1$ centered at a comoving distance $\chi_g$, e.g., $\Delta z = 0.1$ as in Ref.~\cite{zhou2026cavendish}. The Limber approximation leads to $C^{gX}_{\ell} = \chi_g^{-2} \, W_X(\chi_g) \, P_{gi}\left({\color{black}k_\ell}, z \right)$, {\color{black} with $k_\ell=(\ell+{1\over2})/\chi_g$. The} expectation of Eq.~(\ref{equ:fb_esti}) becomes
\begin{equation} \label{equ:fb_expectation}
\la \widehat{f_ib_i}(\ell) \ra 
= \mA_i\, { W_{X_i}(\chi_g) \over W_\kappa(\chi_g) }{ P_{gi}({\color{black}k_\ell})\over P_{gm}({\color{black}k_\ell}) }
= f_ib_i({\color{black}k_\ell})   \;.
\end{equation}
In this limit, each Fourier mode is directly measurable. The kernel function of weak lensing is $W_\kappa(\chi) =  {3\over 2c^2}\Omega_{m0}H_0^2\,  a^{-1} \chi \; N_\kappa(z)$, where $N_\kappa(z) = \int_{z}^\infty dz_s\,  n_s(z_s) \left( 1-{\chi/\chi_s}\right)$ is redshift distribution integral of lensing source $n_s(z_s)$. 
{\color{black}In the case of CMB lensing, the expression is simplified as $N_\kappa(z) = 1-{\chi/\chi_s}$ with $\chi_s=\chi(z\simeq 1100)$. }
The scalar amplitude $\mA_i\equiv f_i\, W_\kappa(\chi_g)/W_{X_i}(\chi_g)$ only relies on cosmological parameters, because of $W_{X_i}\propto f_i$ for the three baryon tracers, as specified below. 

The primary baryon tracer is FRB, as it directly probes free electrons in ionized diffuse gas through dispersion measure \cite{zhang2023physics}, 
\begin{equation} \label{equ:D}
\mD = {3H_0^2\over 8\pi G} {\Omega_{b0} \over m_p}\, \int d\chi\; a^{-1}f_e \left(1+\delta_e\right)   \;.
\end{equation}
Here, the backlight of FRBs traces the electron density fluctuation $\delta_e$ projected along the radial distance $\chi$. In the tomographic measurement using galaxy samples with a narrow redshift distribution, we have
\begin{equation} \label{equ:Ae}
\mA_e = {4\pi G m_p\over c^2} {\Omega_{m0}\over\Omega_{b0}} \chi_g {\,N_\kappa(z_g) \over N_\mD(z_g)}
\end{equation}
where $N_\mD(z) = \int_z^\infty dz_s\, n_\mD(z_s)$ is the redshift distribution integral $n_\mD(z_s)$ of FRBs. 
The clustering of stellar content and neutral hydrogen are two subdominant contributions to the total baryon clustering. 
Given the stellar evolution prescription, the stellar mass $M_*$ is estimated by observables such as photometries and spectral characteristics \cite{bruzual2003stellar, kauffmann2003stellar, zu2015mapping}. We can therefore measure the stellar surface density by $\hat\Sigma_*(\hatn_{\rm pix}) = \Delta\Omega^{-1}  \sum_{\hatn\in{\rm pix}} M_*(\hatn)$ in form, where $\Delta\Omega$ is the angular area of the pixel. Here, we neglect the finite extent of galaxies since the typical value of galaxy size is much smaller than the minimum scale $\sim 1$ Mpc we are interested in. 
The expectation of the estimate after survey systematics correction is {\color{black} $\la\hat\Sigma_*\ra = \int{\rm d}z\, N_*(z)\Sigma_*(z)$, where $N_*(z)$ is the redshift distribution of stellar samples and}
\begin{equation} \label{equ:ss}
\Sigma_*  {\color{black}(z)} 
= {3c H_0^2\over 8\pi G}\, \Omega_{b0} {\chi^2\over H(z)}\, f_* \left(1+\delta_*\right)   \;.    \\
\end{equation}
{\color{black}The projected depth of stellar samples is chosen to be narrow while covering the redshift range of galaxy clustering. This leads to}
\begin{equation} \label{equ:A*}
\mA_* = {4\pi G\over c^2} {\Omega_{m0}\over\Omega_{b0}} {1+z\over \chi_g} {N_{\kappa}(z_g) \over {\color{black}N_*(z_g)} }    \; .
\end{equation} 
For neutral hydrogen, they are traced by the surface brightness temperature of 21cm lines \cite{battye2013h, bull2015late}, 
\begin{equation} \label{equ:Tb}
T_b {\color{black}(z)}
= {9\over 128\pi} {\hbar c^3 A_{10} \over k_B \nu_{21}^2 G m_p}  {H_0^2 (1+z)^2 \over H(z)} \Omega_{b0} f_\mHI\, \left(1+\delta_\mHI\right)   \;.
\end{equation}
It gives 
\begin{equation} \label{equ:AHI}
\mA_\mHI = {64\pi\over 3} {k_B \nu_{21}^2 G m_p\over \hbar c^4 A_{10}}\, {\Omega_{m0}\over\Omega_{b0}} 
{\chi_g\over 1+z} {N_\kappa(z_g)\over N_\mHI(z_g)}    \; ,
\end{equation} 
where {\color{black}the redshift distribution of 21cm sources $N_\mHI(z)$ covers the redshift range of the galaxy clustering.}
The explicit expressions of $\mA_e$, $\mA_*$ and $\mA_\mHI$ in Eqs.~(\ref{equ:Ae}, \ref{equ:A*}, \ref{equ:AHI}) demonstrates that the estimations of these overall factors $\mA_i$ are independent of astrophysics, only relying on the cosmological parameters that are tightly constrained by the CMB and BAO observations.

The solution for baryonic effects shares the same philosophy as the cosmological Cavendish experiment proposed in Ref.~\cite{zhou2026cavendish}, though the two are distinguished by their different objectives. The Cavendish experiment focuses on linear perturbation, where the number of modes is limited. Thus, the precision of gravitational constant measurement is about $\sim 10\%$ with $10^{5}$ FRBs, making the corrections from $f_*b_*$ and $f_\mHI b_\mHI$ subdominant relative to the statistical uncertainties. 
In contrast, to correct the baryonic feedback, forthcoming surveys demand systematic corrections better than $1\%$ accuracy across a wide range of scales $k\sim 0.01 - 1\kunit$. The minor mitigations from $f_*b_*$ and $f_\mHI b_\mHI$ become non-negligible.

\section{\color{black} Baryonic effects of weak lensing in harmonic space}

\begin{figure}
\includegraphics[width=0.71\textwidth]{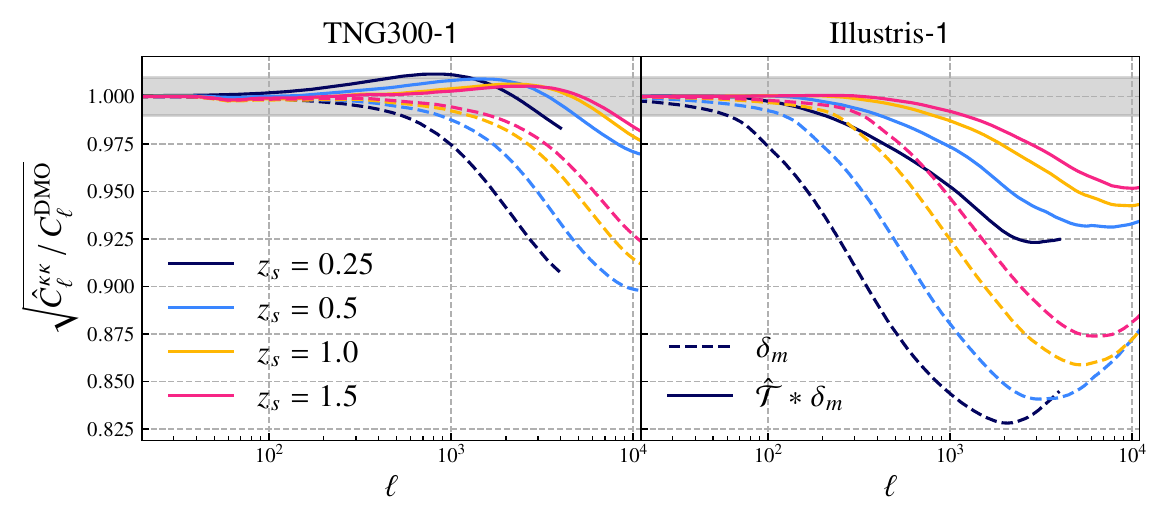}
\caption{ \label{fig:lensing_power_ratio}
\color{black}The ratio of weak lensing angular power spectrum $C_\ell^{\kappa\kappa}$ to its dark-matter-only counterpart $C_\ell^{\rm DMO}$, assuming a thin source plane at $z_s=0.25/0.5/\, 1.0/\, 1.5$ (\textit{dark/blue/yellow/pink} lines). The \textit{dashed} lines indicate the weak lensing convergence $\kappa$ measured from the hydrodynamical simulations TNG300-1 (\textit{left}) and Illustris-1 (\textit{right}), while the \textit{solid} lines show the $\kappa$ results including baryonic corrections with $\hat\mT$. 
For the low source redshift $z_s=0.25$, we exclude the results for $\ell>4000$ due to the limitation of the integration range. 
}
\end{figure}

\begin{figure}
\includegraphics[width=0.71\textwidth]{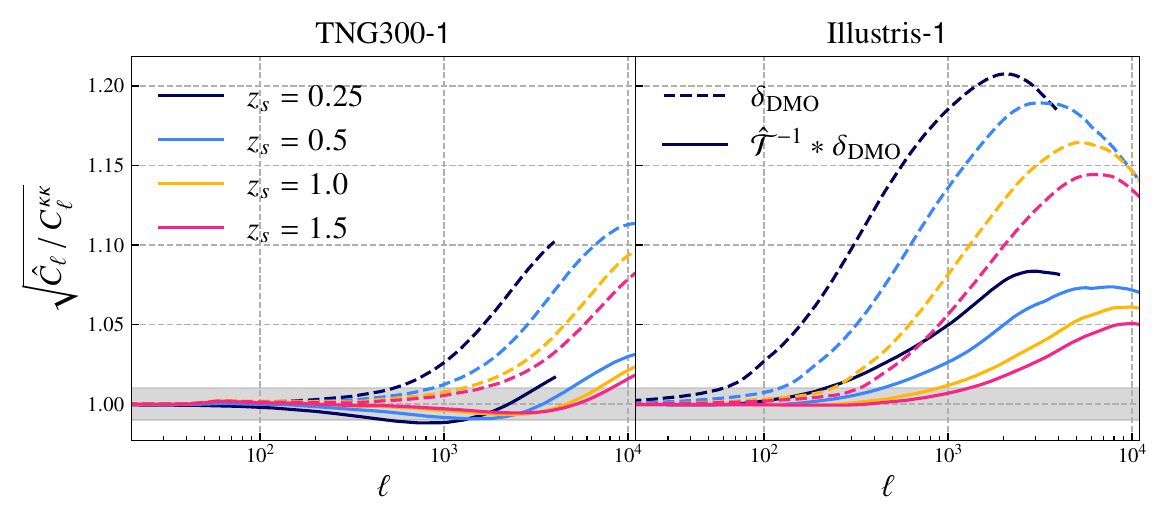}
\caption{ \label{fig:lensing_power_as_model}
\color{black}The ratio of \textit{model} prediction $\hat{C}_\ell$ to the \textit{true} weak lensing angular power spectrum $C_\ell^{\kappa\kappa}$ with baryonic effects, in the same setup as Fig.~\ref{fig:lensing_power_ratio}. The \textit{dashed} lines indicate the DMO model predictions, while the \textit{solid} lines denote the model predictions involving the baryonic correction $\hat\mT$. 
Though not shown here, the residual systematics of $\sim 1\%$ for TNG300-1 and $\sim 5\%$ for Illustris-1 can be further mitigated by marginalizing over the large-$k$ contributions. 
}
\end{figure}

\begin{figure}
\includegraphics[width=0.72\textwidth]{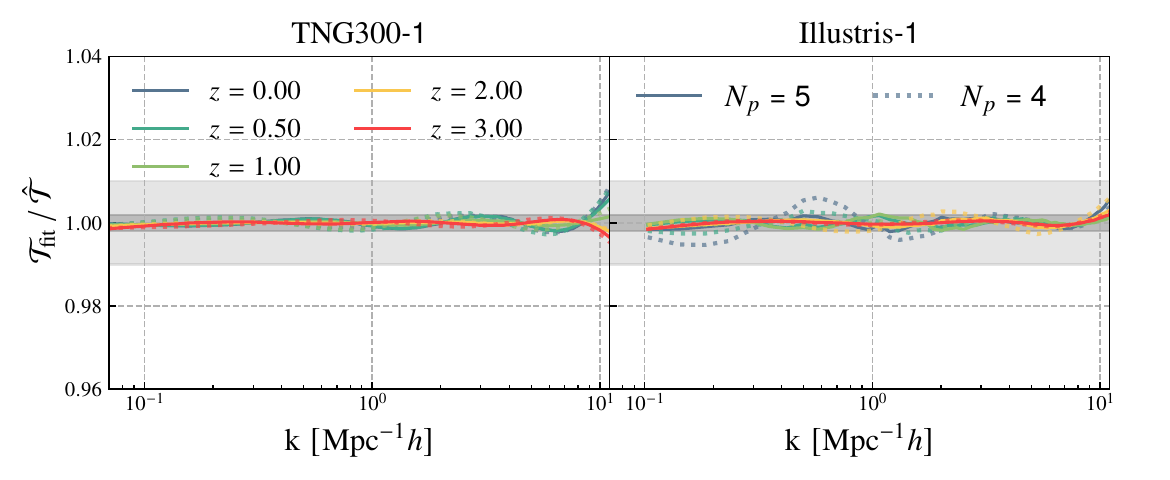}
\caption{ \label{fig:Bspline_reconst}
\color{black}
The ratio of the cubic-spline reconstruction $\mT_{\rm fit}$ to the direct simulation measurement $\hat{\mT}$. The dashed lines indicate the cubic-spline reconstruction based on $N_p=4$ support points over $k\in [0.1, 10]\kunit$, with $< 0.8\%$ deviation within this range. The solid lines denote the result with $N_p=5$ support points, which further reduces the deviation to $< 0.2\%$ (deep shaded region). This demonstrates that the transfer function is smooth and can be accurately described by a few degrees of freedom. 
}
\end{figure}

The weak lensing convergence is a projection of the matter overdensity over a broad redshift range, and baryonic effects in the convergence field mix Fourier modes across redshifts. As a result, contamination from the large-$k$ regime can propagate to low-$\ell$ scales. It is therefore informative to quantify the accuracy achieved by the $\hat{\mT}$ correction in harmonic space without additional treatment of these high-$k$ effects. 
We assess the residual systematics in weak lensing using either nulling baryonic effects at the data level (Fig.~\ref{fig:lensing_power_ratio}) or forward-modeling with a DMO prediction including baryonic corrections (Fig.~\ref{fig:lensing_power_as_model}).

We first measure the weak lensing power spectrum using the TNG300-1 and Illustris-1 simulations, and compare them with their DMO counterparts, as shown in Fig.~\ref{fig:lensing_power_ratio}. 
The weak lensing power spectrum is an integration over comoving distance $\chi$, 
\begin{equation}\label{equ:lensing_cl}
C_\ell^{\kappa\kappa} = \int_0^{\chi_s} d\chi\, \chi^{-2}\, W_\kappa^2(\chi, \chi_s)\, P\left(k_\ell, z(\chi)\right)  \;,
\end{equation}
where $k_\ell = (\ell+{1\over 2})/\chi$ and $W_\kappa(\chi, \chi_s) =  {3\over 2c^2}\Omega_{m0}H_0^2\, (1+z)\, \chi \left( 1-{\chi/\chi_s}\right)$. Without loss of generality, we assume a thin source distribution, with $\chi_s$ denoting the comoving distance to the source plane. By substituting $P=P_{mm}$, $P=\hat\mT^2 P_{mm}$, and $P=P_{\rm DMO}$ into the integral, we obtain the weak lensing power spectrum, the weak lensing power spectrum including baryonic corrections, and the DMO counterpart. 
Here, $\hat\mT$, $P_{mm}$ and $P_{\rm DMO}$ are measured from simulations, and then interpolated over the redshift bins $z\in$\{0.0, 0.1, 0.2, 0.3, 0.4, 0.5, 0.7, 1.0, 1.5, 2.0, 3.0\} in TNG300-1 and $z\in$\{0.0, 0.1, 0.18, 0.26, 0.33, 0.42, 0.5, 0.62, 0.76, 0.89, 1.0, 2.0, 3.0\} in Illustris-1. 
The angular integral covers $0.007 < k_\ell < 15\,\kunit$. We validate the integration against the fitting formula (Appendix~A of Ref.~\cite{osato2021kappatng}) derived from ray-tracing mocks. For TNG300-1 with $z_s \ge 0.5$, the ratio $C_\ell^{\kappa\kappa}/C_\ell^{\rm DMO}$ agrees with the fitting formula to within $<0.5\%$ for $\ell < 10000$. For $z_s=0.25$, deviations start near $\ell\simeq6000$, so we conservatively limit comparisons to $\ell < 4000$.

In the TNG300-1 scenario with mild feedback, baryonic effects are negligible at $\ell\lesssim 1000$. For a typical lensing source at $z_s=0.5$, the suppression on lensing convergence amplitude $|\kappa_\ell| \equiv \sqrt{C_\ell^{\kappa\kappa}}$ reaches the maximum of $\sim 10\%$ at $\ell\sim 10000$. After correction with $\hat\mT$, the residual bias is well controlled to $\lesssim 1\%$ at $\ell< 3000$ and to $\lesssim 2\%$ at $\ell< 10000$, as expected from the accuracy of $\hat\mT$ shown in Fig.~\ref{fig:T-ratio}. This corresponds to an improvement in the systematic bias of the lensing power spectrum from $\lesssim 20\%$ to $\lesssim 4\%$ for all multipoles $\ell\lesssim 10000$. 
For the extremely aggressive feedback in Illustris-1, strong baryonic effects can produce a $\sim 12\%$ suppression around $\ell\sim 1000$ at $z_s=0.5$, and there is also substantial bias for sources at higher redshift. After applying $\hat\mT$ correction, the systematic at $\ell < 1000$ is reduced to $\lesssim 3\%$ for $z_s=0.5$, and to $\lesssim 1\%$ for $z_s=1.0,\,1.5$. Considering the main contribution to the lensing integral arising near $\chi=\chi_s/2$, one expects a residual of $\lesssim 1\%$ at $z_s=0.5$ for $\ell\simeq 1000$, corresponding to $1\kunit$ at $z\sim 0.25$. However, because the projection mixes significant small-scale suppression from lower redshifts, the residual bias becomes larger. Nevertheless, it still represents a reduction by a factor of $\sim 5$ compared to the uncorrected case. 

In Fig.~\ref{fig:lensing_power_as_model}, we further present the ratio of the angular power spectrum $\hat{C}_\ell$ using $P=\hat{\mT}^{-2}P_{\rm DMO}$ to $C_\ell^{\kappa\kappa}$ using $P=P_{mm}$ (solid lines). By definition, $C_\ell^{\kappa\kappa}$ represents the \textit{true} lensing power spectrum including baryonic effects, while $\hat{C}_\ell$ is the \textit{model} prediction with the $\hat\mT$ correction. This quantifies the residual systematics when using relation Eq.~(\ref{equ:T}) as a baryonic correction to the DMO model. The results are similar to those in Fig.~\ref{fig:lensing_power_ratio}. In the TNG300-1 scenario, the residual bias is reduced to $\lesssim 1\%$ at $\ell< 3000$ and to $\lesssim 2\%$ at $\ell< 10000$. In the Illustris-1 scenario, for $\ell < 1000$, it is reduced to $\lesssim 5\%$ at $z_s =0.25$, to $\lesssim 3\%$ at $z_s =0.5$ and to $\lesssim 1\%$ at $z_s=1.0,\, 1.5$. 
Compared with the pure DMO predictions $\hat{C}_\ell \equiv C^{\rm DMO}_\ell$ (dashed lines), the systematic bias is always reduced by a factor of $2\sim 5$ over $\ell<10000$. Therefore, even fully accounting for the mixing of very small-scale Fourier modes, the baryonic correction with $\hat\mT$ is accurate in the mild feedback scenario, and yields a significant improvement in the aggressive feedback scenario. 

Nevertheless, since residual systematics at high-$\ell$ arise from small-scale Fourier modes that are not fully recovered by $\hat\mT$, they can be mitigated by marginalizing over their contributions while retaining the constraining power from larger scales (e.g., unbiased recovery of $k\lesssim 1\,\kunit$). This can be straightforwardly implemented within forward-modeling analyses of weak lensing statistics, using techniques such as lensing counterterms \cite{derose2025lensing}, as the dependencies on scale $k$ and redshift $z$ are separable on the modeling side. The optimal implementation depends on the specific trade-off between statistical uncertainty and systematic bias (e.g., Ref.~\cite{sanchez2026dark}) and remains beyond the scope of this work.

\section{\color{black} Compression of the baryonic transfer function}

In practical analyses, one can exploit the smoothness of $\hat\mT$ in logarithm-$k$ abscissa and the limiting behavior $\mT\rightarrow 1$ as $k\rightarrow 0$. Consequently, $\hat\mT$ has only a few degrees of freedom. This enables a compression of the measured $\hat\mT$ with many $\ell$ bins into a small set of parameters, thereby improving the signal-to-noise ratio by incorporating this prior knowledge. 
To verify this, we reconstruct $\hat\mT$ using cubic-spline defined by several support points $(k_p, \mT_p)$ over $k_p\in[0.1, 10] \,\kunit$. Specifically, we assume the smoothness of $\hat\mT$ allows it to be approximated by a cubic-spline interpolation with a set of support points, implemented using \texttt{Scipy.interpolate.CubicSpline} in logarithm-$k$ abscissa. The ordinates of the support points are treated as free parameters, determined by minimizing  $\sum_i \left[ \hat\mT (k_i) -\mT_{\rm fit}(k_i) \right]^2 $. To enforce the large-scale limit $\mT\rightarrow 1$, we include an additional fixed support point at $k_p=0.08\,\kunit$ with $\mT_p=1$. 
In Fig.~\ref{fig:Bspline_reconst}, we compare the reconstructed $\mT_{\rm fit}$ with the direct measurement $\hat\mT$. For a reconstruction with 4 support points at $k_p= \{0.32,\, 1.0,\, 3.2,\, 10.0\}\, \kunit$, the deviation is already tightly controlled to $< 0.8\%$ across $k\in [0.1, 10]\kunit$. With 5 support points at $k_p= \{0.25,\, 0.6,\, 1.6,\, 4.0,\, 10.0\}\, \kunit$, the deviation is further reduced to $< 0.2\%$. Although the spline representation is not an optimal compression, it suffices to demonstrate that the transfer function is smooth and can be accurately described by a small number of degrees of freedom. 
Therefore, in the context of cosmological constraints, the implementation of the $\hat\mT$ correction is formally equivalent to a parameterization of baryonic physics, but the \textit{parameters} of $\hat\mT$ \textit{model} are directly measured from observations without requiring explicit baryonic physics modeling. Its constraining power is thus expected to be comparable to that of conventional parameterized approaches. 
This conclusion is also consistent with previous studies (e.g., Ref.~\cite{schneider2015new, arico2020modelling, arico2021bacco}), which show that phenomenological models with a small number of parameters can effectively capture baryonic effects on matter clustering. 



\section{Auxiliary results of simulation measurements}

\begin{figure} 
\begin{minipage}{0.42\textwidth}
\includegraphics[width=\textwidth]{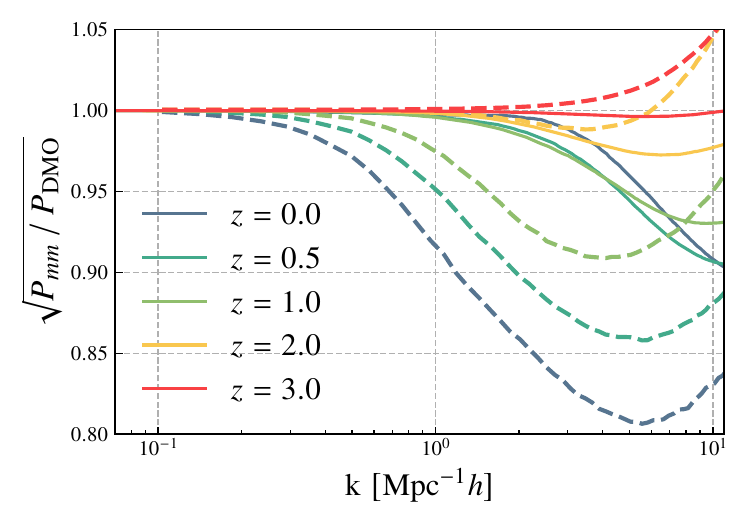}
\end{minipage}  
\begin{minipage}{0.42\textwidth}
\includegraphics[width=\textwidth]{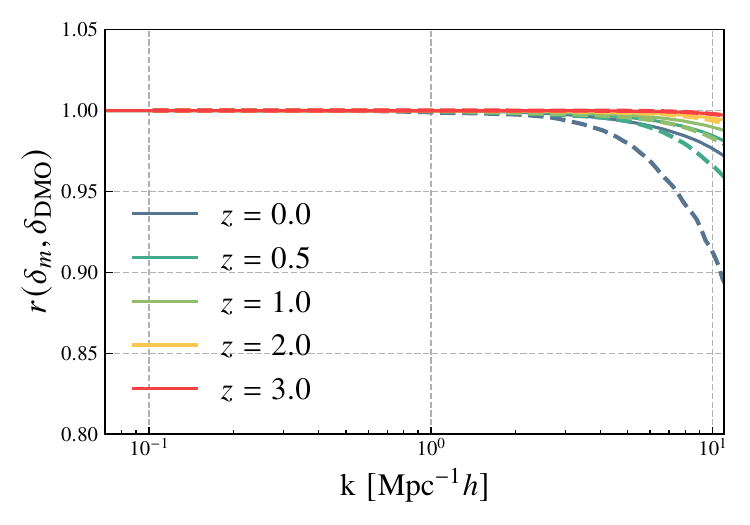}
\end{minipage}
\caption{ \label{fig:SP}
Baryonic suppression (\textit{left} panel) and cross-correlation coefficient (\textit{right} panel) measured in TNG300-1 (\textit{solid} lines) and Illustris-1 (\textit{dashed} lines) simulations. Different colors correspond to different redshifts, covering the range $0<z<3$. 
}
\end{figure}

In Fig.~\ref{fig:SP}, we compare the baryonic suppression $\mT^{-1}(k) = \sqrt{P_{mm}(k)/ P_{\rm DMO}(k)}$ and the cross-correlation coefficients $r(\delta_m, \delta_{\rm DMO})$ measured in simulations. The aggressive AGN feedback employed in the Illustris-1 simulation significantly modifies the matter distribution, resulting in a strong suppression of the matter power spectrum compared to its DMO counterparts. In contrast, the baryonic effects in TNG300-1 are mild. 

\begin{figure}
\includegraphics[width=0.62\columnwidth]{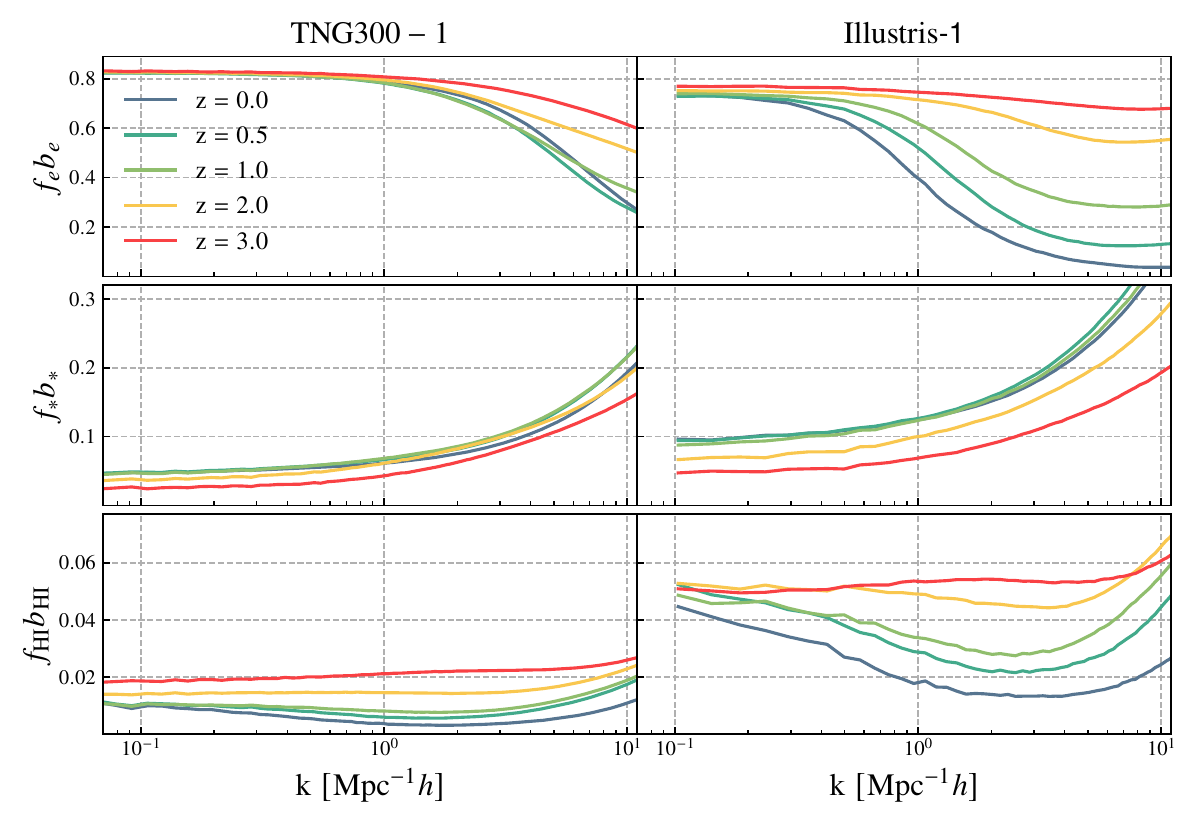}
\caption{ \label{fig:fibi}
Measurements of $f_i b_i$ in hydrodynamical simulation TNG300-1 (\textit{left}) and Illustris-1 (\textit{right}). From top to bottom panels, we show the measurement for free ionized electrons, star+black hole, and neutral hydrogen, respectively. 
Different colors correspond to different redshifts, covering the range $0<z<3$.
}
\end{figure}

\begin{figure}
\includegraphics[width=0.72\columnwidth]{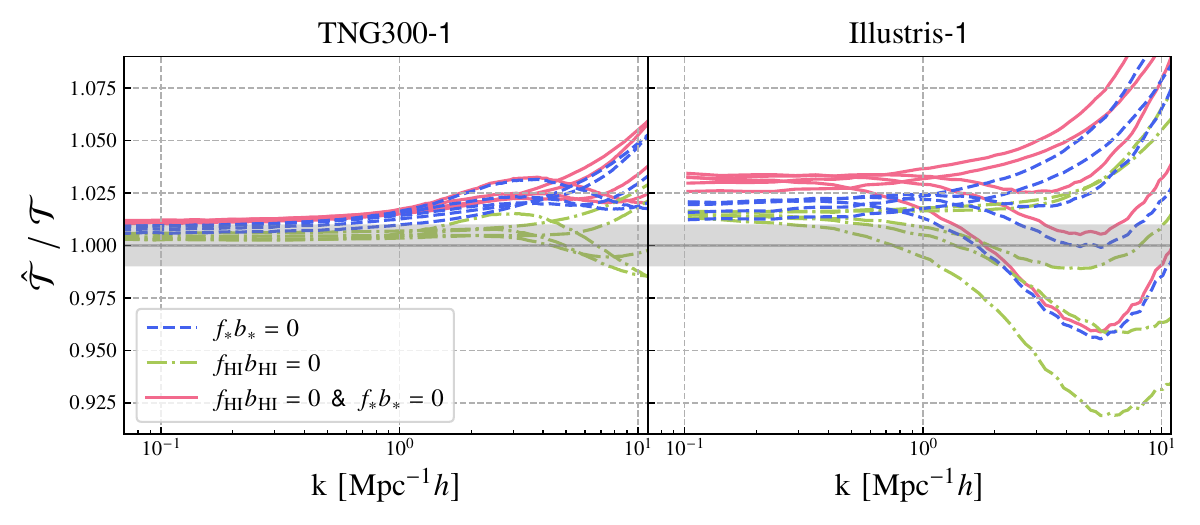}
\caption{ \label{fig:T_null_tracer}
Similar to Fig.~\ref{fig:T-ratio}, but we consider three variants of $\hat\mT$: excluding $f_* b_*$ (blue, dashed), excluding $f_\mHI b_\mHI$ (green, dash-dotted), and excluding both $f_* b_*$ and $f_\mHI b_\mHI$ (red, solid). 
All 5 redshift bins are shown in the same colors. 
}
\end{figure}

In Fig.~\ref{fig:fibi}, we present $\{f_eb_e, f_*b_*, f_\mHI b_\mHI\}$ that are directly measured in TNG300-1 and Illustris-1 simulations. In Fig.~\ref{fig:T_null_tracer}, we present three variants of $\hat\mT$ estimate by excluding $f_*b_*$, $f_\mHI b_\mHI$ or both. These results verify that both stellar content and neutral hydrogen are subdominant contributions compared to ionized electrons, and their contributions to $\mT$ are roughly $\delta\mT_{*} +\delta\mT_{\mHI}  \sim \Omega_b/\Omega_c \left( f_*b_* + (1+\eta) f_\mHI b_\mHI \right) \sim 0.01$. The specific fractions depend on simulation details, with $\sim 1.5\%$ for TNG300-1 and $\sim 3\%$ for Illustris-1. Nevertheless, the systematic shifts of both estimations are non-negligible in forthcoming shear measurement with $<1\%$ precision.

\begin{figure}
\includegraphics[width=0.68\columnwidth]{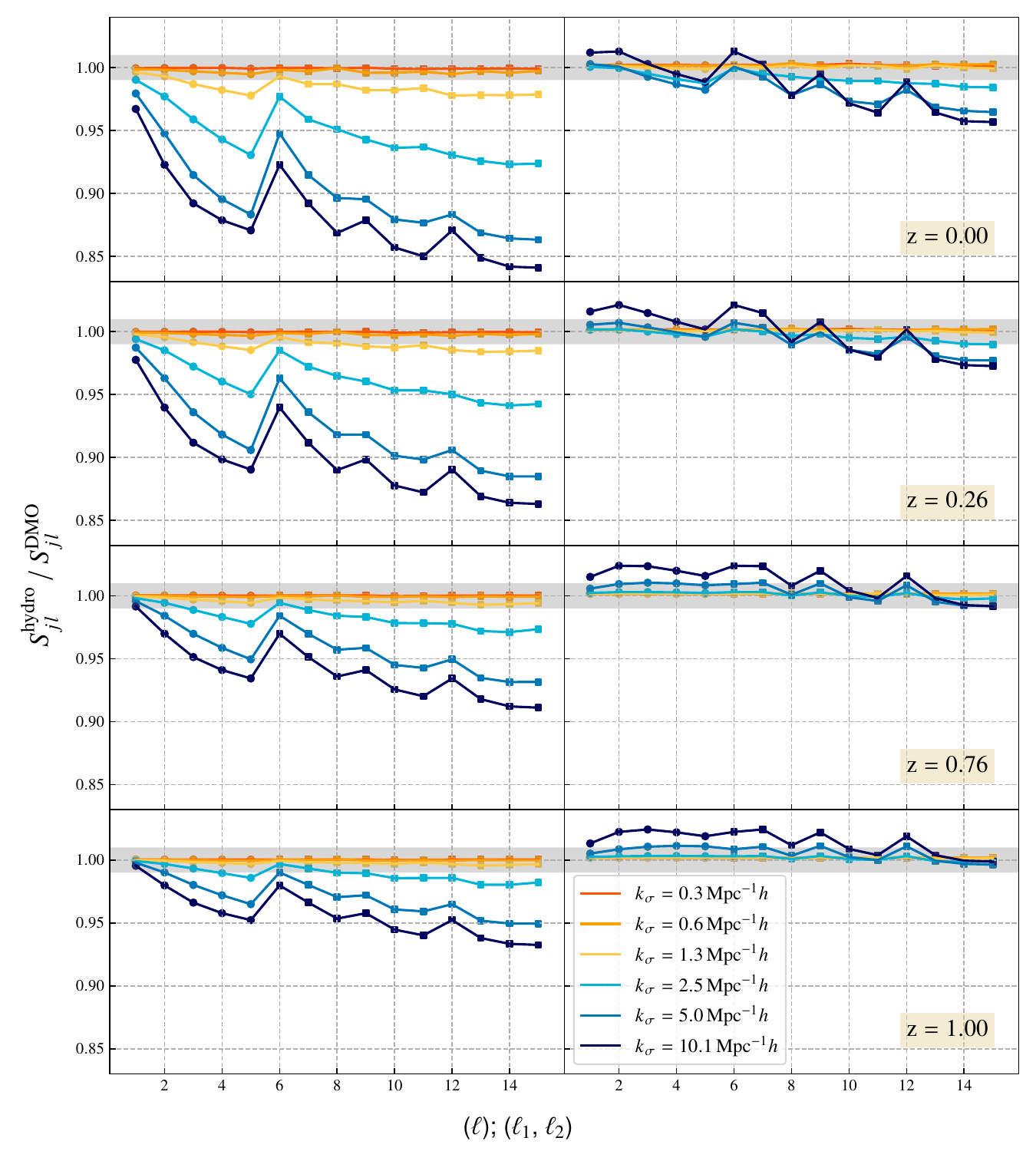}
\caption{ \label{fig:st_coeff_manyz}
Same as Fig.~\ref{fig:st_coeff}, but we present the results at $z=0.0,\, 0.3,\, 0.7,\, 1.0\,$ redshift snapshots in Illustris-1 simulation from top to bottom panels, respectively. 
}
\end{figure}

A direct consequence of assumption ({\bf I}), as also indicated by Eq.~(\ref{equ:def-T}), is that $\mT$ removes baryonic feedback from matter clustering at the field level \cite{sharma2025field}, rather than merely at the power-spectrum level. Consequently, baryonic effects are also eliminated from all summarized statistics. The accuracy of this removal is primarily governed by the validity of Eq.~(\ref{equ:T}), where {\color{black}our results suggest that the accuracy holds for $k\lesssim 1\kunit$. }
In the main text, we take the scattering transform coefficient as an example. The scattering transform of the input field $I_0(\bfx)$ is defined as
\begin{align*}
I_{jl}(\bfx) &=  |I_0 *\psi_{jl}|(\bfx)   \;, \\
I_{j_1l_1;j_2l_2}(\bfx) &=  |I_{j_1l_1} *\psi_{j_2l_2}|(\bfx)   \;, \\
& \cdots  \;,
\end{align*}
where $*$ denotes the convolution operator, and the modulo operator $|...|$ introduces the non-linearity to extract the higher-order correlations. The scattering coefficients of the stochastic field are defined as 
\begin{align*}
S_{0} &= \la I_0 \ra_{\bfx}    \;, \\
S_{jl} &= \la I_{jl} \qquad\;\ra_{\bfx} = \la |I_0 *\psi_{jl}| \ra_{\bfx}     \;, \\
S_{j_1l_1;j_2l_2} &= \la I_{j_1l_1;j_2l_2} \ra_{\bfx} = \la |I_0 *\psi_{j_1l_1}| *\psi_{j_2l_2} | \ra_{\bfx}   \;, \\
& \cdots   \;,
\end{align*}
where $\la\cdots\ra_\bfx$ denotes averaging over all pixels in real space. For simplicity, we present the results of $S_{jl}$ and $S_{j_1l_1;j_2l_2}$, and do not consider the mean of the input field $S_0$ and the higher order coefficients. 
We present the scattering coefficients for Illustris-1 simulation at $z=0.5$ in Fig.~\ref{fig:st_coeff}, and show additional snapshots at {\color{black}$z=0.0,\, 0.26,\, 0.76,\, 1.0$} in Fig.~\ref{fig:st_coeff_manyz}. 
Across $0<z<1$, baryonic feedback induces a suppression in the scattering coefficients, which is effectively removed by the transfer function $\hat\mT$. The effectiveness of this correction depends on the $\hat\mT$ accuracy. Because $\hat\mT$ achieves higher accuracy at high redshift $z=1$ than at low redshift $z=0$ (Fig.~\ref{fig:T-ratio}), the scattering coefficients of the reconstructed field $\hat\delta_{\rm DMO}(\bfx) = \hat\mT* \delta_m(\bfx)$ match the DMO counterpart within $\sim 1\%$ down to resolution $k_\sigma\simeq 5\kunit$ at $z=1$, compared to $k_\sigma\simeq 2.5\kunit$ at $z=0$. These results reinforce the conclusions in the main text.

\begin{figure}
\includegraphics[width=0.68\columnwidth]{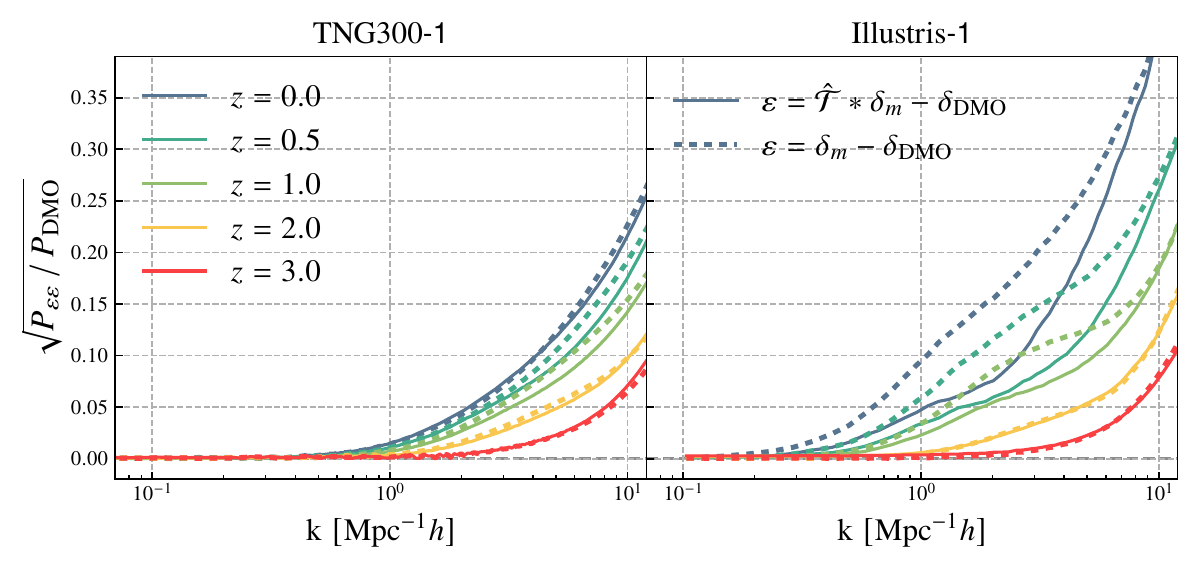}
\caption{ \label{fig:stochastic}
\color{black}The ratio of the residual field power spectrum to the DMO power spectrum. The dashed lines denote the residual field defined as the difference between the matter field and the DMO field, i.e., $\varepsilon = \delta_{m} -\delta_{\rm DMO}$. The solid lines indicate the residual field defined as the difference between the matter field with baryonic correction and the DMO field, i.e., $\varepsilon= \hat\mT*\delta_{m} -\delta_{\rm DMO}$. 
}
\end{figure}

{\color{black}
In addition to higher-order statistics, we measure the power spectrum of the residual field $\varepsilon=\hat\mT*\delta_m-\delta_{\rm DMO}$ and plot $\sqrt{ P_{\varepsilon\varepsilon}/P_{\rm DMO} }$ in Fig.~\ref{fig:stochastic}. For small amplitude and phase shifts, $\alpha_T\equiv1-\hat\mT/\mT\ll1$ and $\alpha_r\equiv1-r(\delta_m,\delta_{\rm DMO})\ll1$, one finds $\sqrt{ P_{\varepsilon\varepsilon}/P_{\rm DMO} } = \sqrt{ 2\alpha_r - 2\alpha_r\alpha_T + \alpha_T^2 } \simeq \sqrt{2}\,\sqrt\alpha_r\, $, so this ratio effectively quantifies the post-correction phase mismatch. For example, for $z=0$ in Illustris-1 (Fig.~\ref{fig:SP}), the phase deviation is as small as $\alpha_r\simeq 0.1\%$ at $k=1\,\kunit$. This corresponds to a residual field amplitude of $\sqrt{ P_{\varepsilon\varepsilon}/P_{\rm DMO} } \simeq 0.05$, in agreement with Fig.~\ref{fig:stochastic}. 
We also plot the ratio for the residual field without baryonic correction, i.e., $\varepsilon = \delta_m - \delta_{\rm DMO}$. For TNG300-1 at $k\lesssim 1\kunit$, the mild baryonic feedback induces a mild suppression, so the residual field is dominated by the intrinsic stochasticity between $\delta_m$ and $\delta_{\rm DMO}$. The amplitude correction with $\hat\mT$ leads to a slight reduction of the residual. In contrast, for Illustris-1, the strong baryonic feedback significantly suppresses the matter field, so the amplitude correction reduces the residual substantially, particularly at $z\lesssim 1$. 
}


\clearpage
\section{\color{black} Auxiliary results of simulation measurements: the massive neutrino and the FLAMINGO suite}

{

In the main text, we have assumed that the matter clustering consists of dark matter and baryons, allowing us to convert the hypothetical DMO field into observables under assumption ({\bf II}). However, massive neutrinos also cluster on large scales and contribute to the total matter field $\delta_m$ observed through weak gravitational lensing. Here, $\delta_m \equiv (\Omega_c\delta_c + \Omega_b\delta_b + \Omega_\nu\delta_\nu )/\Omega_m$ includes the massive neutrino, where $\Omega_\nu \simeq \sum m_\nu / (93.14 h^2\,{\rm eV}) \sim 10^{-3}$ is the neutrino density fraction. Its DMO counterpart is accordingly modified as $\delta_{\rm DMO} \equiv (\Omega_{cb}\delta_{{\rm DMO},c} + \Omega_\nu\delta_{{\rm DMO},\nu} )/\Omega_m$, where $\Omega_{cb} = \Omega_c + \Omega_b$. 
Thus, assumption ({{\bf II}}) is generalized to state that \textit{the $\delta_{{\rm DMO}}$ field can be approximated by the dark species overdensity field, $\delta_{{c\nu}} \equiv (\Omega_c\delta_c  + \Omega_\nu\delta_\nu )/\Omega_{{c\nu}}$, where $\Omega_{{c\nu}} = \Omega_{{c}} + \Omega_{{\nu}}$}. It leads to
\begin{equation}\label{equ:DMO-c-nu}
\delta_{\rm DMO} \simeq \delta_{c\nu} = (\Omega_m\delta_m - \Omega_b\delta_b)/\Omega_{c\nu}  \;.
\end{equation} 
The estimator of the transfer function becomes
\begin{equation} \label{equ:T-nu}
\hat\mT(k) =  {{\Omega_m\over\Omega_{{c\nu}}}} - {{\Omega_b\over\Omega_{{c\nu}}}}
\left[  {{ 1+\eta \over 1+ {{1\over 2}}\eta }} f_e b_e + f_*b_* + (1+\eta) f_\mHI b_\mHI  \right]    \,,
\end{equation}
which is the same as Eq.~(\ref{equ:T}) except for a correction accounting for the mass fraction of dark species, $\Omega_c \rightarrow \Omega_{{c\nu}}$.
An alternative approach is to assume that the dark matter clustering is approximately unaffected by baryonic feedbacks, i.e., $\delta_{{\rm DMO},c} \simeq \delta_c$, and further neglect the neutrino clustering, i.e., $\delta_{{\rm DMO},\nu}\simeq 0$ and $\delta_\nu\simeq 0$, since neutrinos do not cluster below the free-streaming scale $\lambda_{\rm fs} \sim 100 \,{\rm Mpc}/h$. The transfer function resulted from this alternative approximation, denoted as $\hat\mT_c$, only differs from Eq.~(\ref{equ:T-nu}) by an overall factor, i.e., $\hat\mT_c \,/\,\hat{\mT} = (1+\Omega_\nu/\Omega_c) \,/\, (1+\Omega_\nu/\Omega_{cb}) \simeq 1 + \mathcal{O}\left(10^{-3}\right)$. Thus, our baryonic correction is insensitive to the approximation strategy of massive neutrinos. 
We verify this extension to massive neutrinos of Eq.~(\ref{equ:DMO-c-nu}) and (\ref{equ:T-nu}), as well as Eq.~(\ref{equ:def-T}), using hydrodynamical simulations from the FLAMINGO suite.

FLAMINGO is a suite of large-volume hydrodynamical simulations designed for large-scale structure and galaxy cluster surveys \cite{schaye2023flamingo, kugel2023flamingo, helly2026flamingo}, with variants differing in subgrid implementation, cosmology, resolution, and box size. In this work, we use the intermediate-resolution simulations and their DMO counterparts, each with a comoving volume of $1\, {{\rm Gpc}}^3$. The hydrodynamical simulation contains $1800^3$ dark matter particles, $1800^3$ baryon particles, and $1000^3$ neutrino particles, where gas particles are converted into stars and black holes during the late evolution stages. The corresponding DMO simulation has the same number of cold dark matter and neutrino particles.
The non-linear evolution of massive neutrinos is simulated using the $\delta f$ method \cite{elbers2021optimal}, which tracks phase-space perturbations to reduce shot noise. For the baseline cosmology, there is a single massive neutrino species with a mass of $0.06\,{{\rm eV}}$ and two massless species under the normal ordering.
The subgrid physics is calibrated to match the observed galaxy stellar mass function and cluster gas mass fractions at low redshift, while accounting for observational uncertainties \cite{kugel2023flamingo}. Here, the gas fraction in groups and clusters largely indicates the baryonic suppression of matter clustering \cite{van2024contribution, van2026resummation}. AGN feedback is implemented either as thermal winds or kinetic jets, with the baseline model adopting the former. 
The cosmological parameter of the baseline cosmology are taken from the maximum posterior likelihood values from the Dark Energy Survey Year three \cite{abbott2022dark}. Apart from the fiducial simulation, labeled as \texttt{L1\_m9}, we describe the variants used in this work as follows:
\begin{itemize}
\item[-] The \texttt{fgas$\pm$Nsigma} simulations are calibrated analogously to the fiducial model, but the observed cluster gas fractions are shifted up and down by $\pm {{\rm N}}\sigma$, where $\sigma$ is the estimated observational error.
\item[-] The \texttt{M$_*$-1sigma} simulations are similar variants by shifting the observed galaxy stellar mass function. 
\item[-] The \texttt{Jet} simulations implement AGN feedback with jet-like injection. 
\item[-] The \texttt{Planck} simulation adopts the best-fit $\Lambda$CDM cosmology from Planck 2018 results \cite{aghanim2020planck}, adopting the neutrino mass of $\sum m_\nu = 0.06\,{{\rm eV}}$.
\item[-] The \texttt{PlanckNu0p24Var}, \texttt{PlanckNu0p24Fix}, and \texttt{PlanckNu0p48Fix} simulations are variants with higher neutrino masses for the Planck cosmology, with $\sum m_\nu = 0.24\,{{\rm eV}},\, 0.24\,{{\rm eV}},\, 0.48\,{{\rm eV}}$, respectively. For \texttt{PlanckNu0p24Var}, other cosmological parameters are derived from the Planck 2018 likelihood, while in \texttt{PlanckNu0p24Fix} and \texttt{PlanckNu0p48Fix}, they are fixed to the Planck 2018 best-fit values.
\item[-] The \texttt{LS8} simulation adopts a lower power spectrum amplitude with $S_8=0.766$, which is adjusted to better match galaxy clustering and cosmic shear measurements compared to the Planck cosmology. 
\item[-] The \texttt{PlanckDCDM12} and \texttt{PlanckDCDM24} simulations \cite{elbers2025flamingo} are the extensive cosmology of $\Lambda$CDM with decaying cold dark matter. The cold dark matter particles are unstable and decay into dark radiation with decay rates of $\Gamma=12\, {{\rm km}}/s/{{\rm Mpc}}$ and $\Gamma=24\, {{\rm km}}/s/{{\rm Mpc}}$, respectively.
\item[-] The \texttt{NoCooling} simulation \cite{mccarthy2025flamingo} disables radiative cooling, star formation, and feedback. Heating sources, including adiabatic compression, shocks, and photoionization by the UV background radiation, are still included. 
\end{itemize}

In Fig.~\ref{fig:FLAMINGO_suppression}, we present the baryonic suppression of matter clustering at redshifts $z=0.0,\, 0.5,\, 1.0$ in the FLAMINGO simulations. The results span a wide range of baryonic effects, from the zero-feedback scenario in \texttt{NoCooling} to the strong feedback scenario in \texttt{fgas-8sigma}. Compared to the aggressive feedback implementation in Illustris-1 (Fig.~\ref{fig:SP}), the baryonic suppression in FLAMINGO is generally mild at $k\lesssim 1\,\kunit$ and only becomes significant at $k\gtrsim 1\,\kunit$. All suppression strengths do not exceed that of Illustris-1 (see also \cite{schaller2025flamingo}). Although only the strong feedback variants are preferred by current observational constraints (e.g., \cite{bigwood2025kinetic, siegel2025suppression}), we validate our method across all feedback variants, including the unrealistic \texttt{NoCooling} scenario, to provide a comprehensive test. 

In Fig.~\ref{fig:FLAMINGO_r_DMO_m}, we present the cross-correlation coefficient between the total matter field $\delta_m$ and the DMO field $\delta_{{\rm DMO}}$ to validate assumption ({{\bf I}}). The results show that assumption ({{\bf I}}) is accurate at $k\lesssim 1\,\kunit$ and the deviation remains below the $1\%$ level at $k\lesssim 5\,\kunit$, regardless of the choice of baryonic feedbacks and cosmological models. This is consistent with the results shown in Fig.~\ref{fig:T-r2} and the previous study \cite{sharma2025field}.

In Figs.~\ref{fig:FLAMINGO_r_DMO_cnu} and \ref{fig:FLAMINGO_T_DMO_cnu}, we validate assumption ({{\bf II}}) by presenting the cross-correlation coefficients and amplitude ratios between the DMO field $\delta_{{\rm DMO}}$ and the dark species overdensity field $\delta_{{c\nu}}$. Similar to Fig.~\ref{fig:FLAMINGO_r_DMO_m}, the phase of $\delta_{{c\nu}}$ is tightly correlated with $\delta_{{\rm DMO}}$, and the deviations remain below $1-r(\delta_{{\rm DMO}}, \delta_{{\rm c\nu}}) < 1\%$ at $k\lesssim 5\,\kunit$. For the amplitude ratio, the back-reaction of baryonic feedback on the dark species produces varying levels of deviation. However, these deviations are well below $1\%$ for $k< 1\,\kunit$ and remain below $\lesssim 2\%$ up to $k=5\,\kunit$.
The only exception is the \texttt{NoCooling} simulation, where dark matter clustering is monotonically enhanced on small scales compared to its DMO counterpart. This enhancement is caused by shocks generated during mergers, where the kinetic energy of dark matter is transferred to diffuse hot gas, causing dark matter to sink toward the potential well \cite{lin2006influence,jing2006influence}. The absence of feedback allows the gas to closely trace the dark matter halo, exacerbating the energy exchange and thus the back-reaction on dark matter. 
Although the \texttt{NoCooling} picture is incomplete and unrealistic, as cooling and feedback mechanisms also play important roles, it nevertheless presents an extreme test for the validity of assumption ({{\bf II}}) at $k\lesssim 1\,\kunit$.

The measurement of $\hat\mT$ demands the cross-correlations between various baryon tracers and the total matter field. 
The available products of the FLAMINGO suite contain multiple components: dark matter, gas, stars, black holes, and neutrinos. Therefore, apart from neutral gas, all overdensity fields are directly measured. To estimate the neutral hydrogen distribution, we assume that the gas components consist exclusively of hydrogen and helium, since heavier elements comprise $\lesssim 0.5\%$ in the simulations. The neutral hydrogen overdensity is estimated as
$ \widehat{f_{\rm HI}\delta}_{\rm HI} = {1\over 1+\eta}\left( 
    f_{\rm gas} \delta_{\rm gas} - {1+\eta\over 1+\eta/2} f_e\delta_e
\right) $, 
where $f_e\delta_e$ is obtained by inverting the relation 
$ n_e - \bar{n}_e = (1+z)^3\,  {3H_0^2\over 8\pi G} {\Omega_b\over m_p} f_e\delta_e $. 
Here, the physical number density of ionized electrons $n_e$, the gas component overdensity $\delta_{\rm gas}$, and the gas mass fraction $f_{\rm gas}$ are direct products of the simulation. 
Notice that this estimation leads to 
$ \hat{b}_b \equiv  
{ 1+\eta \over 1+ {1\over 2}\eta } f_e b_e + f_*b_* + (1+\eta) \widehat{f_{\rm HI}b}_{\rm HI} 
= f_{\rm gas} b_{\rm gas} + f_*b_* $, 
where the baryon bias combination $\hat{b}_b$ reduces to two terms: the gas component and stellar content. We retain all terms $\{f_eb_e,\,  f_*b_*,\, f_{\rm HI}b_{\rm HI}\}$ in the expression since they are all direct observables. 

To measure the clustering bias of baryon tracers, we use subhalo samples from the SOAP catalog, which is based on the structure finder HBT-HERONS \cite{han2018hbtPlus, forouhar2025assessing, mcgibbon2025soap}. We identify all luminous subhalos as galaxies, with number density $\bar{n}_g\sim 0.03\, h^3{\rm Mpc}^{-3}$, and measure the clustering bias as $b_i(k) \equiv P_{\rm gi}(k)/P_{\rm gm}(k)$. For the \texttt{NoCooling} simulation, no star formation occurs and therefore no galaxies exist; for the decaying dark matter simulations, no structure finder products are available. In these cases, we use the combined field $\delta_{cb} \equiv (\Omega_c\delta_c + \Omega_b\delta_b  )/\Omega_{cb}$ as a proxy for galaxies, since $\delta_{cb}$ is a good tracer of galaxy clustering \cite{villaescusa2014cosmology, castorina2014cosmology}. We emphasize that under the validity of assumption ({\bf I}), the estimated $\hat\mT$ is independent of the redshift-tomography tracer choice. Although the matter field begins to decorrelate from the DMO field at scales $k>1\,\kunit$ and the stochastic component of the biased tracer may affect results (Fig.~\ref{fig:FLAMINGO_r_DMO_m}), the resulting deviation is expected to be $\lesssim 1\%$ even up to $k\sim 5\,\kunit$.

In Fig.~\ref{fig:FLAMINGO_T_precision}, we present the ratio of the estimated transfer function $\hat\mT$ to the ground truth $\mT = \sqrt{P_{\rm DMO} /P_{mm}}$. 
As discussed in the main text, the primary driver of deviations is the violation of assumption ({{\bf II}}), which accounts for the $1\% \sim 2\%$ discrepancy in FLAMINGO results. Particularly in the \texttt{NoCooling} scenario, $\hat\mT$ overestimates the true value by $\sim 3\%$ around $k\sim 5\,\kunit$ because of the severe baryonic back-reaction. 
Nevertheless, the overall accuracy is consistent with validation results of both TNG300-1 and Illustris-1 simulations (Fig.~\ref{fig:T-ratio}). The estimator $\hat\mT$ achieves accuracy better than $1\%$ at $k\lesssim 1\,\kunit$ across all feedback and cosmological variants. At small scales, the deviations remain below $\lesssim 3\%$ for all scales shown here, excluding only the unrealistic \texttt{NoCooling} scenario.

\clearpage
\begin{figure}
\includegraphics[width=0.98\columnwidth]{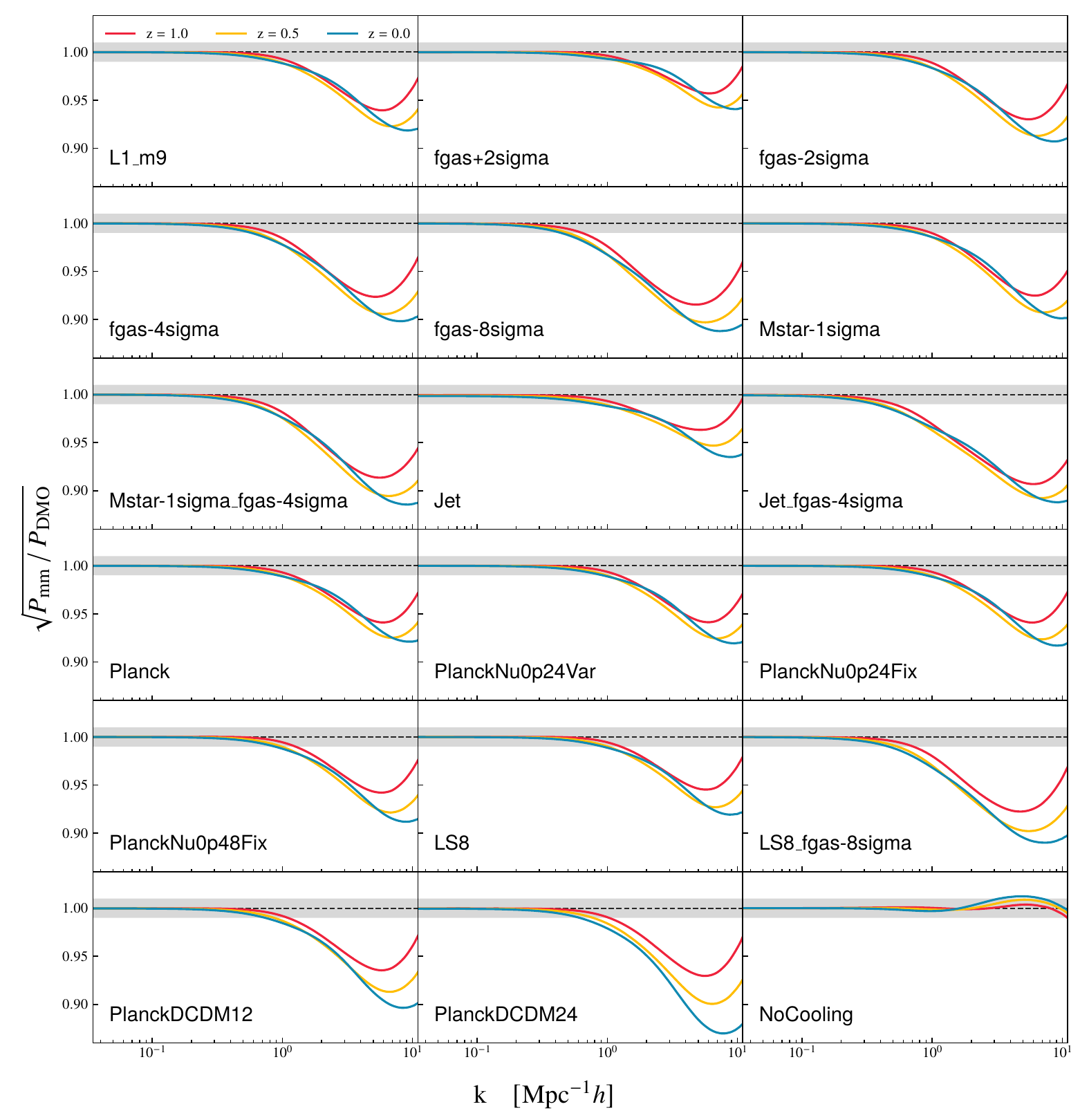}
\caption{ \label{fig:FLAMINGO_suppression}
\color{black}Validation in the FLAMINGO simulations: baryonic suppression of matter clustering across a wide range of subgrid implementations and cosmological models.
Each panel corresponds to a simulation variant. The red, yellow and blue curves indicate redshifts $z=0.0,\,0.5$ and $1.0$ respectively. The gray shaded region indicates the $\pm 1\%$ level.
Implementation details for each variant are described in the main text and cited references.
}
\end{figure}

\begin{figure}
\includegraphics[width=0.98\columnwidth]{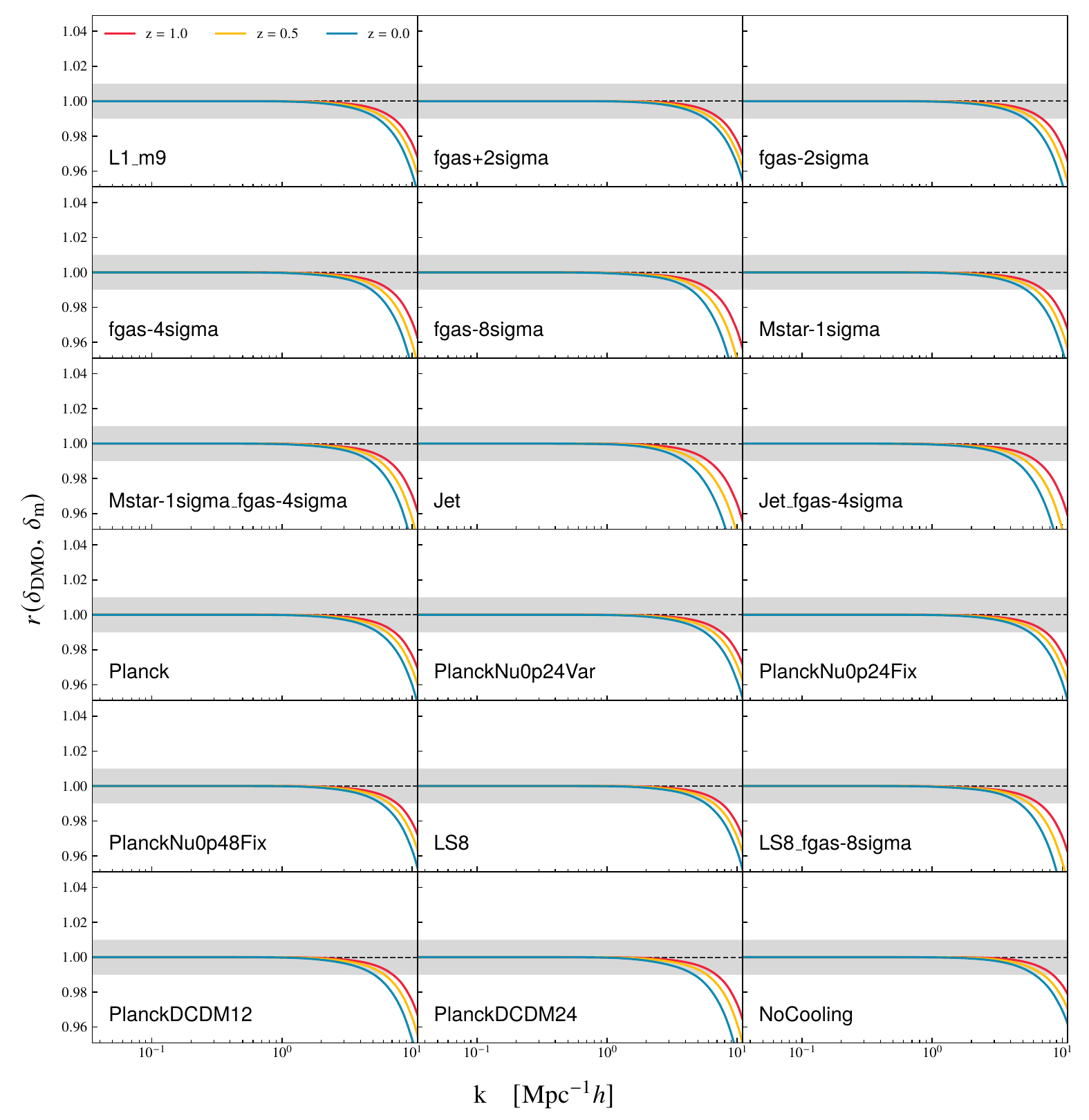}
\caption{ \label{fig:FLAMINGO_r_DMO_m}
\color{black}Validation in the FLAMINGO simulations: cross-correlation coefficient between the total matter field $\delta_m \equiv (\Omega_c\delta_c + \Omega_b\delta_b + \Omega_\nu\delta_\nu )/\Omega_m$ and the DMO field $\delta_{\rm DMO} \equiv (\Omega_{cb}\delta_{{\rm DMO},c} + \Omega_\nu\delta_{{\rm DMO},\nu} )/\Omega_m$. 
Each panel corresponds to a simulation variant. The red, yellow and blue curves indicate redshifts $z=0.0,\,0.5$ and $1.0$ respectively. The gray shaded region indicates the $\pm 1\%$ level.
The results show that assumption ({\bf I}) is accurate at $k\lesssim 1\,\kunit$, and deviations remain below $1\%$ at $k\lesssim 5\,\kunit$ regardless of baryon physics or cosmology.
}
\end{figure}

\begin{figure}
\includegraphics[width=0.98\columnwidth]{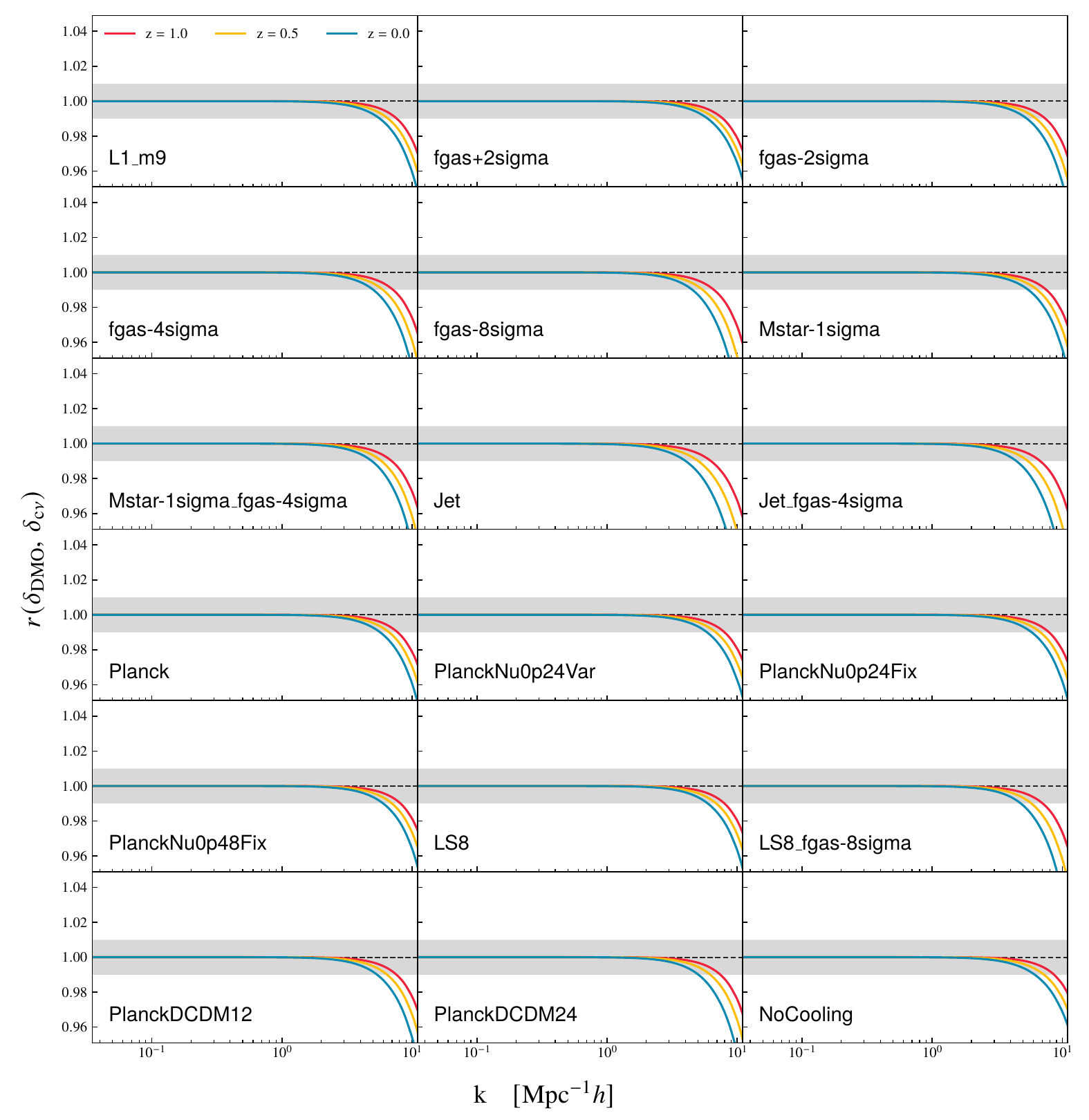}
\caption{ \label{fig:FLAMINGO_r_DMO_cnu}
\color{black}Validation in the FLAMINGO simulations: cross-correlation coefficient between the dark-species overdensity $\delta_{c\nu} \equiv (\Omega_c\delta_c + \Omega_\nu\delta_\nu)/\Omega_m$ and the DMO overdensity $\delta_{\rm DMO}$.
Same figure settings as Fig.~\ref{fig:FLAMINGO_r_DMO_m}.
For assumption ({\bf II}) of Eq.~\ref{equ:DMO-c-nu}, the phase between $\delta_{c\nu}$ and $\delta_{\rm DMO}$ is within $1-r < 1\%$ at $k\lesssim 5\,\kunit$.
}
\end{figure}

\begin{figure}
\includegraphics[width=0.98\columnwidth]{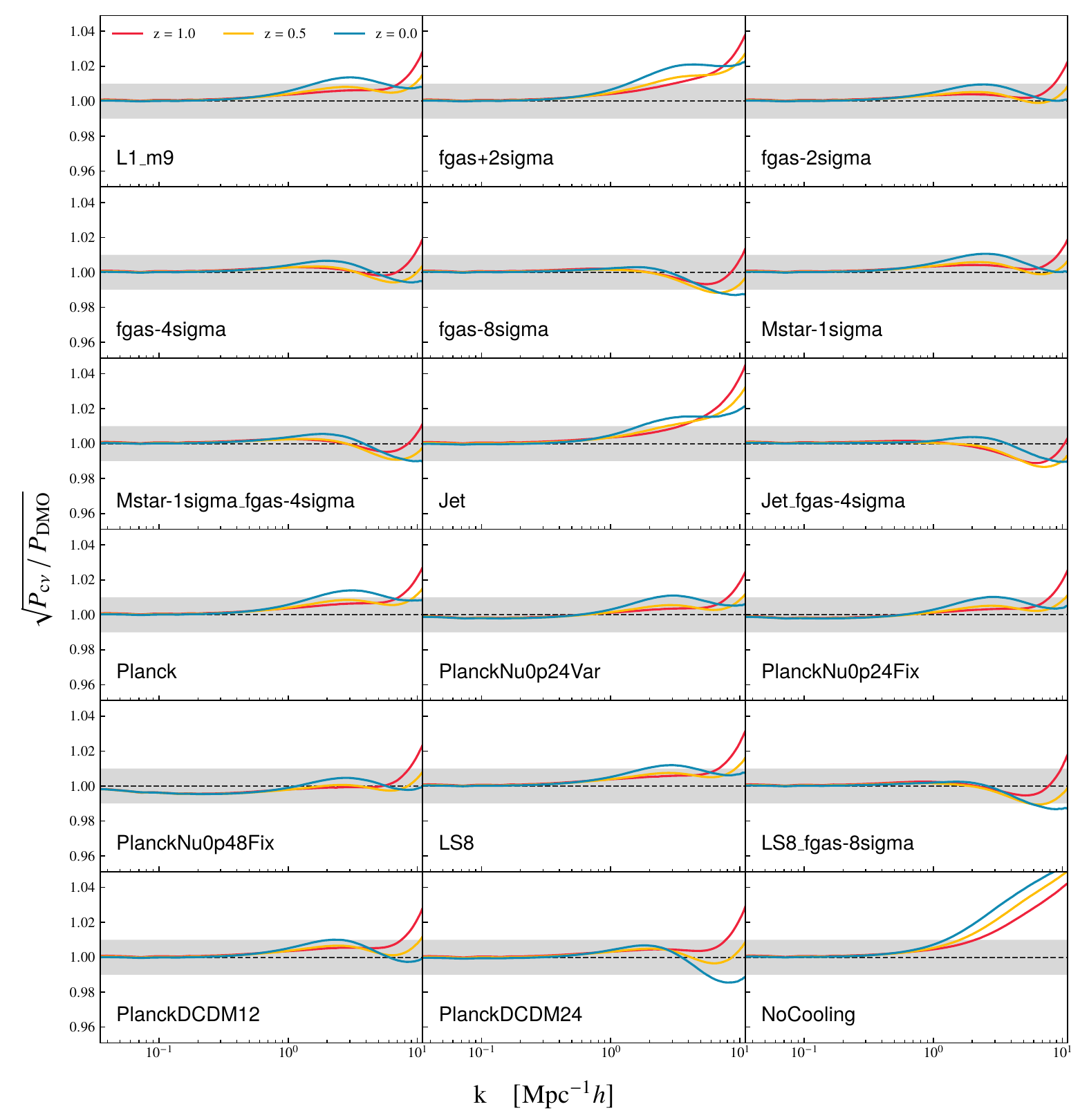}
\caption{ \label{fig:FLAMINGO_T_DMO_cnu}
\color{black}Validation in the FLAMINGO simulations: ratio of the field amplitudes between the dark-species overdensity $\delta_{c\nu} \equiv (\Omega_c\delta_c + \Omega_\nu\delta_\nu)/\Omega_m$ and the DMO overdensity $\delta_{\rm DMO}$.
Same figure settings as Fig.~\ref{fig:FLAMINGO_r_DMO_m}.
For assumption ({\bf II}) of Eq.~\ref{equ:DMO-c-nu}, the amplitude deviation between $\delta_{c\nu}$ and $\delta_{\rm DMO}$ is well below $1\%$ at $k\lesssim 1\,\kunit$, and remains $\lesssim 2\%$ at $k\lesssim 5\,\kunit$, except for the unrealistic \texttt{NoCooling} scenario.
In the \texttt{NoCooling} scenario, the non-radiative hot gas closely traces the dark matter and gains kinetic energy transferred from dark matter during halo mergers, which substantially alters the dark matter distribution. This scenario provides a lower bound on the impact of merging shocks on the validity of assumption ({\bf II}).
}
\end{figure}

\begin{figure}
\includegraphics[width=0.98\columnwidth]{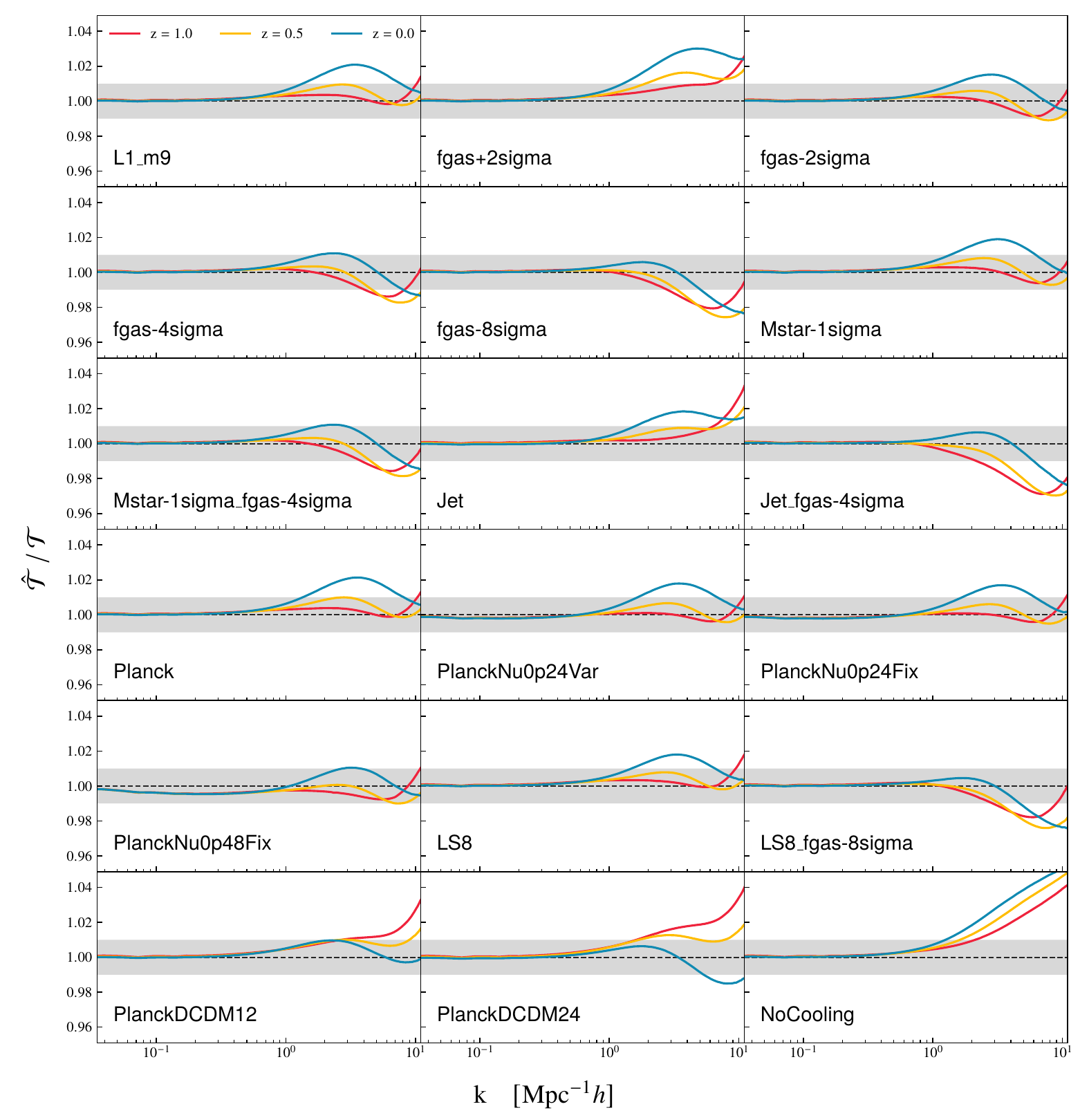}
\caption{ \label{fig:FLAMINGO_T_precision}
\color{black}Validation in the FLAMINGO simulations: ratio of the estimated transfer function $\hat\mT$ to the ground truth $\mT = \sqrt{P_{\rm DMO}/P_{mm}}$.
Same figure settings as Fig.~\ref{fig:FLAMINGO_r_DMO_m}.
The estimator $\hat\mT$ attains accuracy better than $1\%$ at $k\lesssim 1\,\kunit$ across all feedback and cosmological variants, and deviations remain below $\sim 3\%$ at $k\lesssim 10\,\kunit$, except in the unrealistic \texttt{NoCooling} scenario.
As $k\gtrsim 1\kunit$, the primary source of the $1\%$--$2\%$ discrepancy in $\hat\mT$ is violation of assumption ({\bf II}). The remaining $\lesssim 1\%$ deviation likely arises from the violation of assumption ({\bf I}), where stochasticity in the biased tracers (i.e., galaxy) plays a non-negligible role on small scales in $f_ib_i$ estimation.
Overall, the accuracy is consistent with validation results of TNG300-1 and Illustris-1 simulations (Fig.~\ref{fig:T-ratio}).
}
\end{figure}

}



\end{document}